\documentclass{emulateapj}
\usepackage{apjfonts}
\newcommand{\tq}{$t_{Q}$}
\newcommand{\tQ}{t_{Q}}
\newcommand{\etal}{et al.}
\newcommand{\NH}{$N_{\rm H}$}
\newcommand{\NHI}{$N_{\rm H\,I}$}
\newcommand{\nh}{N_{\rm H}}
\newcommand{\nhi}{N_{\rm H\,I}}
\newcommand{\Mdot}{\dot{M}}
\newcommand{\Lbol}{L_{\rm bol}}
\newcommand{\dEdt}{\epsilon_r \Mdot c^{2}}
\newcommand{\LB}{L_{B}}

\newcommand{\Lm}{L_{\rm min}}
\newcommand{\LBo}{L_{B,{\rm obs}}}
\newcommand{\LBm}{L_{B,{\rm min}}}

\newcommand{\Lcut}[1]{10^{#1}\,L_{\sun}}
\newcommand{\EV}[1]{\langle #1 \rangle}
\newcommand{\dlgL}{{\rm d}\log(L)}
\newcommand{\eEdd}{l}
\newcommand{\Lp}{L_{\rm peak}}

\newcommand{\meanNH}{\bar{N}_{H}}
\newcommand{\sigNH}{\sigma_{N_{H}}}
\newcommand{\dtdL}{{\rm d}t/{\rm d}\log{L}}
\newcommand{\fdtdL}{\frac{{\rm d}t}{{\rm d}\log{L}}}

\newcommand{\vvir}{V_{\rm vir}}

\newcommand{\qeos}{q_{\rm EOS}}
\newcommand{\fgas}{f_{\rm gas}}
\newcommand{\zgal}{z_{\rm gal}}

\newcommand{\mbhf}{M^{f}_{\rm BH}}
\newcommand{\dphi}{{\rm d}\Phi/{\rm d}\log L}
\newcommand{\fdphi}{\frac{{\rm d}\Phi}{{\rm d}\log L}}
\newcommand{\nLP}{\dot{n}(\Lp)}
\newcommand{\nLp}{\nLP}
\newcommand{\nstar}{\dot{n}_{\ast}}
\newcommand{\lstar}{L_{\ast}}
\newcommand{\sstar}{\sigma_{\ast}}
\newcommand{\rhobh}{\rho_{\rm BH}}
\newcommand{\lessim}{\lesssim}
\newcommand{\reducechi}{\chi^{2}/\nu}

\shorttitle{Quasar Origins \&\ Evolution}
\shortauthors{Hopkins \etal}
\slugcomment{Submitted to ApJ, June 14, 2005}
\begin{document}

\title{A Unified, Merger-Driven Model for the Origin of Starbursts, Quasars,
the Cosmic X-Ray Background, Supermassive Black Holes and Galaxy Spheroids}
% Relation Between Supermassive Black Holes and Spheroids in Galaxies}
\author{Philip F. Hopkins\altaffilmark{1}, 
Lars Hernquist\altaffilmark{1}, 
Thomas J. Cox\altaffilmark{1}, 
Tiziana Di Matteo\altaffilmark{2}, 
Brant Robertson\altaffilmark{1}, 
Volker Springel\altaffilmark{3}}
\altaffiltext{1}{Harvard-Smithsonian Center for Astrophysics, 
60 Garden Street, Cambridge, MA 02138, USA}
\altaffiltext{2}{Carnegie Mellon University, 
Department of Physics, 5000 Forbes Ave., Pittsburgh, PA 15213}
\altaffiltext{3}{Max-Planck-Institut f\"{u}r Astrophysik, 
Karl-Schwarzchild-Stra\ss e 1, 85740 Garching bei M\"{u}nchen, Germany}

\begin{abstract}

We develop an evolutionary model for starbursts, quasars, and
spheroidal galaxies in which supermassive black holes play a dominant
role.  In this picture, mergers between gas-rich galaxies drive
nuclear inflows of gas, producing intense starbursts and feeding the
growth of supermassive black holes.  During this phase, the black hole
is heavily obscured (a ``buried'' quasar), but feedback energy from
its growth expels the gas, rendering the black hole briefly visible as
a bright, optical source (a ``visible'' quasar), and eventually
halting accretion (a ``dead'' quasar).  The self-regulated growth of
the black hole accounts for the observed correlation between black
hole mass and stellar velocity dispersion in spheroidal galaxies.  We
show that the quasar lifetime and obscuring column density depend on
both the instantaneous and peak luminosities of the quasar, and
determine this dependence using a large set of simulations of galaxy
mergers varying the host galaxy properties, orbital geometry, and gas
physics.

We use our fits to the lifetime and column density to deconvolve
observed quasar luminosity functions and obtain the evolution of the
formation rate of quasars with a certain peak luminosity, $\dot
n(L_{\rm peak}, z)$.  In our model, quasars spend extended periods of
time at luminosities well below their peaks, and so $\dot n(L_{\rm
peak}, z)$ has a maximum, falling off at both brighter and fainter
luminosities, corresponding to the ``break'' in the observed quasar
luminosity function.  We obtain self-consistent fits to hard and soft
X-ray and optical quasar luminosity functions for a model in which
$\dot n(L_{\rm peak}, z)$ varies with redshift according to pure peak
luminosity evolution.  From this form for $\dot n(L_{\rm peak}, z)$,
and our simulation results for the luminosity dependence of the quasar
lifetime and obscuring column, we are able to reproduce many
observable quantities, including: the column density distribution of
both optical and X-ray selected quasar samples, the luminosity
function of broad-line quasars in X-ray samples and the broad-line
(Type I, Type II) fraction as a function of luminosity, the mass
function of active black holes, the observed distribution of Eddington
ratios at both low and high redshift, the present-day mass function of
relic, inactive supermassive black holes and total black hole mass
density, and the spectrum of the cosmic X-ray background.  In each
case, our predictions agree well with observations, matching them to
higher precision than previous tunable models for quasar lifetimes and
obscuration similarly fit to the luminosity function.  We provide a
library of Monte Carlo realizations of our modeling for comparison
with a wide range of observations, using various selection criteria.

\end{abstract}

\keywords{quasars: general --- galaxies: nuclei --- galaxies: active --- 
galaxies: evolution --- cosmology: theory}

\section{Introduction}
\label{sec:intro}

The measurement of anisotropies in the cosmic microwave background
(e.g.\ Spergel et al.\ 2003) combined with observations of high
redshift supernovae (e.g.\ Riess et al.\ 1998, 2000; Perlmutter et
al.\ 1999) have established a ``standard model'' for the Universe, in
which the energy density is dominated by an unknown form driving
accelerated cosmic expansion, and most of the mass is non-baryonic, in
a ratio of roughly 5:1 to ordinary matter.  On small scales, it is
believed that structure formed through gravitational instability.  In
the currently favored cold dark matter (CDM) paradigm, objects grow
hierarchically, with smaller ones forming first and then merging into
successively larger bodies.  As baryons fall into dark matter
potential wells, the gas is shocked and then cools radiatively to form
stars and galaxies, in a ``bottom-up'' progression (White \& Rees
1978).

Even with the many successes of this picture, the processes underlying
galaxy formation and evolution are poorly understood.  For example,
there has yet to be an ab initio calculation, starting from an initial
state prescribed by the standard model, resulting in a population of
objects that reproduces observed galaxies.  However, from the
same initial conditions, computer simulations have yielded a new,
successful interpretation of the Lyman-alpha forest in which
absorption in caused by density fluctuations in the intergalactic
medium (e.g.\ Cen et al.\ 1994; Zhang et al.\ 1995; Hernquist et
al.\ 1996), over many orders of magnitude in column density (e.g.\ Katz
et al.\ 1996a), explicitly related to growth of structure in a CDM
universe (e.g.\ Croft et al.\ 1998, 1999, 2002; McDonald et al.\ 2000,
2004; Hui et al.\ 2001; Viel et al.\ 2003, 2004).  This suggests that
the difficulties with understanding galaxy formation and evolution lie
not in the initial conditions or with the description of dark matter,
but rather with the physics that has been used to model the baryons.

Observations have revealed regularities in the structure of galaxies
that point to some of this ``missing'' physics.  Supermassive black
holes appear to reside at the centers of most galaxies (e.g.\ Kormendy
\& Richstone 1995; Richstone et al.\ 1998; Kormendy \& Gebhardt 2001)
and the masses of these black holes are correlated with either the
mass (Magorrian et al.\ 1998; McLure \& Dunlop 2002; Marconi \& Hunt
2003) or the velocity dispersion (i.e.\ the $M_{\rm BH}$-$\sigma$
relation: Ferrarese \& Merritt 2000; Gebhardt et al.\ 2000; Tremaine et
al.\ 2002) of spheroids, demonstrating a direct link between the origin
of galaxies and supermassive black holes.  Simulations which follow
the self-regulated growth of black holes in galaxy mergers (Di Matteo
et al.\ 2005; Springel et al.\ 2005a) have shown that the energy
released through this process has a global impact on the structure of
the merger remnant.  If this conclusion applies to spheroid formation
in general, the simulations demonstrate that models for the origin and
evolution of galaxies must account for black hole growth and feedback
in a fully {\it self-consistent} manner.

Analytical and semi-analytical 
modeling \citep{SR98,Fabian99,WL02,WL03,BN05} suggests
that, beyond a certain threshold, feedback energy from black holes can
expel gas from the centers of galaxies, shutting down accretion onto
them and limiting their masses.  However, these calculations usually
ignore the impact of this process on star formation and therefore do
not explain the link between black hole growth and spheroid formation,
and furthermore make simplifying assumptions about the dynamics of such accretion.  For
example, the duration of black hole growth is a free parameter, which
is fixed either using observational estimates or assumed to be similar
to e.g.\ the dynamical time of the host galaxy or the $e$-folding time
for Eddington-limited black hole growth $t_{S}=M_{\rm
BH}/\Mdot=4.5\times10^{7}\,l^{-1}\,(\epsilon_r/0.1)\,$yr for accretion
with radiative efficiency $\epsilon_r=L/\Mdot c^{2}\sim0.1$ and
$l=L/L_{\rm Edd}\lesssim1$ \citep{Salpeter64}.  Moreover, these
studies have adopted idealized models for quasar light curves, usually
corresponding to growth at a constant Eddington ratio or
on-off, ``light bulb,'' scenarios.  As we discuss below, less
restrictive modeling suggests that this phase is actually more
complex.

Efforts to model quasar accretion and feedback more self-consistently
\citep[e.g.,][]{CO97,CO01,Granato04} by treating the hydrodynamical
response of gas to black hole growth have generally been restricted to
idealized geometries, such as spherical symmetry, employing simple
models for star formation and galaxy-scale quasar fueling.  However,
these works have made it possible to estimate duty cycles of quasars
and shown that the objects left behind have characteristics similar to
those observed, with quasar feedback being a critical element in
reproducing these features (e.g.\ Sazonov et al.\ 2005; Kawata \&
Gibson 2005; Cirasuolo et al.\ 2005; for a review, see Ostriker \&
Ciotti 2005).

\citet{SDH05b} have incorporated black hole growth and feedback into
simulations of galaxy mergers and included a multiphase model for star
formation and pressurization of the interstellar gas by supernovae
\citep{SH03} to examine implications of these processes for galaxy
formation and evolution.  Di Matteo et al.\ (2005) and Springel et
al.\ (2005a,b) have shown that gas inflows excited by gravitational
torques during a merger both trigger starbursts and fuel rapid black
hole growth.  The growth of the black hole is determined by the gas
supply and terminates as gas is expelled by feedback, halting
accretion, leaving a dead quasar in an ordinary galaxy.  The
self-regulated nature of black hole growth in mergers explains
observed correlations between black hole mass and properties of normal
galaxies \citep{DSH05}, as well as the color distribution of
ellipticals \citep{SDH05a}.  These results lend support to the view
that mergers have played an important role in structuring galaxies, as
advocated especially by Toomre \& Toomre (1972) and Toomre (1977).
(For reviews, see, e.g., Barnes \& Hernquist 1992; Barnes 1998;
Schweizer 1998.)

Subsequent analysis by Hopkins et al.\ (2005a,b,c,d) has shown that the
merger simulations can account for quasar phenomena as a phase of
black hole growth.  Unlike what has been assumed in e.g.\
semi-analytical studies of quasars, the simulations predict
complicated evolution for quasar lifetimes, fueling rates for black
hole accretion, obscuration, and quasar light curves.  The light
curves were studied by \citet{H05a,H05b}, who showed that the
self-termination process gives observable lifetimes $\sim10^{7}\,$yr
for bright optical quasars and predicts a large population of obscured
sources as a natural stage of quasar evolution, as implied by
observations (for a review, see Brandt \& Hasinger 2005).
\citet{H05b} analyzed simulations over a range of galaxy masses and
found that the quasar light curves and lifetimes are always
qualitatively similar, with both the intrinsic and observed quasar
lifetimes being decreasing functions of luminosity, with longer
lifetimes at all luminosities for higher-mass (higher peak luminosity)
systems.  The dependence of the lifetime on luminosity led
\citet{H05c} to suggest a new interpretation of the quasar luminosity
function, in which the steep bright-end consists of quasars radiating
near the Eddington limit and is directly related to the distribution
of intrinsic peak luminosities (or final black hole masses) as has
been assumed previously
\citep[e.g.,][]{SB92,HL98,HM00,KH00,Somerville01,
Tully02,WL03,V03,HQB04,Croton05}, but where the
shallow, faint-end of the luminosity function describes black holes
growing towards or declining from peak phases of quasar activity, with
Eddington ratios generally between $l \sim0.01$ and 1.  The ``break''
in the luminosity function corresponds directly to the {\em peak} in
the distribution of intrinsic quasar properties.  As argued by
\citet{H05c,H05d} this new interpretation of the luminosity function
can self-consistently explain various properties of both the quasar
and galaxy populations, connecting the origin of galaxy spheroids,
supermassive black holes, and quasars.

%\clearpage
\begin{figure}
    \centering
    \plotone{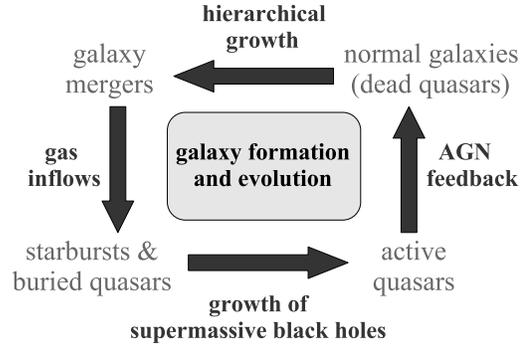}
    \caption{Schematic representation of a ``cosmic cycle'' for
    galaxy formation and evolution regulated by black hole growth in
    mergers.
    \label{fig:cosmiccycle}}
\end{figure}
%\clearpage

Motivated by these results, and earlier work by many others which we
summarize below, in this paper we consider a picture for galaxy
formation and evolution, illustrated schematically as a ``cosmic
cycle'' in Figure~\ref{fig:cosmiccycle}, in which starbursts, quasars,
and the simultaneous formation of spheroids and supermassive black
holes represent connected phases in the lives of galaxies.  Mergers
are expected to occur regularly in a hierarchical universe,
particularly at high redshifts.  Those between gas-rich galaxies drive
nuclear inflows of gas, triggering starbursts and fueling the growth
of supermassive black holes.  During most of this phase, quasar
activity is obscured, but once a black hole dominates the energetics
of the central region, feedback expels gas and dust, making the black
hole visible briefly as a bright quasar.  Eventually, as the gas is
further heated and expelled, quasar activity can no longer be
maintained and the merger remnant relaxes to a normal galaxy with a
spheroid and a supermassive black hole.  In some cases, depending on
the gas content of the progenitors, the remnant may also have a disk
(Springel \& Hernquist 2005; Robertson et al.\ 2005a).  The remnant
will then evolve passively and would be available as a seed to repeat
the above cycle.  As the Universe evolves and more gas is consumed,
the mergers involving gas-rich galaxies will shift towards lower
masses, explaining the decline in the population of the brightest
quasars from $z\sim 2$ to the present, and the remnants that are
gas-poor will redden quickly owing to the termination of star
formation by black hole feedback (Springel et al.\ 2005a), so that they
resemble elliptical galaxies, surrounded by hot X-ray emitting halos
(e.g.\ Cox et al.\ 2005).

There is considerable observational support for this scenario, which
has led the development of this picture for the co-evolution of
galaxies and quasars over recent decades.  Infrared (IR) luminous
galaxies are thought to be powered in part by starbursts (e.g.\ Soifer
et al.\ 1984a,b; Sanders et al.\ 1986, 1988a,b; for a review, see e.g.\
Soifer et al.\ 1987), and the most intense examples locally,
ultraluminous infrared galaxies (ULIRGs), are invariably associated
with mergers (e.g.\ Allen et al.\ 1985; Joseph \& Wright 1985; Armus et
al.\ 1987; Kleinmann et al.\ 1988; Melnick \& Mirabel 1990; for reviews,
see Sanders \& Mirabel 1996 and Jogee 2004).  Radio observations show
that ULIRGs have large, central concentrations of dense gas (e.g.\
Scoville et al.\ 1986; Sargent et al.\ 1987, 1989), providing a fuel
supply to feed black hole growth.  Indeed, some ULIRGs have ``warm''
IR spectral energy distributions (SEDs), suggesting that they harbor
buried quasars (e.g.\ Sanders et al.\ 1988c), an interpretation
strengthened by X-ray observations demonstrating the presence of two
non-thermal point sources in NGC6240 \citep{Komossa03}, which are
thought to be supermassive black holes that are heavily obscured at
visual wavelengths (e.g.\ Gerssen et al.\ 2004; Max et al.\ 2005,
Alexander et al.\ 2005a,b).  These lines of evidence, together with the
overlap between bolometric luminosities of ULIRGs and quasars,
indicate that quasars are the descendents of an infrared luminous
phase of galaxy evolution caused by mergers (Sanders et al.\ 1988a), an
interpretation supported by observations of quasar hosts (e.g.\
Stockton 1978; Heckman et al.\ 1984; Stockton \& MacKenty 1987;
Stockton \& Ridgway 1991; Hutchings \& Neff 1992; Bahcall et al.\ 1994,
1995, 1997; Canalizo \& Stockton 2001).

However, many of the physical processes that connect the phases of
evolution in Figure~\ref{fig:cosmiccycle} are not well understood.
Early simulations showed that mergers produce objects resembling
galaxy spheroids (e.g.\ Barnes 1988, 1992; Hernquist 1992, 1993a) and
that if the progenitors are gas-rich, gravitational torques funnel gas
to the center of the remnant (e.g.\ Barnes \& Hernquist 1991, 1996),
producing a starburst (e.g.\ Mihos \& Hernquist 1996), but these works
did not explore the relationship of these events to black hole growth
and quasar activity.  While a combination of arguments based on time
variability and energetics suggests that quasars are produced by the
accretion of gas onto supermassive black holes in the centers of
galaxies (e.g.\ Salpeter 1964; Zel'dovich \& Novikov 1964; Lynden-Bell
1969), the mechanism that provides the trigger to fuel quasars
therefore remains uncertain.  Furthermore, there have been no
comprehensive models that describe the transition between ULIRGs and
quasars that can simultaneously account for observed correlations like
the $M_{\rm BH}$-$\sigma$ relation.

Here, we study these relationships using numerical simulations of
galaxy mergers that account for the consequences of black hole growth.
In our simulations, black holes accrete and grow throughout a merger
event, producing complex, time-varying quasar activity.  Quasars reach
a peak luminosity, $\Lp$, during the ``blowout'' phase of evolution
where feedback energy from black hole growth begins to drive away the
gas, eventually slowing accretion.  Prior to and following this brief
period of peak activity, quasars radiate at instantaneous
luminosities, $L$, with $L < \Lp$.  However, we show that even with
this complex behavior, the global characteristics that determine the
observed properties of quasars, i.e.\ lifetimes, light curves, and
obscuration, can be expressed as functions of $L$ and $\Lp$,
allowing us to make predictions for quasar populations that agree well
with observations, supporting the scenario sketched in
Figure~\ref{fig:cosmiccycle}.

In \S~\ref{sec:methods}, we discuss our methodology and show how the
quasar lifetimes and obscuration from our simulations can be expressed
as functions of the instantaneous and peak luminosities of quasars.
We also define a set of commonly adopted models for the
quasar lifetime and obscuration against which we compare our
predictions throughout. In \S~\ref{sec:LF}, we apply our models to the
quasar luminosity function, using the observed luminosity function to
determine the distribution of quasar peak luminosities, and show that
this allows us to simultaneously reproduce the hard X-ray, soft X-ray,
and optical quasar luminosity functions at all redshifts $z\lesssim3$,
and the distribution of column densities in both optical and X-ray
samples. In \S~\ref{sec:BLqso}, we determine the time in our
simulations when quasars will be observable as broad-line objects, and
use this to predict the broad-line luminosity function and fraction of
broad-line objects in quasar samples, as a function of luminosity, as
well as the mass function of low-redshift, active broad-line
quasars. In \S~\ref{sec:eddington}, we estimate the distribution of
Eddington ratios in our simulations as a function of luminosity, and
infer Eddington ratios in observed samples at different redshifts.  In
\S~\ref{sec:smbh}, we use our modeling to predict both the mass
distribution and total density of present-day relic supermassive black
holes, and describe their evolution with redshift.  In \S~\ref{sec:xrb},
we similarly apply this model to predict the integrated cosmic X-ray
background spectrum, accounting for the observed spectrum from
$\sim1-100$\,keV.  In \S~\ref{sec:discussion}, we discuss the
primary qualitative implications of our results and propose
falsifiable tests of our picture.  Finally, in \S~\ref{sec:finis},
we conclude and suggest directions for future work.

Throughout, we adopt a $\Omega_{\rm M}=0.3$, $\Omega_{\Lambda}=0.7$,
$H_{0}=70\,{\rm km\,s^{-1}\,Mpc^{-1}}$ ($h=0.7$) cosmology.

\section{The Model: Methodology}
\label{sec:methods}

Our model of quasar evolution has several elements, which we summarize
here and describe in greater detail below.

\begin{itemize}

\item In what follows, a ``quasar'' is taken to mean the course of
black hole activity in a {\em single} merger event. We use the term
``quasar lifetime'' to refer to the time spent by such a quasar at a
given luminosity or fraction of the quasar peak luminosity, integrated
over all black hole activity in a single merger event. This is not
meant to suggest that this would constitute the entire accretion
history of a black hole -- a given black hole may have multiple
``lifetimes'' triggered by different mergers, with each merger in
principle fueling a distinct ``quasar'' with its own lifetime.  There
is no a priori luminosity threshold for quasar activity -- the time
history can include various epochs at low luminosities and accretion
rates.

\item We model the galaxy mergers using hydrodynamical simulations,
varying the orbital parameters of the encounter, the internal
properties of the merging galaxies, prescriptions for the gas physics,
initial ``seed'' black hole masses of the merging systems, and
numerical resolution of the simulations.  The black hole accretion
rate is determined from the surrounding gas (smoothed over the scale
of our spatial resolution, reaching $20\,$pc in the best cases), i.e.\
the density and sound speed of the gas, and its motion relative to the
black hole, using Eddington-limited, Bondi-Hoyle-Lyttleton accretion
theory.  The black hole radiates with a canonical efficiency
$\epsilon_{r}=0.1$ corresponding to a standard \citet{SS73} thin disk,
and we assume that $\sim5\%$ of this radiated luminosity is deposited
as thermal energy into the surrounding gas, weighted by the SPH
smoothing kernel (which has a $\sim r^{-2}$ profile) over the scale of
the spatial resolution. This scale is such that we cannot resolve the
complex accretion flow immediately around the black hole, but we adopt
this prescription because: (1) it reproduces the observed slope and
normalization in the $M_{\rm BH}-\sigma$ relation (Di Matteo et al.\
2005), (2) it follows from observations, based on estimates of the
energy contained in highly-absorbed UV portion of the quasar SED
\citep[e.g.,][]{Elvis94,Telfer02}, (3) it follows from theoretical
considerations of momentum coupling to dust grains in the dense gas
very near the quasar \citep{Murray05} and hydrodynamical simulations
of small-scale radiative heating from quasar accretion \citep{CO01},
and (4) even if the feedback is initially highly collimated, a driven
wind or shock in a dense region such as the center of the merging
galaxies will rapidly isotropize, so long as it is decelerated by
gravity and the surrounding medium, allowing the high sound speed
within the shock to equalize angle-dependent pressure variations
\citep[e.g.,][]{KM90}, and furthermore initial local distortions will
be washed away in favor of triaxial structure determined by the
large-scale density gradients \citep{BKS91}, as occurs in our
simulations.

\item For each of our merger simulations, we compute the bolometric
black hole luminosity and column density along $\sim1000$ lines of
sight to the black hole(s) (evenly spaced in solid angle), as a
function of time from the beginning of the simulation until the system
has relaxed for $\sim1\,$Gyr after the merger.

\item We bin different merger simulations by $\Lp$, the peak
bolometric luminosity of the black hole in the simulation, and the
conditional distributions of luminosity, $P(L|\Lp)$, and column
density, $P(\nh | L,\,\Lp)$, are computed using all simulations that
fall into a given bin in $\Lp$. The final black hole mass (black hole
mass at the end of the individual merger -- subsequent mergers and
quasar episodes could further increase the black hole mass) is
approximately $M_{\rm BH}^{f}\approx M_{\rm Edd}(\Lp)$ (but not
exactly, see \S~\ref{sec:detailsCompare}), so we obtain similar
results if we bin instead by $M_{BH}^{f}$. Our calculation of
$M_{BH}^{f}(\Lp)$ allows us to express our conditional distributions
of luminosity and column density in terms of either peak luminosity or
final black hole mass.  Critically, we find that expressed in terms of
$\Lp$ or $M_{\rm BH}^{f}$, there is no systematic dependence in the
quasar evolution on the varied merger simulation properties -- this
allows us to calculate a large number of observables in terms of $\Lp$
or $M_{\rm BH}^{f}$ without the large systematic uncertainties
inherent in attempting to directly estimate e.g.\ quasar light curves
in terms of host galaxy mass, gas fraction, multiphase pressurization
of the interstellar medium, orbital parameters and merger stage, and
other variables.

\item The observed quasar luminosity function is the convolution of
the time a given quasar spends at some observed luminosity with the
rate at which such quasars are created.  Knowing the distributions
$P(L|\Lp)$ and $P(\nh | L,\, \Lp)$, we can calculate the time spent by
a quasar with some $\Lp$ at an observed luminosity in a given
waveband. We use this to fit to observational estimates of the
bolometric quasar luminosity function $\phi(L)$, de-convolving these
quantities to determine the function $\nLp$; i.e.\ the rate at which
quasars of a given peak luminosity must be created or activated
(triggered in mergers) in order to reproduce the observed bolometric
luminosity function.

\item Given these inputs, we determine the joint distribution in
instantaneous luminosity and black hole mass, column density
distribution, peak luminosity and final black hole mass, as a function
of redshift, i.e.\ $n(L,\,L_{\nu},\,M_{\rm BH},\,\nh ,\,\Lp,\,M_{\rm
BH}^{f}\ |\ z)$, at all redshifts where the observed quasar luminosity
function can provide the necessary constraint. From this joint
distribution, we can compute, for example, luminosity functions in
other wavebands, conditional column density distributions, active
black hole mass functions and Eddington ratio distributions, and relic
black hole mass functions and cosmic backgrounds.  We can compare each
of these results to those determined using simpler models for either
the quasar lifetime or column density distributions; in
\S~\ref{sec:altmodels} we describe a canonical set of such
models, to which we compare throughout this paper.

\end{itemize}

\subsection{The Simulations}
\label{sec:sims}

The simulations were performed with {\small GADGET-2}
\citep{Springel2005}, a new version of the parallel TreeSPH code
{\small GADGET} \citep{SYW01}.  {\small GADGET-2} is based on a fully
conservative formulation \citep{SH02} of smoothed particle
hydrodynamics (SPH), which maintains simultaneous energy and entropy
conservation when smoothing lengths evolve adaptively (for a
discussion, see e.g., Hernquist 1993b, O'Shea et al.\ 2005).  Our
simulations account for radiative cooling, heating by a UV background
(as in Katz et al.\ 1996b, Dav\'e et al.\ 1999), and incorporate a
sub-resolution model of a multiphase interstellar medium (ISM) to
describe star formation and supernova feedback \citep{SH03}.  Feedback
from supernovae is captured in this sub-resolution model through an
effective equation of state for star-forming gas, enabling us to
stably evolve disks with arbitrary gas fractions (see, e.g.\ Springel
et al.\ 2005b; Robertson et al.\ 2004).  In order to investigate the
consequences of supernova feedback over a range of conditions, we
employ the scheme of \citet{SDH05b}, introducing a parameter $\qeos$
to interpolate between an isothermal equation of state ($\qeos=0$) and
the full multiphase equation of state ($\qeos=1$) described above.

Supermassive black holes (BHs) are represented by ``sink'' particles
that accrete gas at a rate $\Mdot$ estimated using an
Eddington-limited version of Bondi-Hoyle-Lyttleton accretion theory
(Bondi 1952; Bondi \& Hoyle 1944; Hoyle \& Lyttleton 1939).  The
bolometric luminosity of the black hole is $\Lbol=\dEdt$, where
$\epsilon_r=0.1$ is the radiative efficiency.  We assume that a small
fraction (typically $\approx 5\%$) of $\Lbol$ couples dynamically to
the surrounding gas, and that this feedback is injected into the gas
as thermal energy, as described above.

We have performed several hundred simulations of colliding galaxies,
varying the numerical resolution, the orbit of the encounter, the
masses and structural properties of the merging galaxies, initial gas
fractions, halo concentrations, and the parameters describing star
formation and feedback from supernovae and black hole growth.  This
large set of simulations allows us to investigate merger evolution for
a wide range of galaxy properties and to identify any systematic
dependence of our modeling.  The galaxy models are described in
\citet{SDH05b}, and we briefly review their properties here.

The progenitor galaxies in our simulations have virial
velocities $\vvir=80,$ $113,$ $160,$ $226,$ $320,$ ${\rm and}\ 500\,{\rm
km\,s^{-1}}$. We consider cases with gas equation of state parameters
$\qeos=0.25$ (moderately pressurized, with a mass-weighted temperature
of star-forming gas $\sim10^{4.5} {\rm K}$) and $\qeos=1.0$ (the full,
``stiff'' Springel-Hernquist equation of state, with a mass-weighted
temperature of star-forming gas $\sim10^{5} {\rm K}$), and initial
disk gas fractions (by mass) of $\fgas=0.2,$ $0.4,$ $0.8,$ ${\rm and}\ 1.0$.
Finally, we scale these models with redshift, altering the physical
sizes of the galaxy components and the dark matter halo concentration
in accord with cosmological evolution \citep{Mo1998}.  Details are
provided in \citet{Robertson05b}, and here we consider galaxy models
scaled appropriately to resemble galaxies of the same $\vvir, \fgas,
{\rm and}\ \qeos$ at redshifts $\zgal=0,$ $2,$ $3,$ ${\rm and}\ 6$.

For each simulation, we generate two stable, isolated disk galaxies,
each with an extended dark matter halo with a \citet{Hernquist90}
profile, motivated by cosmological simulations (e.g.\ Navarro et
al.\ 1996; Busha et al.\ 2004) and observations of halo properties
(e.g.\ Rines et al.\ 2002, 2002, 2003, 2004), an exponential disk of
gas and stars, and (optionally) a bulge.  The galaxies have masses
$M_{\rm vir}=V_{\rm vir}^{3}/(10GH_{0})$ for $\zgal=0$, with the
baryonic disk having a mass fraction $m_{\rm d}=0.041$, the bulge
(when present) has a mass fraction $m_{\rm b}=0.0136$, and the rest of
the mass is in dark matter typically with a concentration parameter
$c=9.0$.  The disk scale-length is computed based on an assumed spin
parameter $\lambda=0.033$, chosen to be near the mode in the observed
$\lambda$ distribution \citep{Vitvitska02}, and the scale-length of
the bulge is set to $0.2$ times the resulting value.  In \citet{H05a},
we describe our analysis of simulation A3, one of our set with
$\vvir=160\, {\rm km\,s^{-1}},\ \fgas=1.0,\ \qeos=1.0,\ {\rm and}\
\zgal=0$, a fiducial choice with a rotation curve and mass similar to
the Milky Way, and \citet{H05b,H05c,H05d} used a set of simulations
with the same parameters but varying $\vvir=80,$ $113,$ $160,$ $226,$ ${\rm
and}\ 320\,{\rm km\,s^{-1}}$, which we refer to below as runs
A1, A2, A3, A4, and A5, respectively.

Typically, each galaxy is initially composed of 168000 dark matter
halo particles, 8000 bulge particles (when present), 24000 gas and
24000 stellar disk particles, and one BH particle. We vary the
numerical resolution, with many of our simulations using instead twice
as many particles in each galaxy, and a subset of simulations with up
to 128 times as many particles.  We vary the initial seed mass of the
black hole to identify any systematic dependence of our results on
this choice.  In most cases, we choose the seed mass either in accord
with the observed $M_{\rm BH}$-$\sigma$ relation or to be sufficiently
small that its presence will not have an immediate 
dynamical effect.  Given the
particle numbers employed, the dark matter, gas, and star particles
are all of roughly equal mass, and central cusps in the dark matter
and bulge profiles are reasonably well resolved (see Fig 2. in
Springel et al.\ 2005b). The galaxies are then set to collide from a
zero energy orbit, and we vary the inclinations of the disks and the
pericenter separation.

%\clearpage
\epsscale{0.9}
\begin{figure*}
    \centering
    \plotone{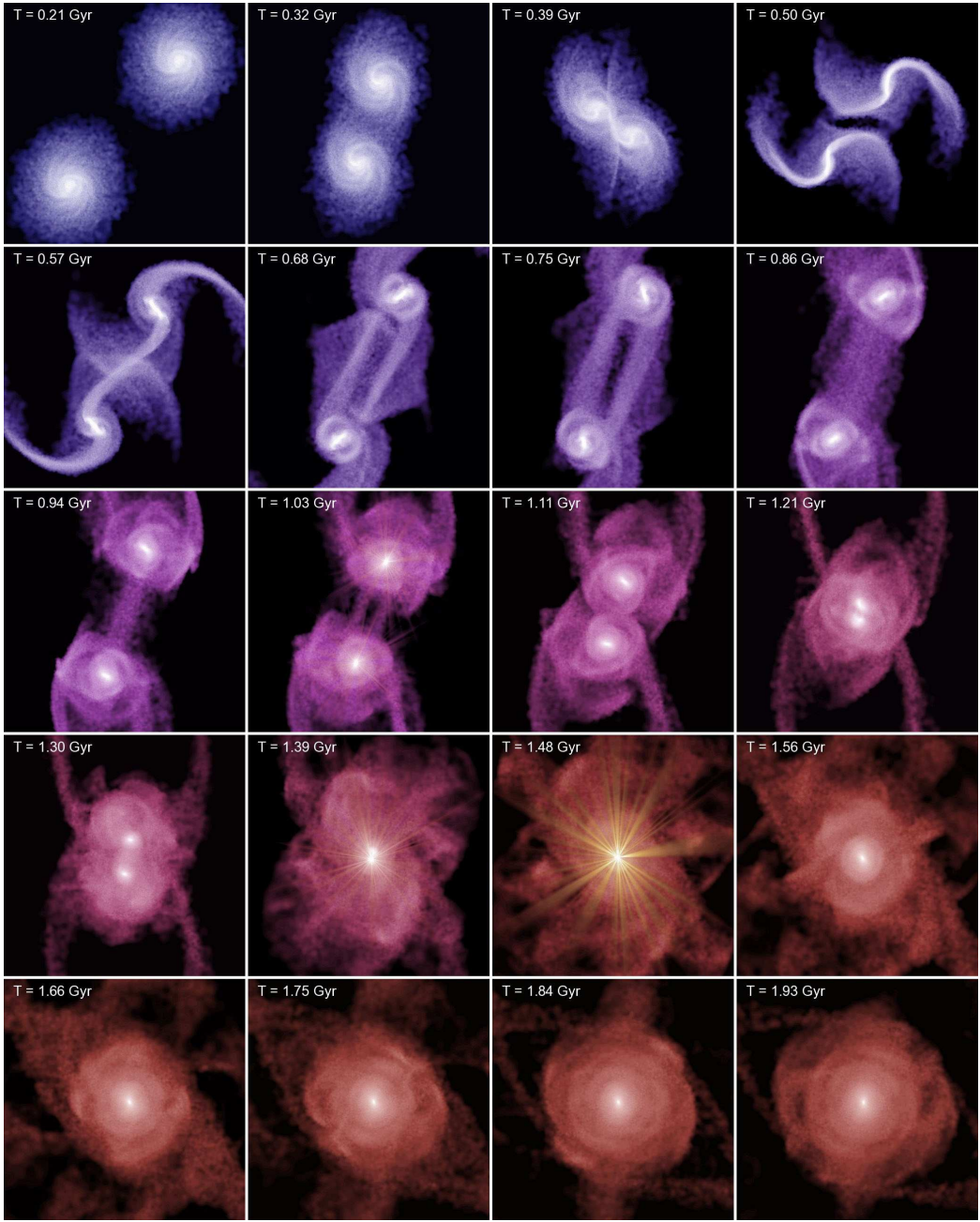}
    \caption{
    Time sequence from one of our merger simulations 
    ($V_{\rm vir}=160\,{\rm km\,s^{-1}}$, 
    initial gas fraction 20\%).  Each panel is $80\,h^{-1}{\rm kpc}$ 
    on a side and shows the simulation time in the 
    upper left corner.  Brightness of individual pixels gives the 
    logarithm of the projected stellar mass density, while color hue indicates
    the baryonic gas fraction, from 20\% (blue) to less than 5\% (red).
    At $T=1.03$, $1.39$ and $1.48\,{\rm Gyr}$, when
    the black hole could be seen as an optical quasar,
    nuclear point sources are shown, providing a representation of the
    relative luminosities of stars and the quasar at these times. 
    \label{fig:sim.example}}
\end{figure*}
\epsscale{1}
%\clearpage

A representative example of the behavior of the simulations is
provided in Figure~\ref{fig:sim.example}, which shows the time
sequence of a merger involving two bulge-less progenitor galaxies with
virial velocities of $160\, {\rm km\,s^{-1}}$ and initial gas
fractions of 20\%.  During the merger, gas is driven to the galaxy
centers by gravitational tides, fueling nuclear starbursts and black
hole growth. The quasar activity is short-lived and peaks twice in
this merger, both during the first encounter and the final coalescence
of the galaxies. To illustrate the bright, optically observable
phase(s) of quasar activity which we identify below, we have added 
nuclear point sources in
the center at the position(s) of the black hole(s) at
times $T=1.03$, $1.39$ and $1.48\,{\rm Gyr}$, generating a
surface density in correspondence to the relative luminosities of
stars and quasar at these times.  At other times, the accretion
activity is either obscured or the black hole accretion rate is
negligible. To make the appearance of the quasar visually more
apparent, we have put a small part of its luminosity in ``rays''
around the quasar. These rays are artificial and are only a visual
guide.

\subsection{Column Densities \&\ Quasar Attenuation}
\label{sec:NH}

From the simulation outputs, we determine the obscuration of the black
hole as a function of time during a merger by calculating the column
density to a distant observer along many lines of sight.  Typically,
we generate $\sim1000$ radial lines-of-sight (rays), each with its
origin at the black hole location and with directions uniformly spaced
in solid angle ${\rm d}\cos{\theta}\,{\rm d}\phi$.  For each ray, we
begin at the origin and calculate and record the local gas properties
using the SPH formalism and move a distance along the ray $\Delta
r=\eta h_{\rm sml}$, where $\eta \leq 1$ and $h_{\rm sml}$ is the
local SPH smoothing length.  The process is repeated until a ray is
sufficiently far from the origin ($\gtrsim 100$ kpc) that the column
has converged.  We then integrate the gas properties along a
particular ray to give the line-of-sight column density and mean
metallicity.  We have varied $\eta$ and find empirically that gas
properties along a ray converge rapidly and change smoothly for
$\eta=0.5$ and smaller.  We similarly vary the number of rays and find
that the distribution of line-of-sight properties converges for
$\gtrsim 100$ rays.

{}From the local gas properties, we use the multiphase model of the
ISM described in \citet{SH03} to determine the mass fraction in
``hot'' (diffuse) and ``cold'' (molecular and HI cloud core) phases of
dense gas and, assuming pressure equilibrium, we obtain the local
density of the hot and cold phases and their corresponding volume
filling factors.  The resulting values are in rough agreement with
those of \citet{MO77}.  Given a temperature for the warm, partially
ionized component of the hot-phase $\sim8000\,{\rm K}$, determined by
pressure equilibrium, we further calculate the neutral fraction of
this gas, typically $\sim0.3-0.5$.  We denote the neutral and total
column densities as \NHI\ and \NH, respectively.  Using only the
hot-phase density allows us to place an effective lower limit on the
column density along a particular line of sight, as it assumes a given
ray passes only through the diffuse ISM, with $\gtrsim 90\%$ of the
mass of the dense ISM concentrated in cold-phase ``clumps.'' Given the
small volume filling factor ($<0.01$) and cross section of cold
clouds, we expect that the majority of sightlines will pass only
through the ``hot-phase'' component.

Using $\Lbol=\dEdt$, we model the intrinsic quasar continuum SED
following \citet{Marconi04}, based on optical through hard X-ray
observations
\citep[e.g.,][]{Elvis94,George98,VB01,Perola02,Telfer02,Ueda03,VBS03},
with a reflection component generated by the PEXRAV model
\citep{MZ95}.  This yields, for example, a B-band luminosity
$\log{(\LB/L_{\sun})}=0.80-0.067\mathcal{L}+0.017\mathcal{L}^{2}-0.0023\mathcal{L}^{3}$,
where $\mathcal{L} = \log{(\Lbol/L_{\sun})} - 12$, and we take
$\lambda_{B}=4400\,$\AA, but as we model the entire intrinsic SED we
can determine the bolometric correction in any frequency interval.

We then use a gas-to-dust ratio to determine the extinction along a
given line of sight at optical frequencies.  Observations suggest that
the majority of reddened quasars have reddening curves similar to that
of the Small Magellanic Cloud (SMC; Hopkins et al.\ 2004, Ellison et
al.\ 2005), which has a gas-to-dust ratio lower than the Milky Way by
approximately the same factor as its metallicity \citep{Bouchet85}.
Hence, we consider both a gas-to-dust ratio equal to that of the Milky
Way, $(A_{B}/\nhi)_{\rm MW}=8.47\times10^{-22}\,{\rm cm^{2}}$, and a
gas-to-dust ratio scaled by metallicity, $A_{B}/\nhi =
(Z/0.02)(A_{B}/\nhi)_{\rm MW}$. In both cases we use the SMC-like
reddening curve of \citet{Pei92}. The form of the correction for hard
X-ray (2-10 keV) and soft X-ray (0.5-2 keV) luminosities is similar to
that of the B-band luminosity. We calculate extinction at X-ray
frequencies (0.03-10 keV) using the photoelectric absorption cross
sections of \citet{MM83} and non-relativistic Compton scattering cross
sections, similarly scaled by metallicity.  In determining the column
density for photoelectric X-ray absorption, we ignore the inferred
ionized fraction of the gas, as it is expected that the inner-shell
electrons which dominate the photoelectric absorption edges will be
unaffected in the temperature ranges of interest.  We do not perform a
full radiative transfer calculation, and therefore do not model
scattering or re-processing of radiation by dust in the infrared. 

For a full comparison of quasar lifetimes and column densities
obtained varying our calculation of \NH, we refer to \citet{H05b} (see
their Figures 1, 5, \& 6), and note their conclusion that, after
accounting for clumping of most mass in the dense ISM in cold-phase
structures, the column density does not depend sensitively on our
assumptions for the small-scale physics of the ISM and obscuration --
typically, the uncertainties in the resulting quasar lifetime as a
function of luminosity are a factor $\sim2$ at low luminosities in the
B-band, and smaller in e.g.\ the hard X-ray.  Because our
determination of the quasar luminosity functions is similar using the
hard X-ray data alone or the hard X-ray, soft X-ray, and optical data
simultaneously, the added uncertainties in our calculation of $\nLp$
in \S~\ref{sec:fullLF} below owing to the uncertainty in our $\nh$
calculation are small compared to the uncertainties owing to
degeneracies in the fitting procedure and uncertain bolometric
corrections.

\subsection{The \NH\ Distribution as a Function of Luminosity}
\label{sec:NHfunction}

Next, we consider the distribution of column densities as a function
of both the instantaneous and peak quasar luminosities.  For each
simulation, we consider \NH\ values at all times with a given
bolometric luminosity $L$ (in some logarithmic interval in $L$), and
determine the distribution of column densities at that $L$ weighted by
the total time along all sightlines with a given \NH. At each $L$, we
approximate the simulated distribution and fit it to a lognormal form,
\begin{equation}
P(N_{H}) = \frac{1}{\sigNH \sqrt{2 \pi}}\, \exp\left[
\frac{-\log^2(N_{H}/\meanNH )}{2 \sigNH^{2}}\right]. 
\end{equation}
This provides a good fit for all but the brightest luminosities, where
quasar feedback becomes important driving the ``blowout'' phase, and
the quasar sweeps away surrounding gas and dust to become optically
observable.

%\clearpage
\begin{figure*}
    \centering
    \plotone{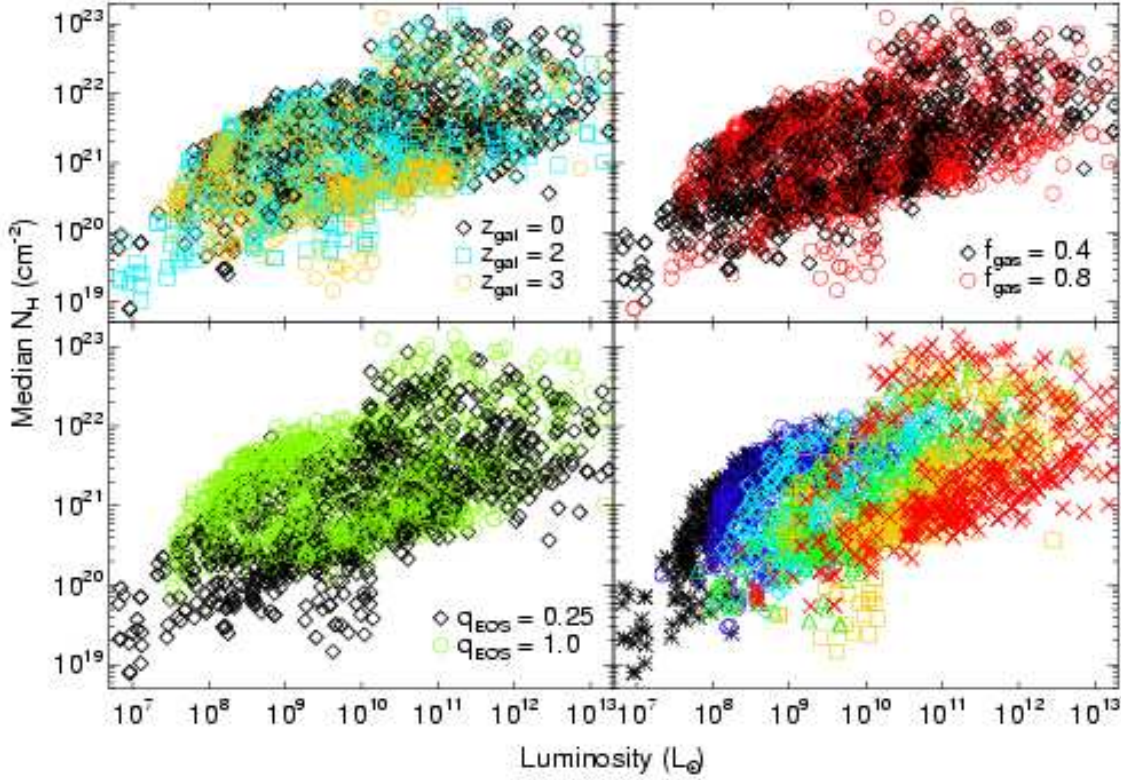}
    \caption{The median fitted total (neutral and ionized) column
    density $\meanNH$ at each luminosity $L$ in the snapshots from our
    series of simulations described in \S~\ref{sec:methods}. We compare changing
    concentrations and halo properties with redshift $\zgal$ (upper
    left), gas fractions $\fgas$ (upper right), the equation of state
    parameter $\qeos$ (lower left), and virial velocity $\vvir$ (lower
    right).  At lower right, simulations with $\vvir=80, 113, 160,
    226, 320, {\rm and}\ 500\,{\rm km\,s^{-1}}$ are shown as black
    asterisks, purple dots, red diamonds, green triangles, yellow
    squares, and red crosses, respectively. Other than a possible
    weak sensitivity to $\qeos$, the column density distribution as a
    function of luminosity shows no systematic dependence on any of
    the varied simulation parameters.
    \label{fig:NH.systematics}}
\end{figure*}
%\clearpage

We show the resulting median column density $\meanNH$ at each
luminosity $L$ in Figure~\ref{fig:NH.systematics}.  In the upper left
panel, simulations with $\zgal=0$ are shown in black, those with
$\zgal=2$ in blue, and those with $\zgal=3$ in yellow.  In the upper
right, simulations with $\fgas=0.4$ are shown in black, those with
$\fgas=0.8$ in red.  In the lower left, simulations with $\qeos=0.25$
are shown in black, those with $\qeos=1.0$ in green. And in the lower
right, simulations with $\vvir=80, 113, 160, 226, 320, {\rm and}\
500\,{\rm km\,s^{-1}}$ are shown as black asterisks, purple dots, red
diamonds, green triangles, yellow squares, and red crosses,
respectively.  Simulations with other values for these parameters (not
shown for clarity, but see e.g. Hopkins et al.\ [2005d]) show similar
trends.

While the increase in typical $\nh$ values with luminosity
appears to contradict observations suggesting that the obscured
fraction decreases with luminosity, this is because the relationship
shown above is dominated by quasars in growing, heavily obscured
phases.  In these stages, the relationship between column density and
luminosity is a natural consequence of the fact that both are fueled
by strong gas flows into the central regions of the galaxy -- more gas
inflow means higher luminosities, but also higher column densities.
During these phases, the lognormal fits to column density as a
function of instantaneous and peak luminosity presented in this
section are reasonable approximations, but they break down in the
brightest, short-lived stages of merger activity when the quasar
rapidly heats the surrounding gas and drives a powerful wind, lowering
the column density, resulting in a bright, optically observable
quasar. Including in greater detail the effects of quasar blowout
during the final stages of its growth in \S~\ref{sec:BLqso}, we find
that this modeling actually predicts the observed decrease in obscured
fraction with luminosity.

The relationship between \NH\ and $L$ shows no strong systematic
dependence on any of the simulation parameters considered. At most,
there is weak sensitivity to $\qeos$, in the sense that the
simulations with $\qeos=1.0$ have slightly larger column densities at
a given luminosity than those with $\qeos=0.25$.  We derive an
analytical model relating both the observed column density and quasar
luminosity to the inflowing mass of gas in \citet{H05d}, by
assuming that while it is growing, the black hole mass is proportional
to the inflowing gas mass in the galaxy core (which ultimately
produces the Magorrian et al.\ [1998] relation between black hole and 
bulge mass), and assuming Bondi accretion, with obscuration along a
sightline through this (spherically symmetric) gas inflow.  Such a
model gives the observed correlation between $N_{\rm H}$ and $L$, and
explains the weak dependence of the column density-luminosity relation
on the ISM gas equation of state. The assumptions above give a
relationship of the form
\begin{equation}
\nh\sim f_{0}\,\frac{1}{m_{H} R_{c}}\,\Bigl( \frac{c_{s}}{c}\Bigr)\,\Bigl( \frac{c L}{G^{2}}\Bigr)^{1/3}, 
\end{equation}
where $f_{0}\sim50$ is a dimensionless factor depending on the
radiative efficiency, mean molecular weight, density profile, and
assumed $M_{\rm BH}-\sigma$ relation; $m_{H}$ is the mass of hydrogen;
$R_{c}$ the radius of the galaxy core ($\sim100\,{\rm pc}$); and
$c_{s}$ the effective sound speed in the central regions of the galaxy. A
$\qeos=1.0$ equation of state, with a higher effective temperature,
results in a factor of $\approx2$ larger sound speed in the densest
regions of the galaxy than a $\qeos=0.25$ equation of state
\citep{SDH05b}, explaining the weak trend seen.  In any event,
the dependence is small compared to the intrinsic scatter for either
equation of state in the value of $\meanNH$ at a given luminosity, and
further weakens at high luminosity, so it can be neglected. 
What may appear to be a systematic offset in $\meanNH$ with
$\vvir$ is actually just a tendency for larger $\vvir$ systems to be
at higher luminosities; there is no significant change in the
dependence of \NH\ on $L$.

We use our large set of simulations to improve our fits (relative to
those of Hopkins et al.\ 2005d) to the \NH\ distribution as a function
of instantaneous and peak luminosities. Looking at individual
simulations, there appears to be a ``break'' in the power-law
scaling of $\meanNH$ with $L$ at $L\sim\Lcut{11}$. We find that the
best fit to the median column density $\meanNH$ is then
\begin{equation}
  \meanNH = \left\{ \begin{array}{ll}
      10^{22.8}\, {\rm cm^{-2}} \Bigl( \frac{L}{\Lp}\Bigr)^{0.54}   & \mathrm{ if\ } L < \Lcut{11} \\
      10^{21.9}\, {\rm cm^{-2}} \Bigl( \frac{L}{\Lcut{11}}\Bigr)^{0.43} & \mathrm{ if\ } L > \Lcut{11}.
\end{array}
    \right.
\end{equation}
Either of these two relations provides an acceptable fit to the
plotted $\meanNH$ distribution if applied to the entire luminosity
range ($\reducechi\approx2.8,\,3.2$ for the first and second
relations, respectively), but their combination provides a
significantly better fit ($\reducechi\approx1.5$), although it is
clear from the large scatter in $\meanNH$ values that any such fit is
a rough approximation.  Despite the complicated form of this equation,
it is, in practice, similar to our $\meanNH\propto L^{0.35}$
fit from previous work and $\meanNH\propto L^{1/3}$ analytical scaling
over the range of relevant luminosities, but is more accurate by a
factor $\sim2-3$ at low ($\lesssim\Lcut{9}$) luminosities. For
comparison, however, we do consider this simpler form for $\nh(L)$ as
well as our more accurate fit above in our subsequent analysis, and
find that it makes little difference to most observable quasar
properties.  At the highest luminosities, near the peak luminosities
of the brightest quasars, the scatter about these fitted median
$\meanNH$ values increases, and as noted above the impact of the
quasar in expelling surrounding gas becomes important and column
densities vary rapidly.  We consider this ``blowout'' phase in more
detail in \S~\ref{sec:BLqso}.

We find that any dependence of $\sigNH$ (the fitted lognormal
dispersion) on $L$ or $L_{\rm peak}$ is not statistically significant,
with approximately constant $\sigNH\approx 0.4$ for individual
simulations.  We similarly find no systematic dependence of $\sigNH$ on
any of our varied simulation parameters. However, it is important to
note that while the dispersion in \NH\ for an individual simulation is
$\sigNH\approx 0.4$, the dispersion in $\meanNH$ across all simulations at a
given luminosity is large, $\sim1$ dex.  Thus, we fit the effective
$\sigNH$ at a given luminosity for the {\em distribution} of quasars
and find it is $\sigNH\approx 1.2$. Although we have slightly revised our
fits for greater accuracy at low luminosities, we note that this
relation is shallower than the relation $\nh\propto L$ roughly
expected if $M_{\rm BH}$ is constant ($L\propto\rho\propto\nh$) or
$L\propto M_{\rm BH}$ always, and strongly contrasts with unification
models which predict static obscuration, or evolutionary
models in which \NH\ is independent of $L$ up to some threshold
\citep[e.g.,][]{Fabian99}.

\subsection{Quasar Lifetimes \&\ Sensitivity to Simulation Parameters}
\label{sec:detailsCompare}

We define the luminosity-dependent quasar lifetime $\tQ=\tQ(\Lm)$ as
the time a quasar has a luminosity above a certain reference
luminosity $\Lm$; i.e.\ the total time the quasar shines at
$L\geq\Lm$.  For ease of comparison across frequencies, we measure the
lifetime in terms of the bolometric luminosity, $L$, rather than e.g.\
the B-band luminosity.  Knowing the distribution of column densities
\NH\ as a function of luminosity and system properties (see
\S~\ref{sec:NHfunction}), we can then analytically or numerically
calculate the distribution of observed lifetimes at any frequency if
we know this intrinsic lifetime.  Below $\sim1$ Myr, our estimates of
\tq\ become uncertain owing to the effects of quasar variability and
our inability to resolve the local small-scale physics of the ISM, but
this is significantly shorter than even the most rapid timescales
$\sim10$\,Myr of substantial quasar evolution.

%\clearpage
\begin{figure}
    \centering
    \plotone{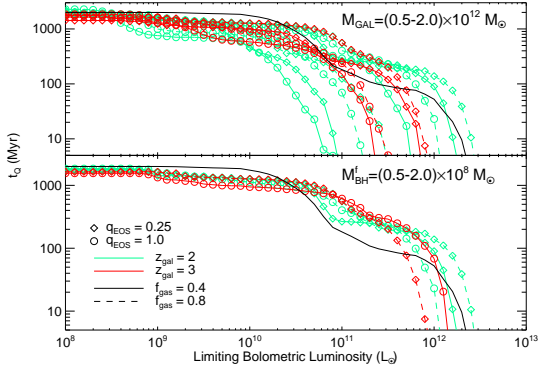}
    \caption{Integrated intrinsic quasar lifetime above a given
    reference bolometric luminosity, $t_{Q}(L)$, as a function of
    luminosity for simulations with host galaxies with total mass
    (top panel) $M_{\rm gal}=0.5-2.0\times10^{12}\,M_{\sun}$, and
    simulations with final black hole masses (bottom panel)
    $\mbhf=0.5-2.0\times10^{8}\,M_{\sun}$ (i.e.\ similar peak
    luminosity $\Lp\sim\Lcut{12}$).  The simulations cover a range in
    equation of state parameter $\qeos$, initial disk gas fraction
    $\fgas$, galaxy redshift (for scaling of halo properties) $\zgal$,
    and virial velocities $\vvir=113-160\,{\rm km\,s^{-1}}$. The black
    line in both cases is for a merger involving Milky Way-like galaxy
    models, which we refer to as A3, with $\fgas=1,\,\qeos=1,\,\zgal=0,\ 
    {\rm and}\ \vvir=160\,{\rm km\,s^{-1}}$.
    \label{fig:tQ.systematics}}
\end{figure}
%\clearpage

As before, we use our diverse sample of simulations to test for
systematic effects in our parameterization of the quasar lifetime.
Figure~\ref{fig:tQ.systematics} shows the quasar lifetime as a
function of reference luminosity $\Lm$ for both a set of simulations
with similar total galaxy mass, $M_{\rm gal}\approx10^{12}\,M_{\sun}$,
and similar final black hole mass (i.e.\ similar peak quasar
luminosity), $\mbhf\approx10^{8}\,M_{\sun}$.  In each case, the
simulations cover a range in $\qeos,$ $\fgas,$ $\zgal,$ ${\rm and}\ \vvir$.

As Figure~\ref{fig:tQ.systematics} demonstrates, at a given $M_{\rm
gal}$, there is a wide range of lifetimes, with a systematic
dependence on several quantities.  For example, for fixed $M_{\rm
gal}$, a lower $\qeos$ means that the gas is less pressurized and more
easily collapses to high density, resulting in larger $\mbhf$ and
longer lifetimes at higher luminosities. Similarly, higher $\fgas$
provides more fuel for black hole growth at fixed $M_{\rm gal}$.
However, for a given $\mbhf$, the lifetime \tq\ as a function of $\Lm$
is similar across simulations and shows no systematic dependence on
any of the varied parameters. We find this for all final black hole
masses in our simulations, in the range
$\mbhf\sim10^{6}-10^{10}\,M_{\sun}$.  We have further tested this as a
function of resolution, comparing with alternate realizations of our
fiducial A3 simulation with up to 128 times as many particles, and
find similar results as a function of $\mbhf$. 

From Figure~\ref{fig:tQ.systematics}, it is clear that the final black
hole mass or peak luminosity is a better variable to use in describing
the lifetime than the host galaxy mass.  The lack of any systematic
dependence of either the quasar lifetime or \NH($L,\Lp$) on host
galaxy properties implies that our earlier results (Hopkins et al.\
2005a-d) are reliable and can be applied to a wide range of host
galaxy properties, redshifts, and luminosities, although we refine and
expand the various fits of these works and their applications herein.
Furthermore, the large scatter in \tq\ at a given galaxy mass has
important implications for the quasar correlation function as a
function of luminosity, as one cannot associate a single quasar
luminosity with hosts of a given mass (see Lidz et al.\ 2005).

%\clearpage
\begin{figure}
    \centering
    \plotone{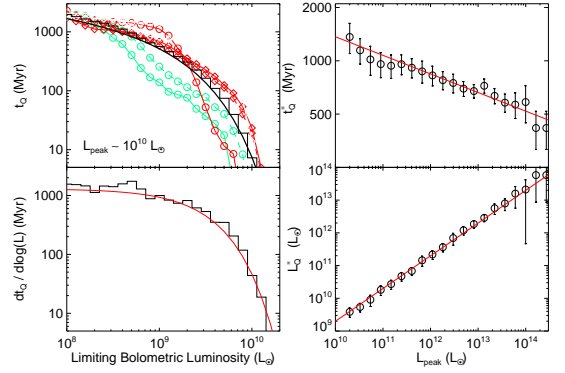}
    \caption{Fits to the quasar lifetime as a
    function of luminosity from our simulations.  Upper left
    shows the intrinsic, bolometric quasar lifetime $t_{Q}$ of a set
    of simulations with $\Lp$ within a factor of 2 of
    $\Lcut{10}$, in the manner of
    Figure~\ref{fig:tQ.systematics}. The black histogram shows the
    geometric mean of these lifetimes, and the black histogram in the
    lower left shows the differential lifetime $\dtdL$ from this
    geometric mean. The black thick line in the upper left and red
    line in the lower left show the best-fit to our analytical
    form, $\dtdL=t^{\ast}_{Q}\,\exp(-L/L^{\ast}_{Q})$.  Upper right
    shows the fitted $t^{\ast}_{Q}$ and resulting errors in each peak
    luminosity (final black hole mass) interval, and the best-fit
    power-law to $t^{\ast}_{Q}(\Lp)$ (red line).  Lower right shows the fitted
    $L^{\ast}_{Q}$ and resulting errors in each peak luminosity (final
    black hole mass) interval, and the best-fit proportionality
    $L^{\ast}_{Q}\propto\Lp$ (red line).
    \label{fig:show.fits}}
\end{figure}
%\clearpage

Although the truncated power-laws we have previously fitted to \tq\
using only the A-series simulations \citep{H05b} provide
acceptable fits to all our runs, we use our new, larger set of simulations
to improve the accuracy of the fits and average over peculiarities of
individual simulations, giving a more robust prediction of the
lifetime as a function of instantaneous and peak luminosity.  For a
given peak luminosity $\Lp$, we consider simulations with an $\Lp$
within a factor of 2, and take the geometric mean of their lifetimes
\tq($L$) (we ignore any points where $\tQ<1$\,Myr, as our calculated
lifetimes are uncertain below this limit). We can then differentiate
this numerically to obtain $\dtdL$ (the time spent in a given
logarithmic luminosity interval), and fit some functions to both
curves simultaneously.  Figure~\ref{fig:show.fits} illustrates this and
shows the results of our fitting. We find that both the integrated
lifetime \tq($L$) and the differential lifetime $\dtdL$ are well
fitted by an exponential,
\begin{equation}
\dtdL=t^{\ast}_{Q}\, \exp [-L/L^{\ast}_{Q}], 
\end{equation}
where both $t^{\ast}_{Q}$ and $L^{\ast}_{Q}$ are functions of $\mbhf$
or $\Lp$.  The best-fit such $\dtdL$ is shown in the figure as a solid
line for simulations with $\Lp\sim\Lcut{10}$, and agrees well
with both the numerical derivative $\dtdL$ (lower left, black
histogram) and the geometric mean $\tQ(L)$ (upper left, black
histogram).  This of course implies
\begin{equation}
\tQ(L)=t^{\ast}_{Q}\,\int^{\Lp}_{L}e^{-L/L^{\ast}_{Q}}\,d\log{L},
\end{equation} 
but we are primarily interested in $\dtdL$ in our subsequent
analysis. 

Although our fitted lifetime involves an exponential, it is in no way
similar to the exponential light curve of constant Eddington-ratio
black hole growth or the model in, e.g., \citet{HL98}, which give
$\dtdL=$\,{\em constant}\,$\sim t_{S}\ll t^{\ast}_{Q}$.  

Our
functional form also has the advantage that, although it should formally be
truncated with $\dtdL=0$ for $L>\Lp$, the values in this regime fall
off so quickly that we can safely use the above fit for all large
$L$. Similarly, at $L\lesssim10^{-4}\,\Lp$, $\dtdL$ falls below the
constant $t^{\ast}_{Q}$ to which this equation
asymptotes. Furthermore, in this regime, the fits above begin to
differ significantly from those obtained by fitting e.g.\ truncated
power-laws or Schechter functions. However, these luminosities are
well below those we generally consider and well below the luminosities
where the contribution of a quasar with some $\Lp$ is significant to
the observed quantities we predict. Moreover, this turndown (i.e.\ the
lower value predicted by an exponential as opposed to a power-law or
Schechter function at low luminosities) is at least in part an
artifact of the finite simulation duration.  The values here are also
significantly more uncertain, as by these low relative accretion
rates, the system is likely to be accreting in some low-efficiency,
ADAF state (e.g.\ Narayan \& Yi 1995), which we do not implement
directly in our simulations. Rather than introduce additional
uncertainties into our modeling when they do not affect our
predictions, we adopt these exponential fits which are
accurate at $L\gtrsim10^{-4}-10^{-3}\,\Lp$.  However, for purposes
where the faint-end behavior of the quasar lifetime is important,
such as predicting the value and evolution of the faint-end quasar
luminosity function slope with redshift, a more detailed examination
of the lifetime at low luminosities and relaxation of quasars after
the ``blowout'' phase is necessary, and we consider these issues
separately in \citet{H05f}.

We also note that in \citet{H05c} we considered several extreme limits
to our modeling, neglecting all times before the final merger and
applying an ADAF correction at low accretion rates (taken
into account a posteriori by rescaling the radiative efficiency
$\epsilon_{r}$ with accretion rate, given the assumption that such low
accretion rates do not have a large dynamical effect on the system
regardless of radiative efficiency), and found that this does not
change our results -- the lifetime at low luminosities
may be slightly altered but the key qualitative point, that the quasar
lifetime increases with decreasing luminosity, is robust against a
wide range of limits designed to decrease the lifetime at low
luminosities.

Figure~\ref{fig:show.fits} further shows the fitted $t^{\ast}_{Q}$
(upper right) and $L^{\ast}_{Q}$ (lower right) as a function of peak
quasar luminosity for each $\Lp$.  We find that
$L^{\ast}_{Q}$, the luminosity above which the lifetime rapidly
decreases, is proportional to $\Lp$,
\begin{equation}
L^{\ast}_{Q}=\alpha_{L}\Lp ,
\end{equation}
with a best fit coefficient $\alpha_{L}=0.20$ (solid line).  The weak
dependence of $t^{\ast}_{Q}$ on $\Lp$ is well-described by a power-law,
\begin{equation}
t^{\ast}_{Q}=t^{(10)}_{\ast}\,\Bigl( \frac{\Lp}{\Lcut{10}}\Bigr)^{\alpha_{T}}, 
\end{equation}
with $t^{(10)}_{\ast}=1.37\times10^{9}\,{\rm yr}$ and
$\alpha_{T}=-0.11$
%, and a double power-law fit (solid line),
%\begin{equation}
%t^{\ast}_{Q}=t^{(10)}_{\ast}\,\left[\Bigl(
%\frac{\Lp}{\Lcut{10}}\Bigr)^{\alpha_{T}}+\Bigl(
%\frac{\Lp}{\Lcut{10}}\Bigr)^{\beta_{T}}\right],
%\end{equation}
%with $t^{(10)}_{\ast}=8.8\times10^{8}\,{\rm yr}$, $\alpha_{T}=-0.71$,
%and $\beta_{T}=-0.07$.  We have carried out our analysis with both
%functions and find no significant differences except for slight
%corrections to the high-mass end of the relic black hole mass
%function (see \S~\ref{sec:smbh}). Unless otherwise specified, we adopt the double
%power-law form for our calculations, as it gives a marginally better
%fit to the simulation lifetimes.

%\clearpage
\begin{figure}
    \centering
    \plotone{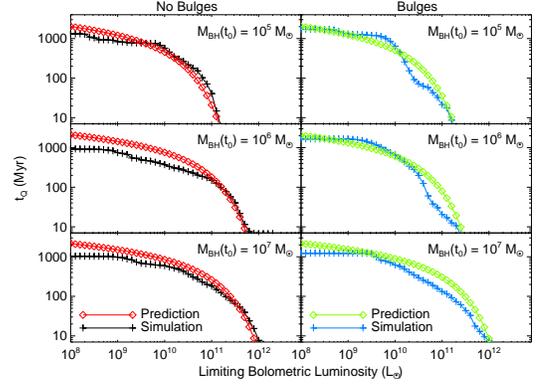}
    \caption{Predicted quasar lifetime as a function of luminosity 
    compared to that obtained in simulations with and without bulges
    and with different initial seed black hole masses. All simulations shown 
    in this plot 
    are initially identical to our fiducial A3 (Milky Way-like) case,
    but with or without an initial stellar bulge and with an initial
    seed black hole mass as labeled. Diamonds show the predicted
    quasar lifetime $t_{Q}$, a function of the peak luminosity of each
    simulated quasar, determined from the fits shown in
    Figure~\ref{fig:show.fits}. Crosses show the lifetime determined
    directly in the simulations.
    \label{fig:check.MBHbulge}}
\end{figure}
%\clearpage

The presence or absence of a stellar bulge in the progenitors can have
a significant impact on the quasar light curve (Springel et al.
2005b), primarily affecting the strength of the strong accretion phase
associated with initial passage of the merging galaxies (e.g.\ Mihos
\& Hernquist 1994).  Likewise, the seed mass of the simulation black
holes could have an effect, as black holes with smaller initial masses
will spend more time growing to large sizes, and more massive black
holes may be able to shut down early phases of accretion in mergers in
minor ``blowout'' events.  In Figure~\ref{fig:check.MBHbulge}, we show
various tests to examine the robustness of our fitted quasar lifetimes
to these variations.  We have re-run our fiducial Milky Way-like A3
simulation both with (right panels) and without (left panels) initial
stellar bulges in the merging galaxies and varying the initial black
hole seed masses from $10^{4}-10^{7}\,M_{\sun}$.  In each case we
compare the lifetime $t_{Q}$ determined directly from the simulations
(crosses) to that predicted from our fits above (diamonds), based only
on the peak luminosity (final black hole mass) of the simulated
quasar. Again, we find that varying these simulation parameters can
have a significant effect on the final black hole mass, but that the
quasar lifetime as a function of peak luminosity is a robust quantity,
independent of initial black hole mass or the presence or absence of a
bulge in the quasar host.

We can integrate the total radiative output of our model quasars,
\begin{equation}
E_{\rm rad}=\int^{\Lp}_{L_{\rm min}} L\,\fdtdL {\rm d}\log L, 
\end{equation}
and using our fitted formulae and $L_{\rm min}\ll L^{\ast}_{Q}$
we find
\begin{equation}
E_{\rm rad}=L^{\ast}_{Q}\, t^{\ast}_{Q}\,\log{e}\,(1-e^{-\Lp/L^{\ast}_{Q}}). 
\label{eq:Erad}
\end{equation}
Knowing $E_{\rm rad}=\epsilon_{r}\mbhf c^{2}$, we can compare the
final black hole mass as a function of peak luminosity to what we
would expect if the peak luminosity were the Eddington
luminosity of a black hole with mass $M_{\rm Edd}$, $L_{\rm
Edd}=\epsilon_{r}M_{\rm Edd} c^{2}/t_{S}$, where $t_{S}$ is the
Salpeter time for $\epsilon_{r}=0.1$. Equating $E_{\rm
rad}=\epsilon_{r}\mbhf c^{2}$ with the value calculated in
Equation~\ref{eq:Erad}, and using the definition of the Eddington mass
at $L=\Lp$ and our fitted $L^{\ast}_{Q}=\alpha_{L}\Lp$, we obtain
\begin{equation}
\frac{\mbhf(\Lp)}{M_{\rm Edd}(\Lp)}=\alpha_{L} \Bigl( \frac{t^{\ast}_{Q}}{t_{S}}\Bigr) \log e 
\approx1.24\, f_{T},
\label{eq:Mbhf}
\end{equation}
where $f_{T}=(\Lp/\Lcut{13})^{-0.11}$ for the power-law fit to
$t^{\ast}_{Q}$. For our calculations explicitly involving black hole
mass, we adopt this conversion unless otherwise noted, as we have
performed our primary calculation (i.e.\ calculated $\nLp$) in terms
of peak luminosity. Moreover, although this agrees well with
the black hole masses in our simulations as a function of peak
luminosity (as it must if the fitted quasar lifetimes are accurate),
this allows us to smoothly interpolate to the highest black hole
masses ($\sim {\rm a\ few}\, \times10^{9}-10^{10}\,M_{\sun}$), which
are of particular interest in examining the black hole population but
for which the number of simulations we have with a given final black
hole mass drops rapidly.

This gives explicitly the modifications to the black hole mass
compared to that inferred from the ``light bulb'' and
``constant Eddington ratio'' models which we outline below in
\S~\ref{sec:altmodels}, in which quasars shine at constant luminosity
or follow exponential light curves, and for which $M_{\rm
BH}^{f}=M_{\rm Edd}(\Lp) / l$, where $l$, the (constant) Eddington
ratio, is generally adopted. The corrections are small, and therefore
most of the black hole mass is accumulated in the bright, near-peak
quasar phase, in good agreement with observational estimates
\citep[e.g.,][]{Soltan82,YT02}; we discuss this in greater detail in
\S~\ref{sec:BLqso} and \S~\ref{sec:smbh}. Furthermore, the increase of
$f_{T}$ with decreasing $\Lp$ implies that lower-mass quasars
accumulate a larger fraction of their mass in slower, sub-peak
accretion after the final merger, while high-mass objects acquire
essentially all their mass in the peak quasar phase. This is seen
directly in our simulations, and is qualitatively in good agreement
with expectations from simulations and semi-analytical models in which
the $M_{\rm BH}-\sigma$ relation is set by black hole feedback in a
strong quasar phase. Compared to the assumption that
$M_{BH}^{f}=M_{\rm Edd}(\Lp)$, this formula introduces a small but
non-trivial correction in the relic supermassive black hole mass
function implied by the quasar luminosity function and $\nLp$ (see
\S~\ref{sec:smbh}).

The predictions of our model for the quasar lifetime and evolution can
be applied to observations which attempt to constrain the quasar
lifetime from individual quasars, for example using the proximity
effect in the Ly$\alpha$ forest \citep{BDO88,HC02,Jakobsen03,YL05} and
multi-epoch observations \citep{MartiniSchneider03}.  However, many
observations designed to constrain the quasar lifetime do so not for
individual quasars, but using demographic or integral arguments based
on the population of quasars in some luminosity interval
\citep[e.g.,][]{Soltan82,HNR98,YT02,YL04,PMN04,Grazian04}. Our
prediction for these observations is similar but slightly more
complex, as an observed luminosity function at a given luminosity will
consist of sources with different peak luminosities $\Lp$, but the
same instantaneous luminosity, $L$. Furthermore, the lifetime being
probed may be either the integrated quasar lifetime above some
luminosity threshold or the differential lifetime at a particular
luminosity.

%\clearpage
\begin{figure*}
    \centering
    \plotone{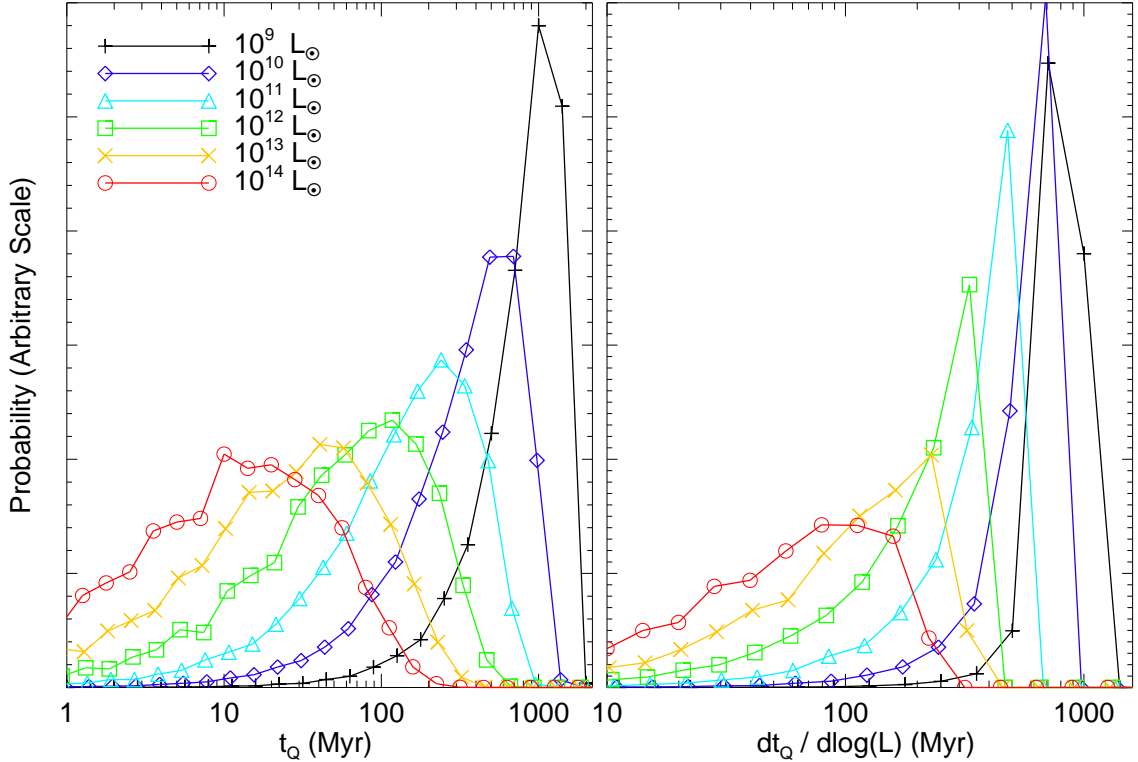}
    \caption{Predicted distribution (fractional number density 
    per logarithmic interval in lifetime) of quasar lifetimes at 
    different bolometric luminosities, for the luminosity function 
    determined in \S~\ref{sec:LF} at $z=0.5$. Left panel plots the distribution of 
    integrated lifetimes $t_{Q}$ (time spent over the course of each  
    quasar lifetime above the given luminosity). Right panel plots 
    the distribution of differential lifetimes $\dtdL$ (time spent 
    by each quasar in a logarithmic interval about the given luminosity). 
   \label{fig:LF.lifetimes}}
\end{figure*}
%\clearpage

For a given determination of the quasar luminosity function using our
model for quasar lifetimes and some distribution of peak luminosities,
we can predict the distribution of quasar lifetimes as a function of
the observed luminosity interval. Figure~\ref{fig:LF.lifetimes} shows
an example of such a result, using the determination of the luminosity
function below in \S~\ref{sec:fullLF}, at redshift $z=0.5$. We
consider several bolometric luminosities spanning the luminosity
function from $10^{9}-10^{14}\,L_{\sun}$, and for each, the
distribution of sources (peak luminosities), and the corresponding
distribution of quasar lifetimes. We show both the distribution of
integrated quasar lifetimes $t_{Q}$ (left panel) and the distribution
of differential quasar lifetimes $\dtdL$ (right panel). The evolution
with redshift is weak, with the lifetime increasing by $\sim1.5-2$ at
a given luminosity at $z=2$.  There is furthermore an ambiguity of a
factor $\sim2$, as some of the quasars observed at a given luminosity
will only be entering a peak quasar phase, whereas the lifetimes shown
are integrated over the whole quasar evolution.  This prediction is
quite different from that of the optical quasar phase which we
describe below in \S~\ref{sec:BLqso} and in \citet{H05a}, as it
considers only the intrinsic bolometric luminosity, but our modeling
and the fits provided above for the bolometric lifetime and column
density distributions should enable the prediction of these
quantities, considering attenuation, in any waveband.  In either case,
it is clear that the lifetime distribution for lower-luminosity
quasars is increasingly more strongly peaked and centered around
longer lifetimes, in good agreement with the limited observational
evidence from e.g.\ Adelberger \& Steidel (2005). This is a
consequence of the fact that in our model quasar lifetimes
decrease with increasing luminosity. The range spanned in the figure
corresponds well to the range of quasar lifetimes implied by the
observations above and others \citep[e.g.][and references
therein]{Martini04}.

\subsection{Alternative Models of Quasar Evolution}
\label{sec:altmodels}

Our modeling reproduces at least the observed hard X-ray quasar
luminosity function by construction, since we use the observed quasar
luminosity functions to determine the birthrate of quasars of a given
$\Lp$, $\nLp$, in \S~\ref{sec:fullLF}. It is therefore useful to
consider in detail the differences in our subsequent predictions
between various models for the quasar lifetime and obscuration, in
order to determine to what extent these predictions are implied by any
model that successfully reproduces the observed quasar luminosity
function, and to what extent they are independent of the observed
luminosity functions and instead depend on the model of quasar
evolution adopted. To this end, we define two models for the
quasar lifetime, and two models for the distribution of quasar column
densities, combinations of which have been commonly used in most
previous analyses of quasars.

For the quasar lifetime, we consider the following two cases:

\bigskip

{\em ``Light-Bulb Model''} \citep[e.g.,][]{SB92,KH00,WL03,HQB04}.  The
simplest possible model for the quasar light curve, the ``feast or
famine'' or ``light-bulb'' model assumes that quasars have only two
states, ``on'' and ``off.'' Quasars turn ``on'', shine at a fixed
bolometric luminosity $L=\Lp$, defined by a ``constant'' Eddington
ratio (i.e.\ $\Lp=l\, M_{\rm BH}^{f}$) and constant quasar lifetime
$t_{Q,\,\rm LB}$. Models where quasars live arbitrarily long with
slowly evolving mean volume emissivity or mean light curve
\citep[e.g.][]{SB92,HM00,KH00} are equivalent to the ``light bulb''
scenario, as they still assume that quasars observed at a luminosity
$L$ radiate at that approximately constant luminosity over some
universal lifetime $t_{Q,\,\rm LB}$ at a particular redshift.  We
adopt $l=0.3$ and $t_{Q,\, \rm LB}=10^{7}\,$yr, as is commonly assumed
in theoretical work and suggested by observations (given this
prior) \citep[e.g.][]{YT02,Martini04, Soltan82,YL04,PMN04,Grazian04},
and similar to the $e$-folding time of a black hole with canonical
radiative efficiency $\epsilon_{r}=0.1$ \citep{Salpeter64} or the
dynamical time in a typical galactic disk or central regions of the
merger.  These choices control only the normalization of $\nLp$, and
therefore do not affect most of our predictions. Where the
normalization (i.e.\ value of the constant $t_{Q}$ or $l$) is
important, we allow it to vary in order to produce the best possible
fit to the observations.

\bigskip

{\em ``Exponential (Fixed Eddington Ratio) Model.''} A somewhat more
physical model of the quasar light curve is obtained by assuming
growth at a constant Eddington ratio, as is commonly adopted in e.g.\
semi-analytical models which attempt to reproduce quasar luminosity
functions \citep[e.g.][]{KH00,WL03,V03}.  In this model, a black hole
accretes at a fixed Eddington ratio $\eEdd$ from an initial mass
$M_{i}$ to a final mass $M_{f}$ (or equivalently, a final luminosity
$L_{f}=\eEdd\,L_{Edd}(M_{f})$), and then shuts off. This gives
exponential mass and luminosity growth, and the time spent in any
logarithmic luminosity bin is constant,
\begin{equation}
dt/\dlgL = t_{S}\,(\ln(10)/\eEdd)
\end{equation}
for $L_{i}<L<L_{f}$. This is true for any exponential light curve;
i.e.\ this model includes cases with an exponential {\em decline} in
quasar luminosity), $f(t)\propto e^{\pm t/t_{\ast}}$, such as that of
\citet{HL98}, with only the normalization $dt/\dlgL =
t_{\ast}\,\ln(10)$ changed, and thus any such model will give
identical results with correspondingly different normalizations. As
with the ``light-bulb'' model, we are free to choose the
characteristic Eddington ratio and corresponding timescale for this
lightcurve, and we adopt $\eEdd=0.3$ (i.e.\ $t_{\ast}\sim10^{8}\,$yr)
in general. Again, however, we allow the normalization to vary freely
where it is important, such that these models have the best chance to
reproduce the observations. For our purposes, models in which this
timescale is determined by e.g.\ the galaxy dynamical time and thus
are somewhat dependent on host galaxy mass or redshift are nearly
identical to this scenario. Further, insofar as the dynamical time
increases weakly with increasing host galaxy mass (as, e.g.\ for a
spheroid with $M_{\rm BH}\propto M_{\rm vir}\sim a\,\sigma^{2}/G$,
where $a$ is the spheroid scale length and $M_{\rm BH}\propto
\sigma^{4}$, such that $t_{\rm dyn}\sim a/\sigma\propto \sigma \propto
M_{\rm vir}^{1/4}$), this produces behavior qualitatively opposite to
our predictions (of increasing lifetime with decreasing instantaneous
luminosity), and yields results which are even more discrepant from
our predictions and the observations than the constant (host-galaxy
independent) case.

\bigskip

A wide variety of ``light-bulb'' or exponential (constant Eddington
ratio) models are possible, allowing for different distributions of
typical Eddington ratios and/or quasar lifetimes (see e.g.\ Steed \&
Weinberg 2003 for an extensive comparison of several classes of such
models), but for our purposes they are essentially identical insofar
as they do not capture the essential qualitative features of our
quasar lifetimes, namely that the quasar lifetime depends on both
instantaneous and peak luminosities, and increases with decreasing
instantaneous luminosity.

We fit both of the simple models above to the
observed quasar luminosity functions in the same manner described in
\S~\ref{sec:LF}, (i.e.\ in the same manner as we fit our more
complicated models of quasar evolution), to determine $\nLp_{\rm LB}$
for the ``light-bulb'' model and $\nLp_{\rm Edd}$ for the ``fixed
Eddington ratio'' model (see Equations~\ref{eqn:nLp.LB} and
\ref{eqn:nLp.Edd}, respectively).
%We use the observed hard X-ray luminosity function in order to minimize uncertainties owing to 
%attenuation, converted to a bolometric luminosity function as above. 
Thus all three models of the quasar light curve, the ``light-bulb'', ``fixed Eddington ratio'', 
and our luminosity-dependent lifetimes model produce an essentially identical 
bolometric luminosity function. 

We also consider two commonly adopted alternative models for the
column density distribution and quasar obscuration:

\bigskip

{\em ``Standard (Luminosity-Independent) Torus''}
\citep[e.g.][]{Antonucci93}.  This is the canonical obscuration model,
based on observations of local, low-luminosity Seyfert galaxies
\citep[e.g.,][]{RMS99}. The column density distribution is derived
from the torus geometry, where we assume the torus inner radius lies
at a distance $R_{\rm T}$ from the black hole, with a height $H_{\rm
T}$, and a density distribution $\rho(\theta)\propto\exp(-\gamma
|\cos\theta |)$, where $\theta$ is the polar angle and the torus lies
in the $\theta=0$ plane. This results in a column density as a
function of viewing angle of
%\begin{equation}
%\begin{split}
%\nh(\theta)= &N_{\rm H,\, 0}\,\exp(-\gamma |\cos\theta |)\,\cos(90-\theta)\,\\
%&\times\sqrt{\Bigl(\frac{R_{\rm T}}{H_{\rm T}}\Bigr)^{2} - \sec^{2}(90-\theta)
%\Bigl(\Bigl(\frac{R_{\rm T}}{H_{\rm T}}\Bigr)^{2}-1\Bigr)}
%\end{split}
%\end{equation}
\begin{eqnarray}
\nh(\theta)&=&N_{\rm H,\, 0}\,\exp(-\gamma |\cos\theta |)\,\cos(90-\theta)\,\nonumber\\
&&\times\sqrt{\Bigl(\frac{R_{\rm T}}{H_{\rm T}}\Bigr)^{2} - \sec^{2}(90-\theta)
\Bigl(\Bigl(\frac{R_{\rm T}}{H_{\rm T}}\Bigr)^{2}-1\Bigr)}
\end{eqnarray}
\citep{Treister04}. Here, $N_{\rm H,\, 0}$ is the column density along
a line of sight through the torus in the equatorial plane and $\gamma$
parameterizes the exponential decay of density with viewing
angle. This is a phenomenological model, and as a result the
parameters are essentially all free.  We adopt typical values, an
equatorial column density $N_{\rm H,\, 0}=10^{24}$\,cm$^{-2}$,
radius-to-height ratio $R_{\rm T}/H_{\rm T}=1.1$, and density profile
$\gamma=4$. This combination of parameters follows \citet{Treister04},
and is designed to fit the observed X-ray column density distribution
and give a ratio of obscured to unobscured quasars $\sim3$, similar to
the mean locally observed value \citep[e.g.][]{RMS99}.

\bigskip

{\em ``Receding (Luminosity-Dependent) Torus''}
\citep[e.g.][]{Lawrence91}.  Many observations suggest that the
fraction of obscured objects depends on luminosity
\citep{Steffen03,Ueda03,Hasinger04,GRW04,sazrev04,Barger05,Simpson05}.
Therefore, some theoretical works have adopted a ``receding torus''
model, in which the torus radius $R_{\rm T}$ (i.e.\ distance from the
quasar) is allowed to vary with luminosity, but the height and other
parameters remain constant. The torus radius is assumed to increase
with luminosity, enlarging the opening angle and thus the fraction of
unobscured quasars. In this case, the column densities are identical
to those shown above, but now $R_{\rm T}/H_{\rm T}=(L/L_{0})^{0.5}$,
where $L_{0}\approx10^{11}\,L_{\sun}$ is the luminosity at which the
ratio of obscured to unobscured quasars is $\approx 3:1$ and the
power-law slope is chosen to fit the dependence of obscured fraction
on luminosity.

\bigskip

Both of these column density distributions represent phenomenological
models with several free parameters, explicitly chosen to reproduce
the observed differences in quasar luminosity functions and column
density distributions. Despite this, it is not clear that these
functional forms represent the best possible fit to the observations
they are designed to reproduce. Furthermore, comparison of our results
in which column density distributions depend on luminosity and peak
luminosity elucidates the importance of proper modeling of the
dependence of column density on quasar evolution.

\section{The Quasar Luminosity Function}
\label{sec:LF}
\subsection{The Effect of Luminosity-Dependent Quasar Lifetimes}
\label{sec:lifetimeLF}

Given quasar lifetimes as functions of both instantaneous and peak
luminosities, the observed quasar luminosity function (in the absence
of selection effects) is a convolution of the lifetime with the
intrinsic distribution of sources with a given $\Lp$. If sources of a
given $L$ are created at a rate $\dot{n}(L,t)$ (per unit comoving
volume) at cosmological time $t_{H}\sim1/H(z)$ and live for some
lifetime $\Delta t_{Q}(L)$, the total comoving number density observed
will be
\begin{equation}
\Delta n=\int^{t_{H}+\Delta t_{Q}(L)}_{t_{H}} \dot{n}(L,t)\,{\rm d}t, 
\end{equation}
which, for a cosmologically evolving $\dot{n}(L, t)$, can be expanded
about $\dot{n}(L,t_{H})$, yielding $\Delta n=\dot{n}(L,t_{H})\,\Delta
t_{Q}(L)$ to first order in $\Delta t_{Q}(L)/t_{H}$. Considering a
complete distribution of sources with some $\Lp$, we similarly obtain
the luminosity function
\begin{equation}
\phi(L)\equiv\fdphi(L)=\int{\frac{{\rm d}t(L,L_{\rm
      peak})}{\dlgL}\,\nLP}\,{\rm d}\log(L_{\rm peak}). 
\label{eqn:LF.int}
\end{equation}
Throughout, we will denote the differential luminosity function, i.e.\
the comoving number density of quasars in some logarithmic luminosity
interval, as $\phi\equiv\dphi$.  Here, $\nLP$ is the comoving number
density of sources created per unit cosmological time per logarithmic
interval in $L_{\rm peak}$, at some redshift, and $\dtdL$ is the 
differential quasar lifetime, i.e.\ the total time that a quasar with a given 
$\Lp$ spends in a logarithmic interval in bolometric luminosity $L$. 
This formulation
implicitly accounts for the ``duty cycle'' (the fraction of active
quasars at a given time), which is proportional to the lifetime at a
given luminosity.  Corrections to this formula owing to finite
lifetimes are of order $(\dtdL)/t_{H}$, which for the luminosities and
redshifts considered here (except for Figure~\ref{fig:LF.highz}), are
never larger than $\sim1/5$ and are generally $\ll 1$, which is
significantly smaller than the uncertainty in the luminosity
function itself.

We next consider the implications of our
luminosity-dependent quasar lifetimes for the relation between the
observed luminosity function and the distribution of peak luminosities
(i.e.\ intrinsic properties of quasar systems).  In traditional 
models of quasar lifetimes and light curves, this relation is
trivial. For example, models in which quasars ``turn on'' 
at fixed luminosity for some fixed lifetime (i.e.\ the ``light-bulb'' model 
defined in \S~\ref{sec:altmodels}) imply 
\begin{equation}
\nLP_{\rm LB} \propto \phi(L=L_{\rm peak}), 
\label{eqn:nLp.LB}
\end{equation}
and models in which quasar light curves are a pure exponential growth
or decay with some cutoff(s) (e.g., exponential or fixed Eddington-ratio models) imply 
\begin{equation}
\nLP_{\rm Edd} \propto \frac{d\phi}{\dlgL}\raisebox{-3pt}{\huge $\mid$}_{L=L_{\rm peak}}.
\label{eqn:nLp.Edd}
\end{equation}
These both have essentially {\em identical} shape to the observed
luminosity function, qualitatively different from our model prediction
that $\nLp$ should turn over at luminosities approximately below the
break in the observed luminosity function (see, e.g. Fig. 1 of
Hopkins et al.\ 2005e). The
luminosity-dependent quasar lifetimes determined from our simulations
imply a new interpretation of the luminosity function,
with $\nLP$ tracing the bright end of the luminosity function similar
to traditional models, but then peaking and turning over below
$\Lp\sim L_{\rm break}$, the break luminosity in standard double
power-law luminosity functions. In our deconvolution of the luminosity
function, the faint end corresponds primarily to sources in
sub-Eddington phases transitioning into or out of the phase(s) of peak
quasar activity. There is also some contribution to the faint-end
lifetime from quasars accreting efficiently (i.e.\ growing
exponentially at high Eddington ratio) early in their activity and on
their way to becoming brighter sources, but this becomes an
increasingly small fraction of the lifetime at lower luminosities. For
example, in Figure 7 of Hopkins et al.\ (2005b), direct calculation of
the quasar lifetime shows that sub-Eddington phases begin to dominate
the lifetime for $L\lesssim0.1\,\Lp$, with $\gtrsim90\%$ of the
lifetime at $L\sim10^{-3}\,\Lp$ corresponding to sub-Eddington
growth. By definition, a ``fixed Eddington ratio'' or ``light bulb''
model is dominated at all luminosities by a fixed, usually large,
Eddington ratio. Even models which assume an exponential decline in
the quasar luminosity from some peak, although they clearly must spend
a significant amount of time at low Eddington ratios, have an
identical $\nLp=\nLp_{\rm Edd}$ (modulo an arbitrary normalization),
and predict far less time at most observable ($\gtrsim10^{-4}\,\Lp$)
low luminosities and accretion rates (because the accretion
rates fall off so rapidly); i.e.\ the population at any observed
luminosity is still dominated by objects near their peak.

%\clearpage
\begin{figure}
    \centering
    \plotone{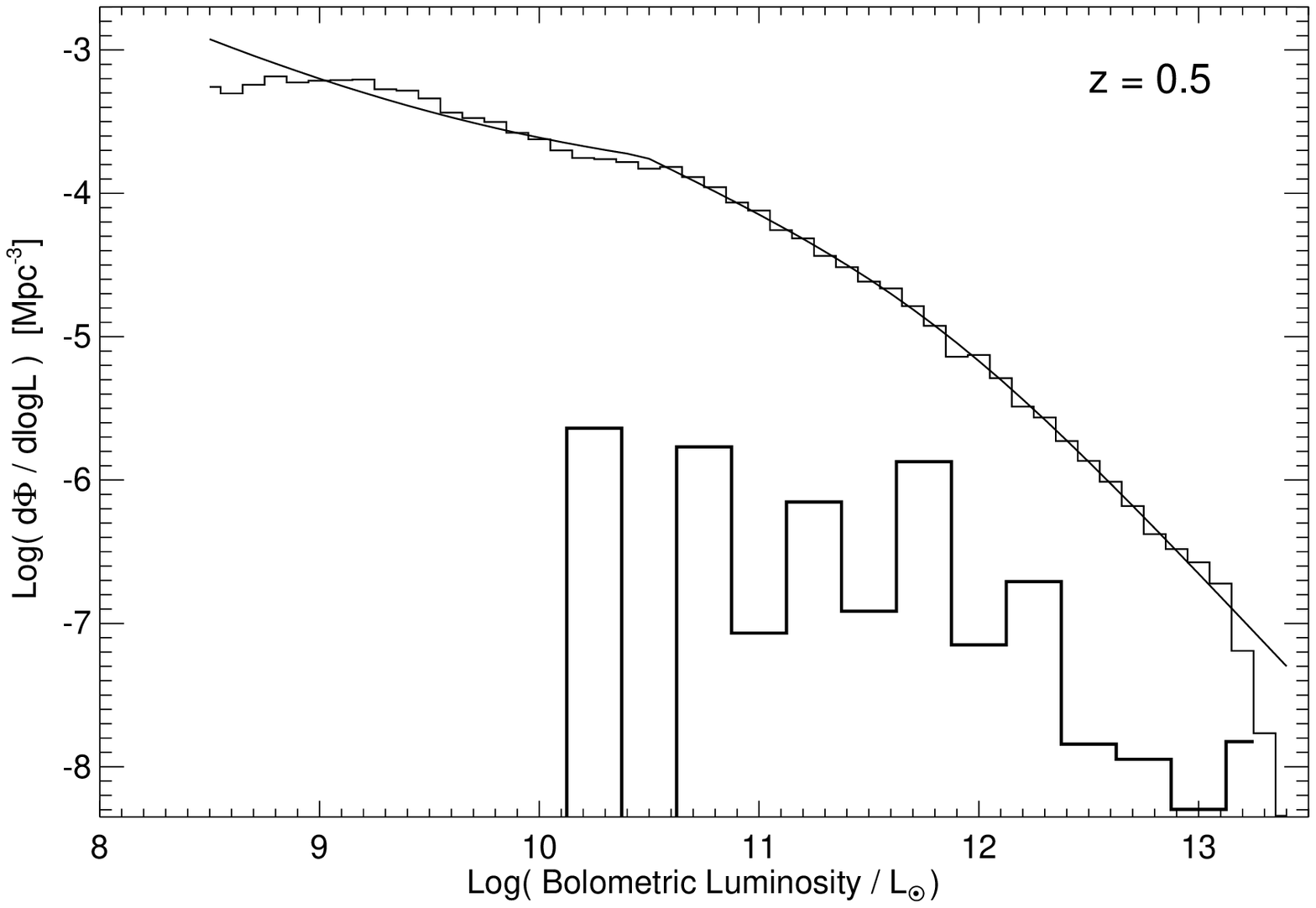}
    \caption{We reproduce (thin histogram) the luminosity function of
    \citet{Ueda03} at redshift $z=0.5$ (thin curve) using the binned
    differential quasar lifetime $\dtdL$ directly from our simulations
    and a fitted distribution of peak luminosities $\nLp$ (thick
    histogram). For each bin in $\log(\Lp)$, we average the binned
    differential lifetime of a set of simulations with peak luminosity
    in the bin. This clearly demonstrates our key qualitative result, 
    that the faint end of the luminosity function is
    reproduced by quasars with {\em peak} luminosity 
    around the break luminosity but observed primarily in sub-Eddington
    states (luminosities $L\ll\Lp$), is not an
    artifact of our fitting formulae or extrapolation to extreme
    luminosities.
    \label{fig:LF.hists}}
\end{figure}
%\clearpage

From our new, large set of simulations, we test this model of the
relationship between the distribution of peak quasar luminosities and
observed luminosity functions, namely our assertion that $\nLp$
should peak around the observed break in the luminosity function, and
turn over below this peak, with the observed luminosity function
faint-end slope dominated by sources with peak luminosities near the
break in sub-Eddington (sub-peak luminosity) states. In particular, we
wish to ensure that this behavior for $\nLP$ is real, and not some
artifact of our fitting functions for the quasar lifetime.

Figure~\ref{fig:LF.hists} shows the best fit $\nLP$ distribution
(solid thick histogram) fitted to the \citet{Ueda03} hard X-ray quasar
luminosity function (solid curve) at redshift $z=0.5$, as well as the
resulting best-fit luminosity function (solid thin histogram).  For
ease of comparison with other quasar luminosities, we rescale the
luminosity function to the bolometric luminosity using the corrections
of \citet{Marconi04}. We determine $\nLP$ by logarithmically binning
the range of $\Lp$, and considering for each bin all simulations with
$\Lp$ in the given range. For each bin, then, we take the average
binned time the simulations spend in each luminosity interval, and
take that to be the quasar lifetime $\dtdL$. We then fit to the
observed luminosity function of \citet{Ueda03}, fitting
\begin{equation}
\phi(L)\approx \sum_{i}\ \dot{n}_{i}(L_{{\rm peak},\, i})\, {\left \langle{\frac
{\Delta t(L,\,L_{{\rm peak},\, i})}{\Delta \log L}}\right \rangle}
\end{equation}
and allowing $\dot{n}_{i}(L_{{\rm peak},\, i})$ to be a free
coefficient for each binned $\Lp=L_{{\rm peak},\, i}$.  Despite our
large number of simulations, the numerical binning process makes this
result noisy, especially at the extreme ends of the
luminosity function. However, the relevant result is clear -- the
qualitative behavior of $\nLP$ described above is unchanged. For
further discussion of the qualitative differences between the $\nLP$
distribution from different quasar models, and the robust nature of
our interpretation even under restrictive assumptions (e.g.\ ignoring
the early phases of merger activity or applying various models for
radiative efficiency as a function of accretion rate), we refer to
\citet{H05c}.

\subsection{The Luminosity Function at Different Frequencies and Redshifts}
\label{sec:fullLF}

Given a distribution of peak luminosities $\nLP$, we can use our model
of quasar lifetimes and the column density distribution as a function
of instantaneous and peak luminosities to predict the luminosity
function at any frequency. From a distribution of \NH\ values and some
a priori known minimum observed luminosity $L_{\nu}^{\rm min}$, the
fraction $f_{\rm obs}$ of quasars with a peak luminosity $L_{\rm
peak}$ and instantaneous bolometric luminosity $L$ which lie above the
luminosity threshold is given by the fraction of \NH\ values below a
critical $N_{H}^{\rm max}$, where $L_{\nu}^{\rm
min}=f_{\nu}L\,\exp{(-\sigma_{\nu}N_{H}^{\rm max})}$. Here,
$f_{\nu}(L)\equiv L_{\nu}/L$ is a bolometric correction and
$\sigma_{\nu}$ is the cross-section at frequency $\nu$. Thus,
\begin{equation}
N_{H}^{\rm max}(\nu,L,L_{\nu}^{\rm min})=\frac{1}{\sigma_{\nu}}\ln{\Bigl( \frac{f_{\nu}(L)L}{L_{\nu}^{\rm min}}\Bigr)},
\end{equation}
and for the lognormal distribution above, 
\begin{equation}
%f_{\rm obs}(\nu,L,L_{\rm peak})=\frac{1}{2} \Bigl[ 1 + {\rm erf}\Bigl( \frac{\log{(N_{H}^{\rm max}(\nu,L)/\meanNH(L,L_{\rm peak}))}}{\sqrt{2}\,\sigNH(L,L_{\rm peak})}\Bigr)\Bigr].
f_{\rm obs}(\nu,L,L_{\rm peak},L_{\nu}^{\rm min})=\frac{1}{2} \Bigl[ 1 + {\rm erf}\Bigl( \frac{\log{(N_{H}^{\rm max}/\meanNH)}}{\sqrt{2}\,\sigNH}\Bigr)\Bigr].
\end{equation}
This results in a luminosity function (in terms of the bolometric luminosity)
%\begin{equation}
%  \begin{split}
%     \phi(\nu,L,L_{\nu}^{\rm min})& =\int{f_{\rm obs}(\nu,L,L_{\rm peak},L_{\nu}^{\rm min})} \\
%     &\quad \times\frac{dt(L, L_{\rm peak})}{\dlgL}\,\nLP\,d\log(L_{\rm peak}),
%     \label{eqn:phifull}
%  \end{split}
%\end{equation}
  \begin{eqnarray}
     \phi(\nu,L,L_{\nu}^{\rm min})&=&\int{f_{\rm obs}(\nu,L,L_{\rm peak},L_{\nu}^{\rm min})} \nonumber\\
     &&\quad \times\frac{dt(L, L_{\rm peak})}{\dlgL}\,\nLP\,d\log(L_{\rm peak}),
     \label{eqn:phifull}
  \end{eqnarray}
where $\phi(\nu,L,L_{\nu}^{\rm min})$ is the number density of sources with 
bolometric luminosity $L$ per logarithmic interval in $L$, with an observed luminosity 
at frequency $\nu$ above $L_{\nu}^{\rm min}$.

Based on the direct fit for $\nLP$ in Figure~\ref{fig:LF.hists}, we wish to 
consider a functional form for $\nLP$ with a well-defined peak and 
falloff in either direction in $\log(\Lp)$. Therefore, we take $\nLP$ to be a 
lognormal distribution, with 
\begin{equation}
\nLP=\nstar\ \frac{1}{\sstar\sqrt{2\pi}}\exp\Bigl[ -\frac{1}{2}\,\Bigl( \frac{\log(\Lp/\lstar)}{\sstar}\Bigr)^{2}\Bigr].
\label{eqn:nLp.lognorm}
\end{equation}
Here, $\nstar$ is the total number of quasars being created
or activated per unit comoving volume per unit time; $\lstar$ is the
center of the lognormal, the characteristic peak luminosity of quasars
being born (i.e.\ the peak luminosity at which $\nLP$ itself peaks),
which is directly related to the break luminosity in the observed
luminosity function; and $\sstar$ is the width of the lognormal in
$\nLp$, and determines the slope of the bright end of the luminosity
function.  Since our model predicts that the bright end of the
luminosity function is made up primarily of sources at high Eddington
ratio near their peak luminosity, i.e.\ essentially identical to
``light-bulb'' or ``fixed Eddington ratio'' models, the bright-end
slope is a fitted quantity, determined by whatever physical processes
regulate the bright-end slope of the active black hole mass function
(possibly feedback from outflows or threshold cooling processes, e.g.\
Wyithe \& Loeb 2003; Scannapieco \& Oh 2004; Dekel \& Birnboim 2004),
unlike the faint-end slope which is a consequence of the quasar
lifetime itself, and is only weakly dependent on the underlying faint-end
active black hole mass or $\nLp$ distribution.

We note that although
this choice of fitting function has appropriate general qualities, it
is ultimately somewhat arbitrary, and we choose it primarily for
its simplicity and its capacity to match the data with a minimum of
free parameters. We could instead, for example, have chosen a double
power-law form with $\nLP =
\nstar/[(\Lp/\lstar)^{\gamma_{1}}+(\Lp/\lstar)^{\gamma_{2}}]$ and
$\gamma_{1}<\gamma_{2}$, but given that the entire faint end of the
luminosity function is dominated by objects with $\Lp\sim\lstar$, the
observed luminosity function has essentially no power to constrain the
faint end slope $\gamma_{1}$, other than setting an upper limit
$\gamma_{1}\lesssim0$. The ``true'' $\nLP$ will, of course, be a
complicated function of both halo merger rates at a given redshift and
the distribution of host galaxy properties including, but not
necessarily limited to, masses, concentrations, and gas fractions.

Having chosen a form for $\nLP$, we can then fit to an observed
luminosity function to determine $(\nstar,\,\lstar,\,\sstar)$. We take
advantage of the capability of our model to predict the luminosity
function at multiple frequencies, and consider both fits to just the
\citet{Ueda03} hard X-ray (2-10 keV) luminosity function, $\phi_{HX}$,
and fits to the \citet{Ueda03}, \citet{Miyaji01} soft X-ray (0.5-2
keV; $\phi_{SX}$), and \citet{Croom04} optical B-band (4400 \AA;
$\phi_{B}$) luminosity functions {\em simultaneously}.  These
observations agree with other, more recent determinations of
$\phi_{HX},\ \phi_{SX},\ \phi_{B}$
\citep[e.g.][respectively]{Barger05,HMS05,Richards05} at most
luminosities, and therefore we do not expect revisions to the observed
luminosity functions to dramatically change our results.  In order to
avoid numerical artifacts from fitting to extrapolated,
low-luminosity slopes in the analytical forms of these luminosity
functions, we directly fit to the binned luminosity function
data. Thus, we fit each luminosity function in all redshift intervals
for which we have binned data.

We find good fits ($\reducechi=68.8/104\approx0.66$) to all luminosity functions at all redshifts
with a pure peak-luminosity evolution (PPLE) model, for
which
\begin{equation}
\lstar=\lstar^{0}\,\exp({k_{L}\,\tau}),\ \nstar=constant,\ \sstar=constant,
\label{eqn:param.evol}
\end{equation}
where $\tau$ is the fractional lookback time ($\tau\equiv H_{0}\int^{z}_{0} dt$) and 
$k_{L}$ is a dimensionless constant fitted with $\lstar,\,\nstar,\,\sstar$.  
It is important to
distinguish this from ``standard'' pure luminosity evolution (PLE)
models \citep[e.g.,][]{Boyle88}, as with $\nLP>0$ and
$\lstar=\lstar(z)$ always, the density of sources, especially as a
function of observed luminosity at some frequency, evolves in a
non-trivial manner. 

We do not find significant improvement in the fits
if we additionally allow $\nstar$ or $\sstar$ to evolve with redshift 
($\Delta\chi^{2}\sim1-2$, depending on the adopted form for the evolution), 
and therefore consider only the simplest parameterization above (Equation~\ref{eqn:param.evol}). 
We also find acceptable fits for a pure density evolution model, with
$\lstar=$\,constant and $\nstar=\nstar^{0}\,\exp{(k_{N}\,\tau)}$ (both
keeping $\sstar$ fixed and allowing it to evolve as well).
%The best-fit values in this case are 
%$(\log\nstar,k_{S},\,\log\lstar,\,\sstar)=
%(-6.11,\,5.17,\,8.80,\,1.25)$ with 
%errors $(0.39,0.45,0.68,0.10)$. 
However, the fits are somewhat poorer ($\reducechi\approx1$), and the resulting
parameters over-produce the present-day density of low-mass supermassive
black holes and the intensity of the X-ray background by an order of
magnitude, so we do not consider them further. In either case, there is a 
considerable degeneracy between the parameters $\sstar$ and $\lstar$, 
where a decrease in $\lstar$ can be compensated by a corresponding 
increase in $\sstar$. This degeneracy is present 
because, as indicated above, the observed luminosity function only
weakly constrains the faint-end slope of $\nLP$. 

The observations shown are insufficient at high redshift to strongly
resolve the ``turnover'' in the total comoving quasar density at
$z\sim2-3$, and thus we acknowledge that there must be corrections to
this fitted evolution at higher redshift, which we address
below. However, as we primarily consider low redshifts, $z
\lesssim 3$,
and show that
the supermassive black hole population and X-ray background are
dominated by quasars at redshifts for which our $\nLp$ distribution is
well determined, this is not a significant source of error in most of
our calculations even if we extrapolate our evolution to $z\gg3$.

%\clearpage
\begin{figure*}
    \centering
    \plotone{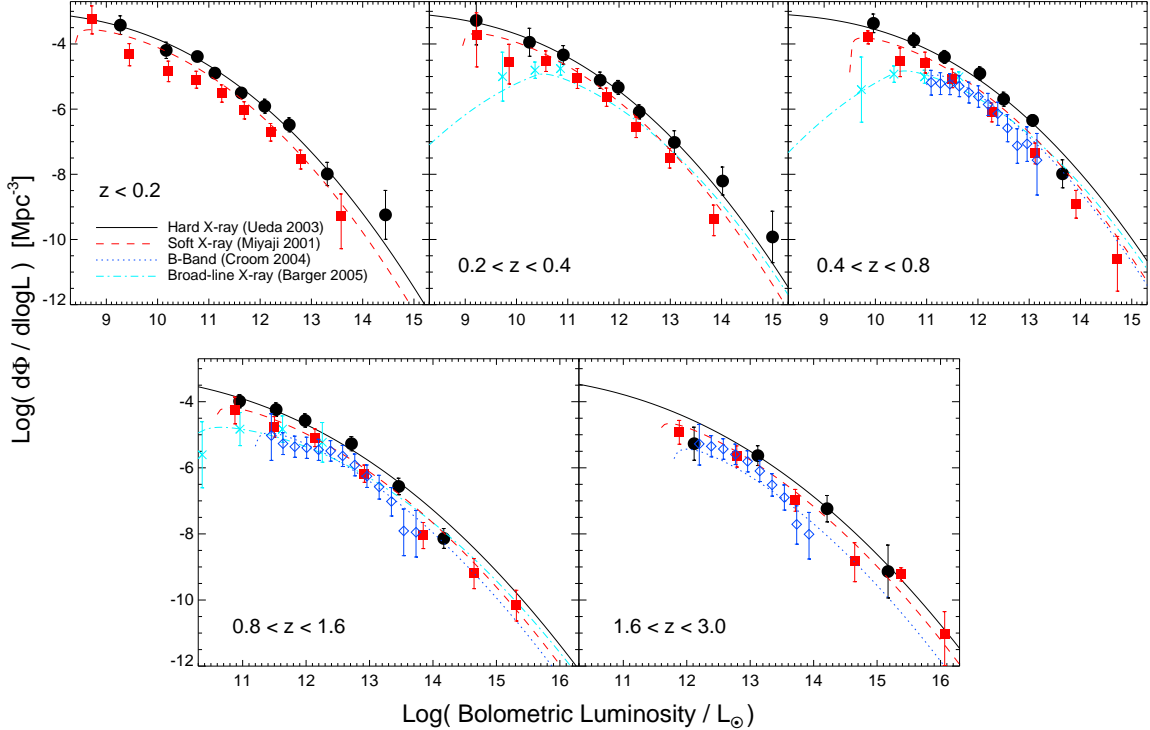}
    \caption{Best-fit luminosity function from the pure
    peak-luminosity evolution $\nLp$ distribution, for redshifts
    $z=0-3$. From our fitted lognormal $\nLP$ distribution, we
    simultaneously reproduce the luminosity function in the hard X-ray
    (2-10 keV; solid black line), soft X-ray (0.5-2 keV; dashed red
    line), and optical B-band (4400 \AA; dotted blue line) at all
    redshifts. Moreover, we reproduce the distribution of broad-line
    quasars in hard X-ray selected samples (cyan dot-dashed line), as
    described in \S~\ref{sec:BLqso}.  All quantities have been rescaled to
    bolometric luminosities for ease of comparison, using the
    corrections of \citet{Marconi04}, with the plotted error bars
    representing both quoted measurement errors and the estimated
    errors in the bolometric corrections. The observations are from
    \citet{Miyaji01} (soft X-ray; red squares), \citet{Ueda03} (hard
    X-ray; black circles), \citet{Croom04} (B-band, blue diamonds),
    and \citet{Barger05} (X-ray selected broad-line quasars; cyan
    crosses).
    \label{fig:LF.all}}
\end{figure*}
%\clearpage

Figure~\ref{fig:LF.all} shows the resulting best-fit PPLE luminosity
functions from the best-fit $\nLP$ distribution, for redshifts
$z=0-3$.  This has the best-fit ($\reducechi=0.67$) values
$(\log\lstar,k_{L},\,\log\nstar,\,\sstar)=
(9.94,\,5.61,\,-6.29,\,0.91)$ 
% (10.19,\,4.91,\,-6.43,\,0.91)$
with corresponding errors $(0.29,0.28,0.13,0.09)$. 
%$(0.48,0.36,0.28,0.09)$.  
Here, $\lstar$ is in solar luminosities and
$\nstar$ in comoving ${\rm Mpc^{-3}\,Myr^{-1}}$.  Fitting to the hard
X-ray data alone gives a similar fit, with the slightly different
values $(\log\lstar,k_{L},\,\log\nstar,\,\sstar)=
(9.54,\,4.90,\,-5.86,\,1.03) \pm (0.66,0.43,0.37,0.13)$,
$\reducechi=0.7$ (note the degeneracy between $\lstar$ and $\sstar$ in
the two fits). Our best-fit value of $k_{L}=5.6$ compares favorably to
the value $\sim6$ found by e.g.\ Boyle et al.\ (2000) and Croom et
al. (2004) for the evolution of the break luminosity in the observed
luminosity function, demonstrating that the break luminosity traces
the {\em peak} in the $\nLp$ distribution at all redshifts. These fits
and the errors were obtained by least-squares minimization over all
data points (comparing each to the predicted curve at its redshift and
luminosity), assuming the functional form we have adopted for $\nLP$.

The agreement we obtain at all redshifts, in each of
the hard X-ray (black solid line), soft X-ray (red dashed line), and
B-band (dark blue dotted line) is good. This is not at all guaranteed
by our procedure, as the fit is highly over-constrained, because we
fit three luminosity functions each at five redshifts to only four
free parameters. Of course, the choice of the functional form for $\nLp$
ensures that we should be able to reproduce at least one luminosity
function and its evolution (e.g.\ the hard X-ray luminosity function,
which is least affected by attenuation), but our modeling of the
column density distributions in mergers allows us to simultaneously
reproduce the luminosity functions in different wavebands without
imposing assumptions about obscured fractions or sources of
attenuation. Expressed as bolometric luminosity functions, $\phi_{B}$,
$\phi_{SX}$, and $\phi_{HX}$ would be identical in the absence of
obscuration, similar to the predicted $\phi_{HX}$ as obscuration is
minimal in the hard X-ray.

%\clearpage
\begin{figure}
    \centering
    \plotone{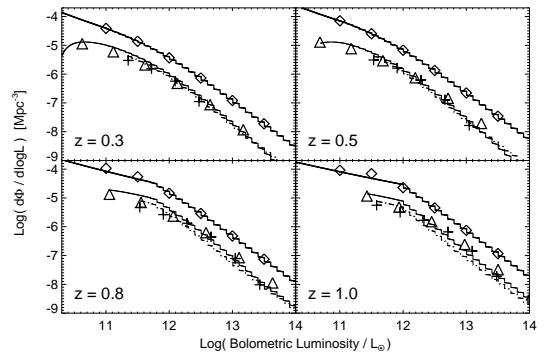}
    \caption{Hard X-ray (thick), soft X-ray (thin), and B-band
    (dot-dash) LFs determined from our model of quasar lifetimes and
    column densities, based on a distribution of intrinsic source
    properties fitted to the observed hard X-ray LF 
    and the limiting magnitudes of observed samples, at the different
    redshifts shown.  All quantities are rescaled to bolometric
    luminosities with the bolometric corrections of \citet{Marconi04}.
    Symbols show the observed LFs for hard X-rays
    \citep[][diamonds]{Ueda03}, soft X-rays
    \citep[][triangles]{Miyaji00}, and B-band
    \citep[][crosses]{Boyle00}. Reproduced from \citet{H05d}.
    \label{fig:LF.lowz}}
\end{figure}
%\clearpage

For redshifts $z\leq1$, we reproduce in our Figure~\ref{fig:LF.lowz},
Fig.~2 of \citet{H05d}, which shows in detail the agreement between
hard X-ray \citep{Ueda03}, soft X-ray \citep{Miyaji00}, and optical
\citep{Boyle00} luminosity functions resulting from the time and
luminosity dependent column density distributions derived from the
simulations. The differential extinction predicted for different
frequencies (and magnitude limits) of observed samples based on the
column density distributions in our simulations accounts for the
different shape of the luminosity function in each band, and the
evolution of the luminosity function with redshift is driven by a
changing $\lstar$, the peak of the $\nLp$ distribution
(Equation~\ref{eqn:param.evol}).  We emphasize that in our analysis,
the key quantity constrained by observations is the fitted $\nLP$
distribution with redshift. All other quantities and distributions are
derived from the basic input physics of our simulations, with no
further assumptions or adjustable factors in our modeling beyond the
prescription for Bondi (Eddington-limited) accretion and
$\sim5\%$ energy deposition in the ISM, which are themselves
constrained by observations and theory as discussed in
\S~\ref{sec:methods} and in Di Matteo et al.\ (2005).

We can, of course, fit the previously defined simpler model of quasar
lifetimes, either a ``light-bulb'' or exponential light curve/fixed
Eddington ratio model, and obtain an identical hard X-ray luminosity
function. We determine these fits (see also Equation~\ref{eqn:nLp.LB}
\& \ref{eqn:nLp.Edd}) and use them throughout when we compare the
predictions of such models (described in \S~\ref{sec:altmodels}) to
those of our simulated quasar lifetimes in our subsequent
analysis. Applying a standard torus model to any model of the
luminosity function reproduces, by design, the mean offset between the
B-band and hard X-ray luminosity functions, as the parameters of this
model are {\it tuned} to reproduce this offset. As many observations
show, the fraction of broad-line quasars increases with luminosity
\citep{Steffen03,Ueda03,Hasinger04,sazrev04,Barger05,Simpson05}, and
so reproducing the relationship between B-band and hard X-ray
luminosity functions requires adding parameters to the standard torus
model which allow luminosity-dependent scalings, i.e.\ the class of
``receding torus'' models. These, again by construction, reproduce the
distinction between hard X-ray and B-band quasar luminosity functions,
including the dependence of this difference on luminosity. These are,
however, phenomenological models designed to fit these
observations. Our simulations, on the other hand, provide a
self-consistent description of the column density, which predicts the
{\em differences} between hard X-ray, soft X-ray, and optical
luminosity functions without the addition of tunable parameters or
model features designed to reproduce these observations.

Our fits are accurate down to low luminosities, as is clear from our
prediction for the X-ray luminosity function at bolometric
luminosities $L\sim\Lcut{9}$. Furthermore, we have calculated the
predicted $z\lessim0.1$ luminosity function in the B-band as well as
in H$\alpha$ emission, using the conversion between the two from
\citet{Hao05} and comparing directly to their luminosity functions for
Seyfert galaxies and low-luminosity active galactic nuclei (AGN) (both
type I and II), and find that our distribution $\nLp$ and model for
quasar lifetimes and obscuration reproduces the complete observed
luminosity function down to a B-band luminosity $M_{B}\sim-16$.
Although our prediction falls below the observed Seyfert luminosity
function at fainter magnitudes, there is no reason to believe that
mergers should be responsible for all nuclear activity at these
luminosities (and indeed alternative fueling mechanisms for such faint
objects likely exist) - it is surprising, in fact, that this picture
reproduces the observed AGN activity to such faint luminosities.

Using the bolometric corrections of \citet{Elvis94} instead of
\citet{Marconi04} results in a significantly steeper cutoff in the
luminosity function at high bolometric luminosities, as the bolometric
luminosity inferred for the brightest observed X-ray quasars is almost
an order of magnitude smaller using the \citet{Elvis94}
corrections. However, this is because the \citet{Elvis94} bolometric
corrections do not account for any dependence on luminosity, and
further the quasars in the sample of Elvis et al.\ (1994) are X-ray
bright \citep{ERZ02}, whereas it has been well-established that the
ratio of bolometric luminosity to hard or soft X-ray luminosity
increases with increasing luminosity
\citep[e.g.,][]{Wilkes94,Green95,VBS03,Strateva05}.  Recent comparisons
between large samples of quasars selected by both optical and X-ray
surveys \citep{RisalitiElvis05} further suggests that this is an
intrinsic correlation, not driven by e.g.\ the dependence of
obscuration on luminosity.  For a direct comparison of the bolometric
luminosity functions resulting from the two corrections, we refer to
\citet{H05d}. Our analysis uses the form for the UV to X-ray flux
ratio, $\alpha_{\rm OX}$, from \citet{VBS03}, but our results are
relatively insensitive to the different values found in the
literature. It is important to account for this dependence, as it
creates a significant difference in the high-luminosity end of the
bolometric quasar luminosity function and implies that a
non-negligible fraction of the brightest quasars are not seen in
optical surveys \citep[see the discussion in][]{Marconi04,Richards05}.

%\clearpage
\begin{figure*}
    \centering
    \plotone{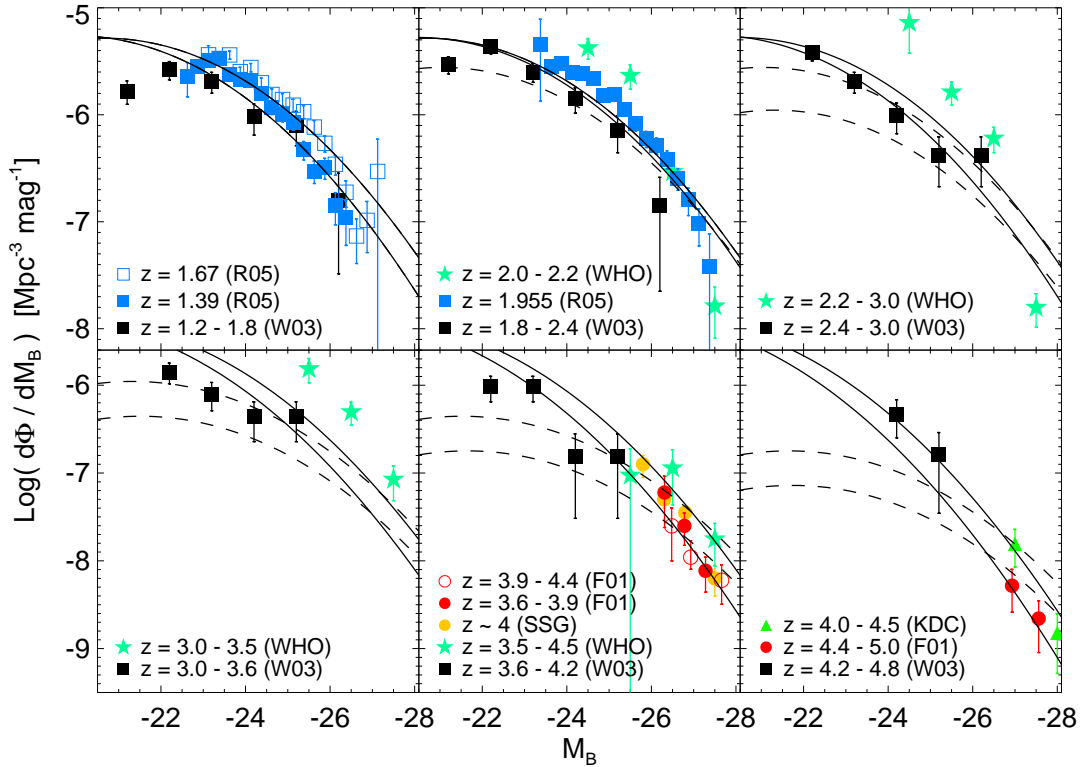}
    \caption{Running our predicted broad-line luminosity function
    (determined in \S~\ref{sec:LF}, \ref{sec:BLqso}) to high redshifts, with either total
    density (dashed lines) or break luminosity ($\lstar$; solid lines)
    decreasing exponentially with redshift above $z=2$. In each panel,
    our prediction is shown for the minimum and maximum redshift of
    the corresponding interval from the COMBO-17 luminosity function
    of \citet{Wolf03} (W03; black squares). Other references for the
    observations shown are: R05 - \citet{Richards05}, WHO -
    \citet{WHO94}, F01 - \citet{Fan01}, SSG - \citet{SSG}, KDC -
    \citet{KDC}.
    \label{fig:LF.highz}}
\end{figure*}
%\clearpage

Finally, our fitted form for the evolution of the break luminosity,
with $\lstar\propto\exp{(k_{L}\tau)}$, cannot continue to arbitrarily
high redshift. At redshifts $z\gtrsim2-3$, this asymptotes
because $\tau\rightarrow 1$, whereas the observed quasar population
declines above $z\sim2$. This difference is not important for most of
our calculated observables, as they are either independent of
high-redshift evolution or evolve with cosmic time in some fashion as
$\propto\int{\nLp\,{\rm d}t}$, with little time and thus negligible
contributions to integrated totals at high redshifts. However, some
quantities, in particular the high-mass end of the black hole mass
function (see \S~\ref{sec:smbh}), which is dominated by the small
number of the brightest quasars at high redshifts, can receive large
relative contributions from these terms. Therefore, it is important in
estimating these quantities to be aware of the turnover in the quasar
density at high redshifts.

We quantify this in Figure~\ref{fig:LF.highz}, where we show the
predicted broad-line luminosity function (where the broad-line phase
is determined below in \S~\ref{sec:smbh}) in six luminosity intervals
from $z\sim1.2-4.8$.  The intervals are those of the COMBO-17
luminosity function from \citet{Wolf03}, but we further compare to the
observed luminosity functions of \citet{WHO94}, \citet{SSG},
\citet{KDC}, \citet{Fan01}, and \citet{Richards05} at the appropriate
(labeled) redshifts.  At each redshift $z>2$, we take the
fitted $\nLp$ distribution above (Equations~\ref{eqn:nLp.lognorm},\
\ref{eqn:param.evol}) and rescale it according to an exponential
cutoff: either pure density evolution (PDE),
$\nLp\rightarrow\nLp\times10^{-\alpha_{\rm PDE}\,(z-2)}$, or pure peak
luminosity evolution (PPLE),
$\lstar\rightarrow\lstar\times10^{-\alpha_{\rm PPLE}\,(z-2)}$.
Fitting to the data gives $\alpha_{\rm PDE}\sim0.65$ and $\alpha_{\rm
PPLE}\sim0.55$, ($\reducechi\approx1.3$ for both) in reasonable
agreement with the density evolution of e.g.\ Fan et al.\ (2001). We
note that this evolution, extrapolated as far as $z\sim6$, is
consistent also with the constraints on $z\sim6$ quasars from
\citet{Fan03}, especially in the PPLE case.

In each panel, we plot the resulting broad-line luminosity function
(see \S~\ref{sec:BLqso}), for both the minimum and maximum redshift of
the redshift bin, and both the PPLE (solid lines) and PDE (dashed
lines) cases.  The degeneracy between these 
possibilities is well-known, as
current observations do not resolve the break in the luminosity
function. Furthermore, the predicted luminosity function should be
considered uncertain especially at low luminosities, as the quasar
lifetime at these luminosities and redshifts can become comparable to
the age of the Universe, at which point our formalism for the
luminosity function as a function of $\nLp$ becomes inaccurate.
However, we are able to make testable predictions, based on
differences between the two models in integrated {\em galaxy}
properties (for example, color-magnitude diagrams of red sequence
galaxies at low masses or the fraction of recently formed spheroids as
a function of mass and redshift), which distinguish the PPLE and PDE
models for the evolution of the quasar luminosity function at
$z\gtrsim2-3$ \citep{H05e}.  Owing to these degeneracies and the poor
constraints on the observed high-redshift luminosity functions, we
have not considered them (those at $z>3$) in our fits to $\nLp$, but
use them here to roughly constrain the turnover in the quasar
density above $z\sim2$ (i.e.\ fitting to $\alpha_{\rm PDE}$ and
$\alpha_{\rm PPLE}$). Which form of the turnover we use makes little
difference in our subsequent analysis, but, as discussed above,
including {\em some} turnover is important in calculating select
quantities such as the extreme high-mass end of the black hole mass
function.

\subsection{The Observed \NH\ Distribution}
\label{sec:NHdistrib}

%\clearpage
\begin{figure*}
    \centering
    \plotone{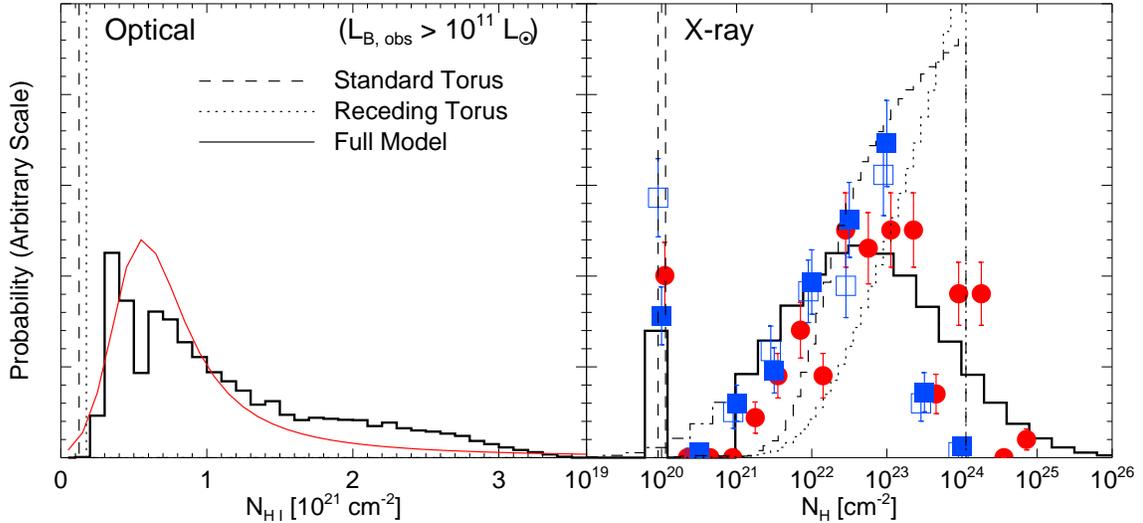}
    \caption{Left panel: Distribution of column densities expected
    from the characteristic quasars $\Lp\sim\lstar$ of the luminosity
    function observed in optical samples, for a standard torus model 
    of quasar obscuration (dashed), 
    a receding torus model (dotted), and the distributions of column densities 
    as a function of instantaneous and peak luminosity in our simulations (solid).  
    The distribution of neutral
    \NHI\ values is obtained requiring an observed B-band luminosity
    $>\Lcut{11}$. The smooth red curve is the best-fit to the
    $E_{B-V}$ distribution of bright SDSS quasars with $z<2.2$, from
    \citet{Hopkins04}, rescaled to column densities and plotted about
    a peak (mode) \NHI\ (undetermined in Hopkins et al.\ 2004) of
    $\nhi\approx0.5\times10^{21}\,{\rm cm^{-2}}$. The $i$-band
    absolute magnitude limit imposed in the observed sample,
    $M_{i}<-22$, corresponds approximately to our plotted B-band limit
    $\LBo>\Lcut{11}$. Reproduced from \citet{H05b}.  Right panel:
    Integrated distribution of total (neutral and ionized) column
    densities expected for a complete hard X-ray sample, from the
    column densities of our simulations and the $\nLp$
    distribution. The distribution below $10^{21}\ {\rm cm^{-2}}$ is
    shown (dot-dashed line) and re-plotted as a single bin at
    $\nh=10^{20}\ {\rm cm^{-2}}$ for our modeled columns.  Data shown are the results of
    \citet{Treister04} (blue squares) and \citet{Mainieri05} (red
    circles), with assumed Poisson errors. Solid squares assume an
    intrinsic photon index $\Gamma=1.9$, for the soft X-ray quasar
    spectrum, open squares $\Gamma=1.7$.
    \label{fig:NH.distrib}}
\end{figure*}
%\clearpage

Given the column density distributions and quasar lifetimes calculated
from our simulations in \S~\ref{sec:methods}, and the quantity $\nLP$
determined above (\S~\ref{sec:fullLF}), we can predict the
distribution of column densities observed in a given sample. This will
depend not only on the range of observed luminosities and the redshift
of the sample, but also on the minimum observed magnitude and
frequency (i.e.\ the selection function) of the sample.  For a nearly
complete sample or estimate of the luminosity function, for example
the hard X-ray luminosity function, at least to $\nh\sim10^{25}\ {\rm
cm^{-2}}$, we can integrate the $\nh(L,\Lp)$ distribution over
the $\nLP$ distribution (weighted by the lifetime at $L$).

Figure \ref{fig:NH.distrib} plots the resulting distribution of column
densities for this analysis. The left panel reproduces and expands
upon a portion of Fig.~3 of \citet{H05b}, showing the distribution of
column densities (scaled linearly) expected from the characteristic
quasars $\Lp\sim\lstar$ of the luminosity function observed in optical
samples, based on the simulated column density distributions as a
function of luminosity and peak luminosity (solid black line).
Specifically, we plot the distribution of neutral \NHI\ values
requiring that the observed B-band luminosity be above some reference
value $\LBm$. The smooth curve shown is the best-fit to the $E_{B-V}$
distribution of bright SDSS quasars with $z<2.2$, from
\citet{Hopkins04}. The curve has been rescaled in terms of the column
density (inverting our gas-to-dust prescription) and plotted about a
peak (mode) \NHI\ (undetermined in Hopkins et al.\ 2004) of
$\nhi\approx0.5\times10^{21}\,{\rm cm^{-2}}$.  The observationally
implied $E_{B-V}$ distribution is determined from fitting to the
distribution of photometric reddening in all SDSS bands (i.e.\ using
the five-band photometry as a proxy for spectral fitting) in Sloan
quasars, relative to the modal quasar colors at each redshift, for
quasars with an absolute magnitude limit $M_{i}<-22$.  The $i$-band
absolute magnitude limit imposed in the observed sample, $M_{i}<-22$,
corresponds approximately to our plotted B-band limit
$\LBo>\Lcut{11}$. This estimate does not account for bright but
strongly reddened quasars having their colors altered to the point
where color selection criteria of quasar surveys will not include
them. However, this effect would only serve to bring our distribution
into better agreement with observations, as it would slightly lower
the high-\NHI\ tail.  We also consider the predictions of a standard
torus model and receding (luminosity-dependent) torus model in the
figure (dashed and dotted lines, respectively). These should not be
taken literally in this case -- they reflect that these
phenomenological models do not predict the distribution of
low/moderate column densities, but rather assume that all lines
of sight not intersecting the torus are ``unobscured,'' and encounter
some constant, small column density (usually chosen to be
$\nh\sim10^{20}\,{\rm cm^{-2}}$).

The right panel of \ref{fig:NH.distrib} shows the integrated
distribution (in $\log\nh$) for a complete hard X-ray
sample, both as predicted from our simulations based on the joint
distribution of column density, luminosity, and peak luminosity
(solid), and for both the standard torus model (dashed) and receding
torus model (dotted) described in \S~\ref{sec:altmodels}. The data
shown are the results of \citet{Treister04} (blue squares) and
\citet{Mainieri05} (red circles), with assumed Poisson errors, from
multiband {\it Chandra} and {\it HST} observations of GOODS
fields. The solid squares are obtained by assuming an intrinsic photon
index for the soft X-ray quasar spectrum of $\Gamma=1.9$, the open
squares assuming $\Gamma=1.7$. For the sake of direct comparison with
observed distributions, objects with $\nh<10^{21}\ {\rm cm^{-2}}$, for
which only an upper limit to the column density would be determined in
X-ray observations, are grouped together and plotted as a single bin
at $\nh=10^{20}\ {\rm cm^{-2}}$. The actual distribution below
$10^{21}\ {\rm cm^{-2}}$ is shown as a dot-dashed line. We note that
our model of the quasar spectrum assumes a photon index $\Gamma=1.9$
in the soft X-ray, but this has no effect on the column densities
calculated from the surrounding gas in our simulations.

The agreement between the observed column density distribution and the
result of our simulations once the same selection effect is applied
supports our model for quasar evolution, and the good agreement
extends to both optical and X-ray samples.  Probing to fainter
luminosities or frequencies less affected by attenuation broadens the
column density distribution, as is seen from the inferred column
density distributions in the X-ray.  This broadening occurs because,
at lower luminosities, observers will see both intrinsically bright
periods extinguished by larger column densities (broadening the
distribution to larger \NH\ values) and intrinsically faint periods
with small column densities (broadening the distribution to smaller
\NH\ values).  The distribution as a function of reference luminosity
is a natural consequence of the dynamics of the quasar
activity. Throughout much of the duration of bright quasar activity,
column densities rise to high levels as a result of the same process
that feeds accretion, producing the well-known reddened population of
quasars \citep[e.g.][]{Webster95,Brotherton01,Francis01,
Richards01,Gregg02,White03,Richards03}, extending to bright quasars
strongly reddened by large \NHI. Furthermore, a significant number of
quasars are extinguished from optical samples or attenuated to lower
luminosities, giving rise to the distinction between luminosity
functions in the hard X-ray, soft X-ray, and optical.

The standard torus model described in \S~\ref{sec:altmodels}, although
unable to predict the distribution of column densities seen in
optically, relatively unobscured quasars, does a fair job of reproducing
the observed distribution of X-ray column densities. The parameters of
the model are, of course, chosen to reproduce the data shown
\citep[the model parameters are taken
from][]{Treister04}. Nevertheless, our prediction is still a
better fit to the observed distribution, with
$\reducechi\approx2$ as opposed to $\reducechi\approx7$ (although the
absolute values depend on the estimated systematic
errors in the column density estimations). The receding torus model
fares even more poorly in reproducing the observed column density
distributions, and is ruled out at high significance
($\reducechi\approx10$), although this can be alleviated if the
observed samples are assumed to be incomplete above
$\nh\sim10^{23}\,{\rm cm^{-2}}$. This disagreement results because, in
order to match the observed scaling of broad-line fraction with
luminosity (see \S~\ref{sec:BLqso} below), this model assumes a
larger covering fraction for the torus at lower luminosities,
normalized to a similar obscured fraction as the standard torus model
near the break in the observed quasar luminosity function. However,
since quasars with luminosities below the break dominate the total
number counts, this predicts that the cumulative column density
distribution must be significantly more dominated by objects with
large covering angles, giving a larger Compton-thick population,
inconsistent with the actual observed column density distribution.

Although we do not see a significant fraction of extremely
Compton-thick column densities $\nh\gtrsim10^{26}\,{\rm cm^{-2}}$ in
the distributions from our simulations, our model does not rule out
such values. It is possible that bright quasars in unusually massive
galaxies or quasars in higher-redshift, compact galaxies which we have
not simulated may, during peak accretion periods, reach such values in
their typical column densities. Moreover, as our model assumes
$\sim90\%$ of the mass of the densest gas is clumped into cold-phase
molecular clouds, a small fraction of sightlines will pass through
such clouds and measure column densities similar to those shown for
the ``cold phase gas'' in, e.g.\ Figure 2 of \citet{H05a},
$\nh\gtrsim10^{25-26}{\rm cm^{-2}}$.

Furthermore, we have not determined the ``shape'' at any instant of
the obscuration (e.g.\ the dependence of obscuration on radial
direction), as in practice, for most of the most strongly obscured
phases in peak merger activity, the central regions of the merging
galaxies are highly chaotic. Generally, the scale of the obscuration
in the peak merger phases is $\sim100\,$pc, quite different than that
implied by most traditional molecular torus models, but we note that
our resolution limits, $\sim 20\,$pc in the dense central regions of
the merger, prevent our ruling out collapse of gas in the central
regions into a smaller but more dense torus. However, several efforts
to model traditional tori through radiative transfer simulations
\citep[e.g.,][]{GD94,Schartmann05} suggest significant column
densities produced on scales of $\sim100-200\,$pc, comparable to our
predictions, and we note that only the solid angle covered by a torus,
not the absolute torus scale, is constrained in the typical
phenomenological torus model \citep[e.g.][]{Antonucci93}.

Whether the obscuration of bright quasars originates on larger scales
than is generally assumed is observationally testable, either through
direct probes of polarized scattered light tracing the
obscuring/reflecting structure \citep[e.g.,][]{Zakamska05}, or through
correlations between obscuration and e.g.\ host galaxy morphologies
and inclinations \citep[e.g.,][]{Donley05}.  These larger scales
typical of the central regions of a galaxy are widely accepted as the
scales of obscuration in starbursting systems (e.g.\ Soifer et
al. 1984a,b; Sanders et al.\ 1986, 1988a,b; for a review, see e.g.\
Soifer et al.\ 1987), which in our modeling is associated with rapid
obscured quasar growth and precedes the quasar phase.  Thus, it is
natural to associate obscuration with these large scales in any
picture which associates starbursts and rapid black hole growth or
quasar activity, as opposed to the smaller scales $\sim$\,pc
implied by torus models primarily developed to reproduce observations of
quiescent, low-luminosity Type II AGN, which are usually not directly
associated with merger activity. These low-luminosity AGN are in a
relaxed state, suggesting the possibility that the remaining cold gas
in the central regions of our merger remnants will collapse once the
violent effects of the merger and bright quasar phase have passed,
producing a more traditional small torus in a quiescent nucleus.  The
central point is that regardless of the form of obscuration, the
typical magnitude of the obscuration is a strongly evolving function
of time, luminosity, and host system properties, and the observed
column density distributions reflect this evolution.

\section{Broad-Line Quasars}
\label{sec:BLqso}
\subsection{Determining the Broad-Line Phase}
\label{sec:whenBL}

Optical samples typically identify quasars through their colors,
relying on the characteristic non-stellar power-law continua of such
objects. However, observations of X-ray selected AGN show a large
population of so-called Type 2 AGN, most of which have Seyfert-like
luminosities and typical spectra in X-rays and wavelengths longward of
$1\,\mu$m \citep[e.g.,][]{Elvis94}, but are optically obscured to the
point where no broad lines are visible. Their optical continua, in
other words, resemble those of typical galaxies and thus they are not
identified by conventional color selection techniques in optical
quasar surveys. Traditional unification models \citep{Antonucci93}
have postulated a static torus as the explanation for the existence of
the Type 2 population, with such objects viewed through the dusty
torus and thus optically obscured. Moreover, both synthesis models
of the X-ray background
\citep{SW89,Madau94,Comastri95,Gilli99,Gilli01} and recent direct
observations in large surveys \citep[e.g.,][]{Zakamska04,Zakamska05}
indicate the existence of a population of Type 2 quasars, with similar
obscuration but intrinsic (unobscured) quasar-like luminosities.

Observations of both radio-loud
\citep{HGD96,SRL99,Willott00,SR00,GRW04} and radio-quiet
\citep{Steffen03,Ueda03,Hasinger04,sazrev04,Barger05,Simpson05}
quasars, however, have shown that the broad-line fraction increases
with luminosity, with broad-line objects representing a large fraction
of all AGN at luminosities above the ``break'' in the luminosity
function and rapidly falling off at luminosities below the break.
Modifications to the standard torus unification model explain this via
a luminosity-dependent inner torus radius \citep{Lawrence91}, but this
represents a tunable modification to a purely phenomenological
model. Furthermore, as the observations have improved, it has become
clear that even these luminosity-dependent torus models cannot produce
acceptable fits to the broad line fraction as a function of luminosity
\citep[e.g.,][]{Simpson05}. However, we have shown above that the
obscuring column, even at a given luminosity, is an evolutionary
effect, dominated by different stages of gas inflow in different
merging systems giving rise to varying typical column densities,
rather than a single static structure. It is of interest, then, to
calculate when quasars will be observed as broad-line objects, and to
compare this with observations of broad line quasars and their
population as a function of luminosity.

%\clearpage
\begin{figure*}
    \centering
    \plotone{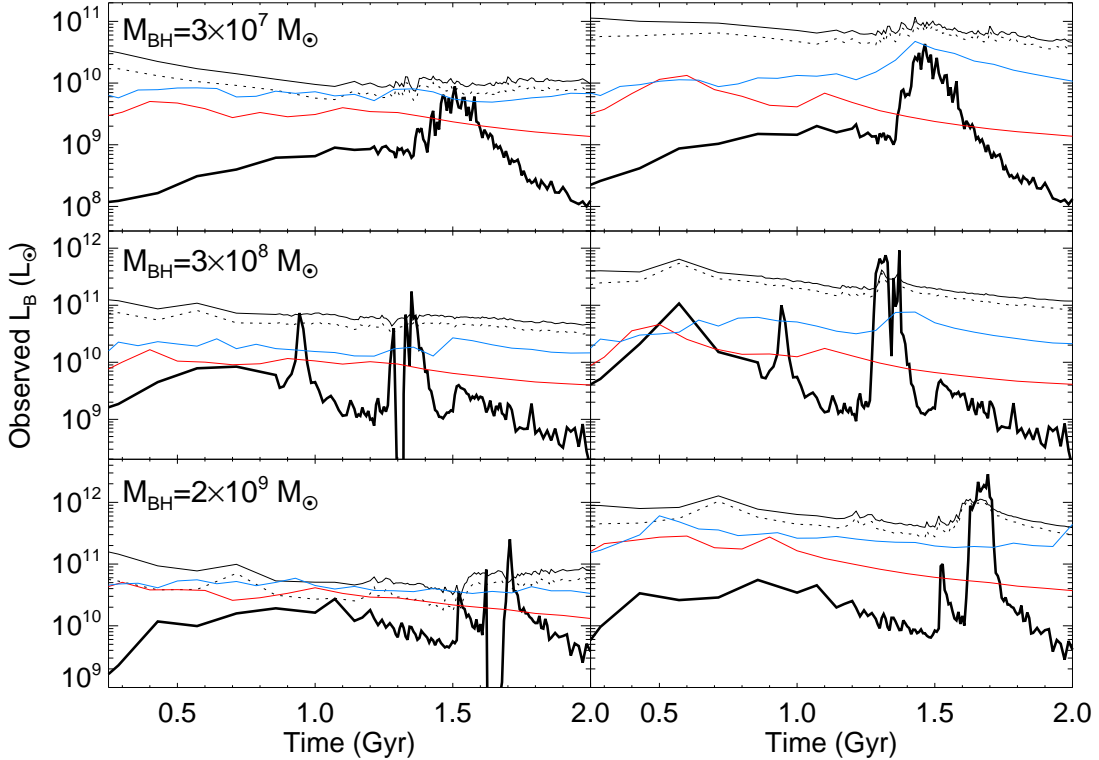}
    \caption{Intrinsic (right panels) and median attenuated (left
    panels) B-band luminosity of the quasar (thick line) and host
    galaxy (integrated over all stars, thin line; ignoring bulge
    stars, dotted line) as a function of time. Results are shown from
    three representative simulations: A2, A3, and A5 (see \S~\ref{sec:sims})
    with $\qeos=1.0,\, \zgal=0$, and virial velocities
    $\vvir=113,\, 160,\ {\rm and}\ 320\, {\rm km\,s^{-1}}$. Each
    quasar should be observable as a broad-line AGN when $L_{\rm
    B,\,QSO}\gtrsim L_{\rm B,\,host}$. Colors show the stellar light curve 
    with different gas fractions $\fgas=1.0$ (black), $\fgas=0.4$ (blue), 
    and $\fgas=0.2$ (red); quasar light curves are similar for each gas fraction.     
    \label{fig:BL.in.sims}}
\end{figure*}
%\clearpage

Figure~\ref{fig:BL.in.sims} shows the B-band luminosity as a function
of time for both the quasars and host galaxies in three representative
simulations: the A2, A3, and A5 cases described in detail in
\S~\ref{sec:sims}. These simulations each have $\fgas=1.0,\,
\qeos=1.0,\, \zgal=0$, with virial velocities $\vvir=113,\, 160\, {\rm
and}\ 320\, {\rm km\,s^{-1}}$, with resulting final black hole masses
$\mbhf=3\times10^{7},\ 3\times10^{8},\ {\rm and}\ 2\times10^{9}\,
M_{\sun}$, respectively.  The thick line in each case shows the quasar
B-band luminosity, and the thin line shows the integrated B-band
luminosity of all stars in the galaxy. New stars are formed
self-consistently in the simulations according to the ISM gas
properties, equation of state and star formation model described in
\citet{SH03}, with the age and metallicity taken from the local
star-forming ISM gas, which is enriched by supernova feedback from
previous star formation. We then use the stellar population synthesis
model of \citet{Bruzual03} to determine the B-band luminosity (the
B-band mass-to-light ratio) of new stars based on the stellar age and
metallicity.  The dotted line shows the result neglecting bulge
particles, which must be initialized at the beginning of the
simulation with random or uniform ages and metallicities instead of
those quantities being determined self-consistently from the
simulation physics. The right panels plot the intrinsic values of
these quantities, and the left panels plot the median observed values
of these quantities, where we have used our method for determining
column densities and dust attenuation (\S~\ref{sec:NH}) to every star
and bulge particle for each line of sight.

Unfortunately, the host galaxy luminosity does {\em not} scale
with instantaneous and peak quasar luminosity as do, for example, the
quasar lifetime and obscuration. Rather, there are important
systematic dependencies, the largest of which is the dependence on host galaxy gas
fraction.  If the host galaxies are more massive, more concentrated,
or have a weaker ISM equation of state pressurization, then they will
more effectively drive gas into the central regions and maintain high
gas densities for longer periods of time, as the deeper potential well
or lack of gas pressure requires more heat input from the quasar
before the gas can be expelled. These conditions will generally
produce a quasar with a larger peak luminosity (final black hole
mass), but also form more new stars, meaning that the B-band relation
between host and quasar luminosity is roughly preserved. 

However, the the black hole consumes only a small fraction of the
available gas (comparison of e.g.\ the stellar mass and black
hole mass suggests the black hole consumes $\sim0.1\%$ of the gas
mass), and so, at least above some threshold $\fgas\lesssim0.1$, the
quasar peak luminosity does not significantly depend on the galaxy gas
fraction (see, e.g.\ Figure 2 of Robertson et al.\ 2005b). But, the
mass of new stars formed {\it during} the merger does strongly depend on the
available gas. For example, simulations which are otherwise identical
but have initial $\fgas=0.2,\ 0.4,\ 0.8,\ 1.0$ (i.e.\ an increasing
fraction of the initial disk mass in gas instead of stars) produce
similar peak quasar luminosity and final total stellar mass (within
$\sim30\%$ of one another), reflecting the conversion of most
gas into stars and the fact that the peak quasar luminosity is
determined more by the depth of the potential well than the total
available gas supply. But, the mass of {\it new}
stars formed in a merger scales
roughly as $M_{\ast,\, \rm new}\propto \fgas$ (as it must if the
initial gas fraction does not change the final total stellar mass),
and since young stellar populations dominate the observed B-band
luminosity (especially during the peak merger and starburst phases
associated with the bright quasar phase of interest), this implies
roughly that $L_{B}\propto\fgas$.

We demonstrate this explicitly in Figure~\ref{fig:BL.in.sims}, where
we show in each panel the host galaxy and stellar B-band light curves
for otherwise identical simulations with different gas fractions,
$\fgas=0.2\ {\rm (red)},\ 0.4\ {\rm (blue)},\ {\rm and}\ 1.0\ {\rm
(black)}$.  In each of these cases, the quasar light curve is nearly
identical (we show only the $\fgas=1.0$ quasar lightcurve, for
clarity, but the others are within $\sim30\%$ of the curve shown at
most times, with no systematic offset).

In order for a quasar to be classified as a ``broad-line'' object, the
optical spectrum must be visible and identified as such in the
observed sample.  This is clearly related to the ratio of quasar to
host galaxy luminosity, but the threshold for classification is not
obvious. In an X-ray or IR-selected sample, optical follow-up should
be able to disentangle host galaxy light and identify quasar
broad-line spectra with fluxes a factor of several fainter than the
host. However, automated optical selection based on color or
morphological criteria might well exclude objects unless the quasar
luminosity is a factor of several greater than that of the host
galaxy.  Therefore, there is significant systematic uncertainty in the
theoretical definition of a broad-line quasar. To first order, based
on the above arguments, we can classify ``broad-line quasars'' as
objects in which the quasar optical luminosity is larger than some
multiple $f_{\rm BL}$ of the host galaxy optical luminosity.  Because
the relevant ratio is different depending on the survey and selection
techniques, we consider the range $f_{\rm BL} = 0.3-3$, with a rough
median $f_{\rm BL}=1$. Furthermore, because our simulations do not
allow us to model the broad-line regions of the quasar or spectral
line structures as influenced by e.g.\ reddening and dust absorption,
we adopt the B-band luminosity of the quasar and host galaxy as
a proxy for optical luminosity and more complex (but often quite
sample-specific) color and morphological selection criteria.

%\clearpage
\begin{figure}
    \centering
    \plotone{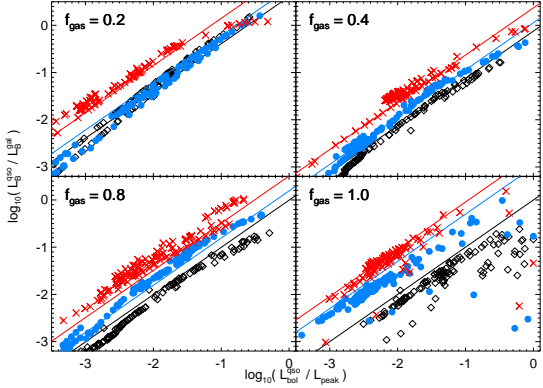}
    \caption{Ratio of observed (attenuated) B-band quasar luminosity to host galaxy luminosity 
    as a function of the ratio of instantaneous to peak quasar bolometric luminosity. Results are 
    from simulations A2 (black diamonds), A3 (blue circles), and A5 (red $\times$'s) 
    (the same simulations shown in Figure~\ref{fig:BL.in.sims})
    with $\qeos=1.0,\, \zgal=0$, and virial velocities
    $\vvir=113,\, 160,\ {\rm and}\ 320\, {\rm km\,s^{-1}}$. 
    Each panel shows the same simulations except for a 
    different initial gas fraction $\fgas = 0.2,\ 0.4,\ 0.8,\ 1.0$ as labeled. Solid lines are the 
    predictions of Equation~\ref{eqn:BL.scaling}.
    \label{fig:test.BL}}
\end{figure}
%\clearpage

In Figure~\ref{fig:BL.in.sims}, the B-band host galaxy luminosity is
quite flat as a function of time, relative to the quasar B-band
luminosity, and is roughly given by $L_{B}^{\rm gal}/L_{\sun}\sim
M_{\ast,\, \rm new}/M_{\sun}$, where $M_{\ast,\, \rm new}$ is the mass
of new stars formed in the merger.  As noted above, this scales
approximately linearly with initial gas fraction at fixed final total
stellar mass $M_{\ast}$, giving $L_{B}^{\rm gal}/L_{\sun}\approx
c_{\rm gal} (M_{\ast}/M_{\sun})\fgas$, where $c_{\rm gal}$ is a
correction of order unity which we can fit from the simulations
(essentially a mean mass-to-light ratio for the newly formed stars).
The bolometric correction of the quasar is usually defined by $L_{\rm
bol}^{\rm qso}=c_{B}L_{B}^{\rm qso}$, and the quasar peak luminosity
is $\Lp=c_{L}\, L_{\rm Edd}(M_{\rm BH}^{f})$, where again $c_{L}$ is a
correction factor of order unity which we can calculate from our form
for the quasar lifetime (see Equation~\ref{eq:Mbhf}) or measure in the
simulations.  

If we require that the quasar B-band luminosity be
larger than a factor $f_{\rm BL}$ of the host galaxy B-band
luminosity, we obtain
\begin{equation}
L_{\rm bol}^{\rm qso}/L_{\sun}>f_{\rm BL} c_{B}\, c_{\rm gal} (M_{\ast}/M_{\sun}) \fgas.
\end{equation}
Dividing this through by $\Lp$, we have 
\begin{equation}
\frac{L_{\rm bol}^{\rm qso}}{\Lp}\gtrsim0.4\,\fgas\,f_{\rm BL}\ 
{\Bigl(}\frac{c_{\rm gal}}{1.0}{\Bigr)}
{\Bigl(}\frac{c_{B}}{12.0}{\Bigr)}
{\Bigl(}\frac{M_{\rm BH}^{f}/M_{\ast}}{0.001}{\Bigr)}^{-1}
{\Bigl(}\frac{c_{L}}{1.24}{\Bigr)}^{-1}.
\end{equation}
We can test this scaling relation against the results of our simulations, 
and do so in Figure~\ref{fig:test.BL}. Rearranging the equations above 
gives
%\begin{equation}
%\begin{split}
%\frac{L_{B}^{\rm qso}}{L_{B}^{\rm gal}} \approx 
%&\ 3.4\,\fgas^{-1}\,\frac{L_{\rm bol}^{\rm qso}}{\Lp}\\
%&\times{\Bigl(}\frac{c_{\rm gal}}{1.0}{\Bigr)}^{-1}
%{\Bigl(}\frac{c_{B}}{12.0}{\Bigr)}^{-1}
%{\Bigl(}\frac{M_{\rm BH}^{f}/M_{\ast}}{0.001}{\Bigr)}
%{\Bigl(}\frac{c_{L}}{1.24}{\Bigr)}, 
%\end{split}
%\label{eqn:BL.predict}
%\end{equation}
\begin{eqnarray}
\frac{L_{B}^{\rm qso}}{L_{B}^{\rm gal}} &\approx 
&\ 3.4\,\fgas^{-1}\,\frac{L_{\rm bol}^{\rm qso}}{\Lp}\nonumber\\
&&\times{\Bigl(}\frac{c_{\rm gal}}{1.0}{\Bigr)}^{-1}
{\Bigl(}\frac{c_{B}}{12.0}{\Bigr)}^{-1}
{\Bigl(}\frac{M_{\rm BH}^{f}/M_{\ast}}{0.001}{\Bigr)}
{\Bigl(}\frac{c_{L}}{1.24}{\Bigr)}, 
\label{eqn:BL.predict}
\end{eqnarray}
which we can compare to our direct calculation of 
$L_{B}^{\rm qso}/{L_{B}^{\rm gal}}$ and $L_{\rm bol}^{\rm qso}/\Lp$ 
for each simulation snapshot. 

Ultimately, we are not interested so much in the intrinsic B-band 
luminosity of the quasar and host galaxy, but rather the observed
luminosities; i.e.\ we are interested in the ratio ${L_{B, \rm\
obs}^{\rm qso}}/{L_{B, \rm\ obs}^{\rm gal}} = ({L_{B}^{\rm
qso}}/{L_{B}^{\rm gal}})\,(\exp\{-(\tau_{Q}-\tau_{G})\})$, where
$\tau_{Q}$ and $\tau_{G}$ are ``effective'' optical depths which we
use to denote the mean attenuation of quasar and host galaxy B-band
luminosities, respectively.  We have considered the distribution of
column densities attenuating the quasar as a function of instantaneous
and peak quasar luminosity in detail in \S~\ref{sec:NHfunction} above;
the attenuation of the host galaxy as a function of luminosity,
observed band, halo mass, and star formation rate are discussed in
detail in \citet{Jonsson05}. Combining these fits gives, roughly,
$(\exp\{-(\tau_{Q}-\tau_{G})\})\sim (M_{\rm
BH}^{f}/10^{8}\,M_{\sun})^{0.16}$, but a better approximation can be
determined directly from the simulations.  

This scaling can be understood roughly using toy models of uniformly
mixed luminous sources within the galaxy described by
\citet{Jonsson05}, after accounting for the fact that the luminosity
(star formation rate) dependent portion of the attenuation scales with
luminosity in a similar manner to our quasar attenuation (compare our
$\tau_{Q}\propto\nh\propto L_{\rm qso}^{0.43-0.54}$ to their
$\tau_{G}\propto L_{\rm B,\ gal}^{0.55}$). The key consequence of this
is that more massive systems (higher bulge and black hole masses) have
their host galaxy light proportionally more attenuated in mergers,
meaning that (as suggested by the comparison of light curves in
Figure~\ref{fig:BL.in.sims}) the quasar is more likely to be observed
with an optical luminosity larger than that of its host.

Figure~\ref{fig:test.BL} plots the ratio of the observed (attenuated)
B-band quasar luminosity to the observed host galaxy B-band luminosity
as a function of the ratio of instantaneous to peak quasar bolometric
luminosity. We show the results for four different gas fractions
$\fgas = 0.2,\ 0.4,\ 0.8,\ 1.0$ as labeled. For each gas fraction, we
consider our simulations A2 (black diamonds), A3 (blue circles), and
A5 (red $\times$'s) (the same simulations shown in
Figure~\ref{fig:BL.in.sims}) with $\qeos=1.0,\, \zgal=0$, and virial
velocities $\vvir=113,\, 160\, {\rm and}\ 320\, {\rm km\,s^{-1}}$,
using the labeled initial gas fraction. The colored lines in each
panel show the predictions of combining the scalings expected for the
intrinsic luminosities (Equation~\ref{eqn:BL.predict}) and
attenuations as above, giving
\begin{equation}
\frac{L_{\rm B,\ obs}^{\rm qso}}{L_{\rm B,\ obs}^{\rm gal}}=7.9\,\frac{1}{\fgas}\,
{\Bigl(}\frac{M_{\rm BH}^{f}}{10^{8}\,M_{\sun}}{\Bigr)}^{0.2}\ \frac{L}{\Lp}, 
\label{eqn:BL.scaling}
\end{equation}
where the colored lines each use the $M_{\rm BH}^{f}$ and $\fgas$ of
the simulation of the corresponding color and panel. This scaling
provides a good estimate of the observed optical quasar-to-galaxy
luminosity ratio, including the complicated effects of attenuation,
evolving mass-to-light ratios, metallicities, and host galaxy
properties, as a function of gas fraction, final black hole mass, and
the ratio of the current to peak quasar luminosity. Although, for
clarity, we have not shown a range of simulations varying other
parameters, we find that this scaling is robust to the large number of
quantities we have considered in our simulations -- there
are systematic offsets in e.g.\ $\Lp$ and $M_{\rm BH}^{f}$ with
changes such as e.g.\ different ISM equations of state, but the
scaling in terms of $\Lp$ and $M_{\rm BH}^{f}$ is unchanged.

Because the ratio of observed quasar and host galaxy B-band
luminosities in our simulations obeys the scaling of
Equation~\ref{eqn:BL.scaling}, we can use it to predict the
properties of ``broad-line'' quasars, defined by $L_{\rm B,\ obs}^{\rm
qso}>f_{\rm BL} L_{\rm B,\ obs}^{\rm gal}$.  To do so,
however, we must assume a typical host galaxy gas
fraction. Unfortunately, because our empirical modeling in terms of
the quasar lifetime as a function of $L$ and $\Lp$ does not have a
systematic dependence on host galaxy gas fraction (see
\S~\ref{sec:detailsCompare}), we have no constraint on this parameter.
It is, however, convenient for several reasons to consider
$\fgas=0.3$ as a typical value for bright quasars.

First, such a gas
fraction is capable of yielding the brightest observed
quasars; second, scaling a Milky-Way like disk with the observed $z=0$
gas fraction $\sim0.1$ to the redshifts of peak quasar activity gives
a similar gas fraction \citep[e.g.,][]{SDH05a}; third, gas fractions
$\gtrsim30\%$ in major mergers are needed to explain the observed
fundamental plane (Robertson et al.\ 2005c, in preparation), kinematic
properties (Cox et al.\ 2005c, in preparation), and central phase
space densities (Hernquist, Spergel \& Heyl 1993) of elliptical
galaxies; fourth, this choice implies that the brightest quasars with
$M_{\rm BH}^{f}\sim10^{10}\,M_{\sun}$ attain observed B-band
luminosities $\sim1000$ times that of their hosts at their peaks, as
is observed \citep[e.g.,][]{MD04}. Finally, and most important, the
assumed $\fgas$ and $f_{\rm BL}$ are degenerate in our predictions for
the broad-line population, as they both enter linearly in the ratio of
host galaxy to quasar B-band luminosity. Therefore, the range of
$f_{\rm BL}=0.3-3$ which we consider (for a fixed median $\fgas=0.3$)
can be equivalently considered, for a fixed median $f_{\rm BL}=1$, to
represent a theoretical uncertainty in the host galaxy gas fraction,
$\fgas=0.1-0.9$; i.e.\ spanning the range from present, relatively
gas-poor Milky-Way like disks to almost completely gaseous
disks. This, then, gives for our ``broad-line'' criterion,
\begin{equation}
\frac{L}{\Lp}\gtrsim 0.2\,{\Bigl(}\frac{f_{\rm BL}}{1.0}{\Bigr)}\,
{\Bigl(}\frac{\fgas}{0.3}{\Bigr)}\,
{\Bigl(}\frac{M_{\rm BH}^{f}}{10^{7}\,M_{\sun}}{\Bigr)}^{-0.2}.
\label{eqn:BL.min}
\end{equation}

The ``broad-line'' phase is thus, as is clear from
Figure~\ref{fig:BL.in.sims} and implicit in our definition of the
broad-line phase, closely associated with the final ``blowout'' stages
of quasar evolution, when the mass of the quasar reaches that
corresponding to its location on the $M_{\rm BH}-\sigma$ relation and
gas is expelled from the central regions of the galaxy, shutting down
accretion \citep{DSH05}. We note that combining the equation above
with our fitted quasar lifetimes gives an integrated time when the
quasar would be observable as a broad line object of $t_{\rm
BL}\sim10-20$\,Myr, in good agreement with the optically observable
bright quasar lifetimes we calculate directly from our quasar light
curves, including the effects of attenuation, and with empirical
estimates of the quasar lifetime which are based directly on
optically-selected, broad-line quasar samples.

The $({M_{\rm BH}^{f}}/{10^{7}\,M_{\sun}})^{0.2}$ term in the above
equation reflects the fact that, below a certain peak luminosity,
quasars are less likely to reach luminosities above that of the host
galaxy, as can be seen in the uppermost panels of
Figure~\ref{fig:BL.in.sims} for a final black hole mass of
$\mbhf=3\times10^{7}$ -- i.e.\ the smallest AGN are
proportionally less optically luminous than their hosts.  This does
not imply that such systems are not inherently broad-line objects, but
only that the host galaxy light will increasingly dominate at lower
luminosities. We also caution against extrapolating this to large or
small $M_{\rm BH}^{f}$, as the attenuation becomes more difficult
to predict at these peak luminosities, and the linear formula above is
not always accurate (see Figure~\ref{fig:test.BL}).

We can use this estimate of the broad-line phase and our model of the
quasar lifetime to calculate the total energy radiated in this bright,
optically observable stage following the calculation of
\S~\ref{sec:detailsCompare}, but with a minimum luminosity determined
by Equation~\ref{eqn:BL.min}. This gives an integrated fraction
$\sim0.3-0.4$ ($\sim\exp\{-0.2\,f_{\rm BL}\,(\fgas/0.3)/\alpha_{L}\}$)
of the total radiant energy emitted during the broad-line phase. Thus,
despite the short duration of this optical quasar stage, a large
fraction of the total radiated energy is emitted (as it represents the
final $e$-folding in the growth of the black hole) when most of the
final black hole mass (\S~\ref{sec:detailsCompare}) is accumulated.
Accounting for the luminosity dependence of our bolometric corrections
(with the optical fraction of the quasar energy increasing with
bolometric luminosity) as well as the small fraction of objects
observable at lower luminosities (with larger typical obscuring column
densities) increases this fraction to as much as $\sim0.6-0.7$ for
bright quasars.  Therefore, despite the fact that the {\em duration}
of the optically observable broad-line quasar phase may be $\sim1/10$
that of the obscured quasar growth phase, the changing quasar
luminosity over this period and non-trivial quasar lifetime as a
function of luminosity implies only small corrections to counting
arguments such as that of \citet{Soltan82}, which rely on the total
observed optical quasar flux density to estimate the relic
supermassive black hole density.

\subsection{The Broad-Line Fraction as a Function of Luminosity}
\label{sec:BLfraction}

By estimating the time that a quasar with some $\Lp$ will be
observable as a broad-line quasar at a given luminosity, we can then
calculate the broad-line quasar luminosity function in the same
fashion as the complete quasar luminosity function in
\S~\ref{sec:fullLF}.  Instead of the full quasar lifetime $\dtdL$, we
consider only the time during which broad-lines would be observed
(i.e.\ that the quasar spectrum would be recognized as opposed to the
host galaxy spectrum), as identified in our simulations
(\S~\ref{sec:whenBL}).

For a sample selected in hard X-rays (i.e.\ the selection function
only being relevant at column densities $\gtrsim10^{24}\ {\rm
cm^{-2}}$), we show the resulting ``broad-line'' luminosity function
in Figure~\ref{fig:LF.all} (cyan dot-dashed lines), and compare it to
the broad-line quasar luminosity function identified in the hard X-ray
luminosity function of \citet{Barger05}. The agreement is good at all
luminosities, and our model explains both the fact that broad-line
quasars dominate the luminosity function at luminosities well above
the ``break'' in the luminosity function, and the downturn in the
broad-line quasar population at luminosities below the
peak. Essentially, the broad-line quasar population more closely
traces the shape of the $\nLP$ distribution, giving rise to the
observed behavior as a dual consequence of luminosity-dependent quasar
lifetimes and the evolutionary nature of quasar obscuration in our
simulations.

%\clearpage
\begin{figure*}
    \centering
    \plotone{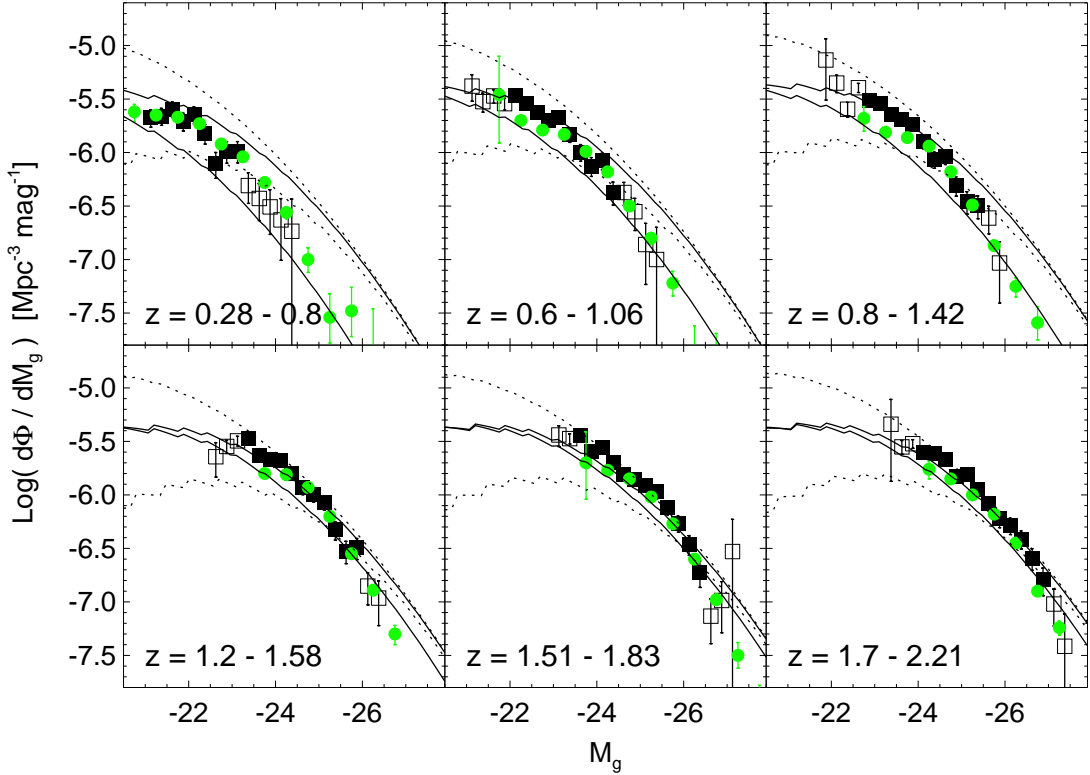}
    \caption{Broad-line quasar luminosity function of
    \citet{Richards05} from the 2dF-SDSS (2SLAQ) survey (black squares) and that
    of \citet{Croom04} (green circles) from the 2dF survey, compared
    to our predicted ``broad-line'' luminosity function from our
    determination of the relative quasar and host galaxy luminosities in our simulations
    (solid line), where we estimate that quasars are observable as ``broad-line'' objects
    when their observed B-band luminosity is greater than a factor $f_{\rm BL}$ of that of the host galaxy, 
    Solid lines are
    shown for the minimum and maximum observed redshift in each bin
    (as labeled), assuming $f_{\rm BL}=1$. Dotted lines show the result for
    $f_{\rm BL}=0.3$ and $f_{\rm BL}=3$, at the mean redshift
    of the bin, i.e.\ corresponding to ``broad-line'' luminosity functions in surveys 
    which are complete to quasars with observed optical luminosity $\sim0.3$
    and $3$ times that of the host galaxy, respectively, or alternatively reflecting 
    nearly complete theoretical uncertainty regarding merging galaxy gas fractions ($\fgas=0.1-0.9$). 
    Open squares are observations with uncertain incompleteness corrections
    in \citet{Richards05}.
    \label{fig:LF.sdss}}
\end{figure*}
%\clearpage

Figure~\ref{fig:LF.sdss} compares our theoretical predictions to the
2dF-SDSS (2SLAQ) g-band luminosity function of broad-line quasars from
\citet{Richards05} (black squares), as well as the B-band luminosity
function from \citet{Croom04} (green circles), at several redshifts
from $z\sim0.3-2$, over which range the surveys are expected to be
relatively complete (for broad-line quasars). The 2dF-SDSS result is
the most recent determination of the broad-line luminosity function,
but compares well with previous determinations by, e.g.,
\citet{Boyle88}, \citet{KK88}, \citet{MZZ88}, \citet{Boyle90},
\citet{BJS91}, \citet{Zitelli92}, \citet{Boyle00}, and
\citet{Croom04}.  Open squares correspond to bins in luminosity which
have been corrected for incompleteness following
\citet{PageCarrera00}, but this correction is uncertain as the bins
are not uniformly sampled. We compare this at each redshift to the
prediction of our determination of the quasar ``broad-line'' phase,
where we estimate that the quasar is observable as a broad line object
when its observed B-band luminosity is greater than a factor $f_{\rm
BL}=1$ of that of the host galaxy. We calculate this for both the
minimum and maximum observed redshift of each bin to show the range
owing to evolution of the luminosity function over each interval in
redshift. The systematic uncertainty in our prediction can be
estimated from the dotted lines, which show the prediction (at the
mean redshift of the bin) if we instead require the observed quasar
B-band luminosity to be above a factor of 0.3 (upper lines) or 3
(lower lines) of the observed host galaxy B-band luminosity, which as
discussed in \S~\ref{sec:whenBL} can alternatively be considered an
uncertainty in host galaxy gas fraction, with $\fgas=0.1$ and
$\fgas=0.9$, respectively.

The agreement at all luminosities and redshifts shown is encouraging,
given the simplicity of our determination of the broad-line phase from
the simulations, but the systematic uncertainties are large,
emphasizing the importance of calculating detailed selection effects
in contrasting e.g.\ ``broad-line'' samples from optical and X-ray
surveys, as opposed to assuming a constant obscured fraction at a
given luminosity based on the ratio of luminosity functions as has
been adopted in previous phenomenological models. The difference
between different choices of $\fgas$ is suppressed at the high
luminosity (and correspondingly high redshift) end of the luminosity
function, because the quasar-to-galaxy B-band luminosity ratio scales
as $\propto (M_{\rm BH}^{f})^{0.2}$; i.e.\ regardless of the choice of
$f_{\rm BL}$, quasars increasingly overwhelm their host galaxy in
large systems near their peak luminosity. However, at low luminosity,
the predictions rapidly diverge, implying that a measurement of the
faint end of the broad-line quasar luminosity function, with a
reliable calibration of $f_{\rm BL}$, can constrain the typical gas
fractions of quasar host galaxies and the evolution of these gas
fractions with redshift.

%\clearpage
\begin{figure}
    \centering
    \plotone{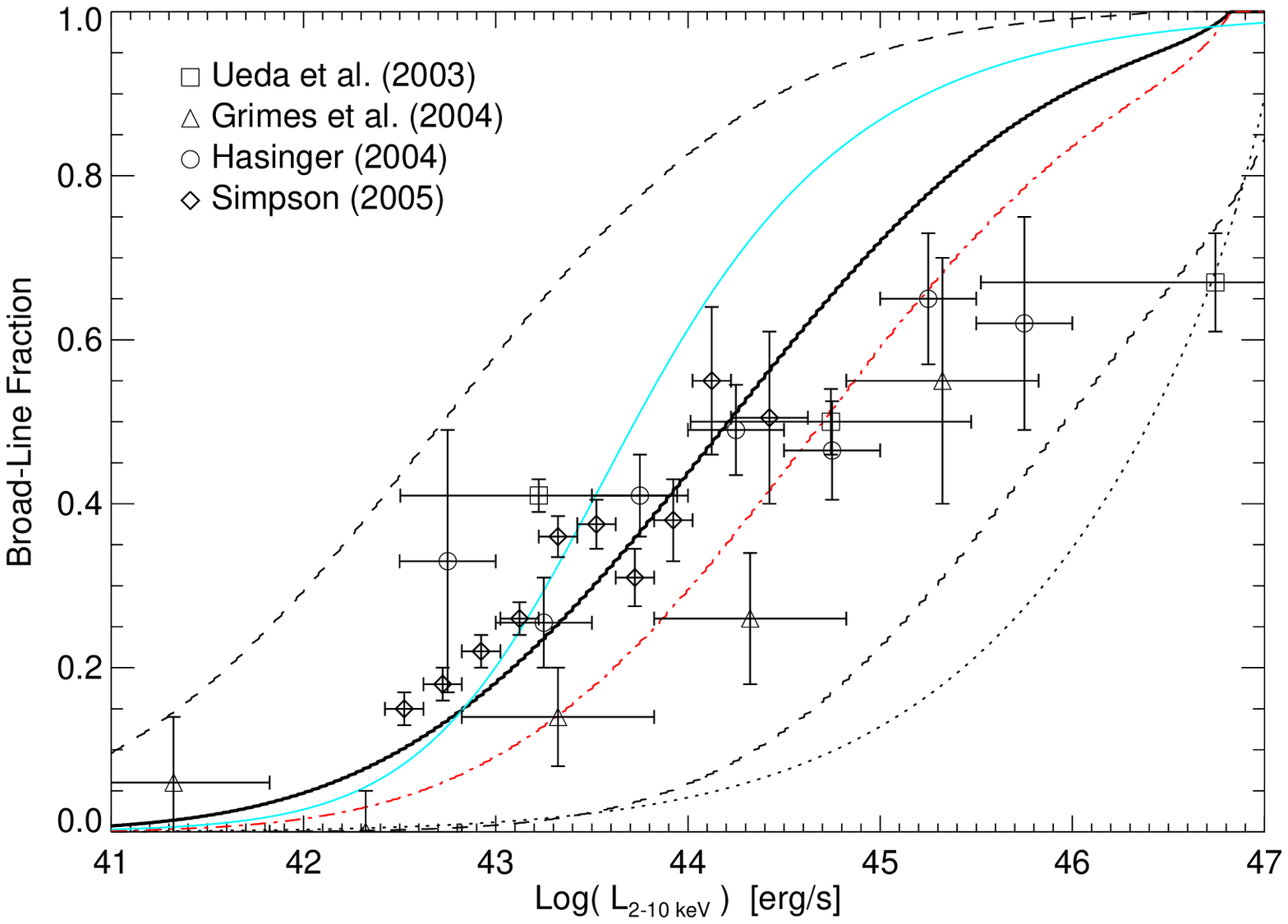}
    \caption{Predicted ``broad-line'' fraction of a complete X-ray
    sample at low $z\lesssim0.3$ redshift, from our simulations (where
    the object is observable as a broad-line quasar when it has an observed B-band 
    luminosity greater than a factor $f_{\rm BL}=1.0$ of that of its host galaxy), is shown (thick black
    line).  The results, changing our $f_{\rm BL}$ to 0.3 and
    3.0 are shown, or equivalently of assuming a host galaxy gas fraction $\fgas=0.1$ 
    or 0.9 instead of $\sim0.3$ (dashed), as are the results assuming a
    ``light bulb'' model where quasars spend a fixed time
    $t_{Q}=20$\,Myr as broad line objects with a luminosity of $\Lp$
    (dotted).  For comparison, the (scaled to 2-10 keV luminosity)
    observations of \citet{Ueda03} (squares), \citet{Hasinger04}
    (circles), \citet{GRW04} (triangles), and \citet{Simpson05}
    (diamonds) are shown. The predicted result at higher redshift
    ($z\gtrsim1$) is shown (red dot-dashed line), offset owing to the
    shift in break luminosity of the luminosity function with
    redshift. The best-fit luminosity-dependent torus model, {\em
    fitted} to the data, is shown as the solid cyan line. The best-fit
    static torus model is a constant broad-line fraction $\sim0.3$ (not shown for clarity). 
    \label{fig:BL.fraction}}
\end{figure}
%\clearpage

By dividing out the predicted luminosity function $\phi_{HX}$, we
can estimate the fraction of ``broad line'' objects observed in reasonably
complete X-ray samples as a function of luminosity. This is shown in
Figure~\ref{fig:BL.fraction}, where for ease of comparison we have
shown the broad-line fraction as a function of hard X-ray (2-10 keV)
luminosity. Our prediction, based on determining the time
a quasar with a given luminosity $L$ and peak luminosity $\Lp$
in our simulations will be observable with a B-band luminosity greater
than a fraction $f_{\rm BL}=1.0$ of the host galaxy observed B-band
luminosity, is shown as the thick black line. This is compared to
the observations of \citet{Ueda03} (squares), \citet{Hasinger04}
(circles), \citet{GRW04} (triangles), and \citet{Simpson05}
(diamonds).  The data from \citet{Hasinger04} has been scaled from
soft X-ray (0.5-2 keV) using our bolometric corrections, and the data
from \citet{GRW04} and \citet{Simpson05} have been converted from
[O{\sc~iii}] luminosity as in \citet{Simpson05} using the mean
correction for Seyfert galaxies \citep{Mulchaey94}, $L_{\rm
[O\,III]}=0.015\times L_{\rm 2-10\,keV}$.  

We also plot as upper and lower dashed lines the results of changing
$f_{\rm BL}$, the fraction of the host galaxy B-band luminosity above
which the quasar B-band luminosity must be observed for identification
as a ``broad-line'' object, considering $f_{\rm BL}=0.3,\ {\rm and}\
3$, respectively. We determine this for the low-redshift
$z\lesssim0.3$ quasar distribution, from which most of the data are
drawn. The red dot-dashed line shows the difference at high redshift,
if just $z\gtrsim1$ quasars are considered (for $f_{\rm BL}=1$). The
broad-line fraction is systematically lower, primarily because the
break luminosity in the luminosity function moves to higher luminosity
with redshift, meaning that at a fixed luminosity below the break, a
smaller fraction of observed objects are at $L\sim\Lp$ in the
``blowout'' phase of peak optical quasar luminosity.  Finally, the
dotted line shows the results assuming a ``light bulb'' model for the
broad-line phase (but still using our $\nLp$ distribution, otherwise
this translates to a constant obscured fraction with luminosity)
lifetimes, with a fixed broad-line lifetime of $t_{Q}=20\,$Myr.

The prediction of the most basic torus model, with constant broad-line
fraction $\sim0.36$, is ruled out to high significance
($\reducechi=18.5,\ 17.2$ if we consider all data points, or if we
consider only the most well-constrained data, from Simpson [2005],
respectively).  Furthermore, the solid cyan line shows the best-fit
luminosity-dependent torus model, in which the broad line fraction is
given by \citep[e.g.,][]{Simpson98,GRW04}
\begin{equation}
f = 1 - 1/\sqrt{1 + 3L/L_{0}}, 
\end{equation}
where $L_{0}$ is the luminosity where the number of broad line objects
is equal to the number of non-broad line objects. This fit is at best
marginally acceptable over a narrow range in luminosities
($\reducechi=14.0,\ 7.3$). Modified luminosity-dependent, receding
torus models have been proposed which give a better fit to the data
by, for example, allowing the torus height to vary with luminosity
\citep[e.g.,][]{Simpson05}, but there is no physical
motivation for these changes, and they introduce such variation
through additional free parameters that allow a curve of
essentially arbitrary slope to be fitted to the data. 

However, the prediction of our model agrees reasonably well
($\reducechi=4.0,\ 1.2$) with the observations over the entire range
covered, a span of six orders of magnitude in luminosity. We emphasize
that our prediction, which matches the data better than standard torus
models that are actually {\em fitted} to the data, is not a fit to the
observations.  Instead, it is derived from the physics of our
simulations, including black hole accretion and feedback which are
critical in driving the ``blowout'' phase which constitutes most of
the time a quasar is visible as a ``broad-line'' object by our
estimation, and from the $\dot n(L_{\rm peak})$ distribution implied
by our model of quasar lifetimes and the {\em bolometric} quasar
luminosity function. The agreement suggests that our choice of the
parameter combination $f_{\rm BL}\fgas=0.3$ is a good approximation.
As noted above, this implies that calibrating $f_{\rm BL}$ for an
observed sample, combined with the mean broad-line fraction and our
modeling, can provide a constraint (albeit model-dependent) on the
host galaxy gas fraction of quasars at a given redshift, which cannot
necessarily be directly measured even with difficult, detailed host
galaxy probes, as gas is rapidly converted into stars throughout the
merger. The uncertainty plotted, while large, actually represents a
larger theoretical uncertainty -- as discussed above, if an
observational sample were well-defined such that it were complete to
broad-line objects with observed optical luminosity above a
fraction $f_{\rm BL}$ of the host galaxy luminosity, the
range we consider would correspond to a range $\fgas=0.1-0.9$ in the
quasar host galaxy gas fraction, which the observations could then
constrain.

In our modeling, the broad line fraction as a function of luminosity
does not depend sensitively on the observed luminosity function, as
evidenced by the relatively similar prediction at high redshift.  The
evolution we do predict with redshift, in fact, agrees well with that
found by \citet{Barger05} over the redshift range $z=0.1-1.2$ (see
also La Franca et al.\ 2005), an aspect of the observations which is
not reproduced in any static or luminosity-dependent torus model but
follows from the evolution of the quasar luminosity function in our
picture for quasar growth. However, we do caution that gas fractions
may systematically evolve with redshift, and as discussed above, a
higher gas fraction will give generally shorter ``broad-line''
lifetimes using our criteria of quasar optical luminosity being higher
than some fraction of the host galaxy luminosity, which will also
contribute to the evolution in the mean ``broad-line'' fraction with
redshift.  Finally, neglecting the role of luminosity-dependent quasar
lifetimes gives unacceptable fits to the data ($\reducechi=66.0,\
77.5$), as the broad-line fraction as a function of luminosity is a
consequence of both the evolution of obscuration and the dependence of
lifetime on luminosity.

Our model for quasar evolution provides a direct physical motivation
for the change in broad line fraction with luminosity and suggests
that it is not a complicated selection effect. As an observational
sample considers higher luminosities (i.e.\ approaches and passes the
``break'' in the observed luminosity function), a comparison of the
luminosity function and the underlying $\nLp$ shows that it is
increasingly dominated by sources near their peak luminosity in the
final stages of Eddington limited growth. The final stages of this
growth expel the large gas densities obscuring the quasar, rendering
it a bright, optically observable broad-line object for a short
time. Therefore, we expect that the fraction of broad-line objects
should increase with luminosity in quasar samples, as
indicated by the observations.

%\clearpage
\begin{figure}
    \centering
    \plotone{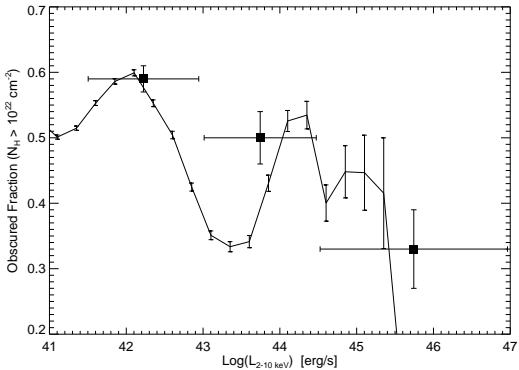}
    \caption{Predicted ``obscured'' fraction (solid line) in an X-ray sample with identical 
    redshift and luminosity range to that of \citet{Ueda03}, as a function of hard X-ray 
    (2-10 keV) luminosity. Vertical error bars show 
    Poisson errors estimated from the total time at a given luminosity across all our simulations 
    (absolute values of the error bars should not be taken literally). The ``obscured'' fraction is defined 
    as the fraction of quasars with X-ray column densities $\nh>10^{22}\,{\rm cm^{-2}}$ in 
    bins of $\Delta\log{L_{2-10\,{\rm keV}}}$. The observations from \citet{Ueda03} are shown 
    as black squares.
    \label{fig:Ueda.BL}}
\end{figure}
%\clearpage

Many observational measures do not consider a direct optical analysis
of the quasar spectrum in estimating the fraction of broad-line
objects as a function of luminosity. For example, \citet{Ueda03} adopt
a proxy, classifying as ``obscured'' any quasars with an X-ray
identified column density $\nh>10^{22}\,{\rm cm^{-2}}$, and as
``unobscured'' quasars below this column density. We can compare to
their observations, using the column density distributions as a
function of luminosity from our simulations, which cover the entire
range in luminosity of the observed sample.  Specifically, we use a
Monte Carlo realization of these distributions, employing our fitted
$\nLp$ distribution at each redshift to produce a list of quasar peak
luminosities and then generating all other properties based on the
probability distribution of a given property in simulations with a
similar peak luminosity. We describe this methodology in detail in
\S~\ref{sec:discussion}, and provide several such mock quasar
distributions at different redshifts. 

In Figure~\ref{fig:Ueda.BL}, we compare our estimated ``obscured'' and
``unobscured'' fractions as a function of hard X-ray luminosity, using
the same definitions as well as redshift and luminosity limits as the
observed sample. The solid line shows our prediction, with vertical
error bars representing Poisson errors, where the number of ``counts''
is proportional to the total time spent by simulations at the plotted
luminosity (the absolute value of these errors should not be taken
seriously). The ``obscured'' fraction is determined in bins of
luminosity $\Delta\log L_{2-10\ {\rm keV}}=0.5$. Despite our large
number of simulations, there is still some artificial ``noise'' owing
to incomplete coverage of the merger parameter space, namely the
apparent oscillations in the obscured fraction. However, the mean
trend agrees well with that observed (black squares), suggesting that
the success of our modeling in reproducing the fraction of ``broad
line'' objects as a function of luminosity is not a consequence of the
definitions chosen above. We do not show the predictions of the
standard and luminosity-dependent torus models, as (because
essentially any line of sight through the torus encounters a column
density $\nh>10^{22}\,{\rm cm^{-2}}$) the predictions of these models
are identical to those shown and compared to the same observations in
Figure~\ref{fig:BL.fraction}.

Our prediction that the fraction of broad-line objects should rise
with increasing luminosity is counterintuitive, given our fitted
column density distributions in which typical (median) column
densities increase with increasing luminosity.  This primarily owes to
the simplicity of our \NH\ fits; we assume the distribution is
lognormal at all times, but a detailed inspection of the cumulative
(time-integrated) column density distribution shows that at bright
(near-peak) luminosities, the distribution is in fact bimodal (see
e.g.\ Figure 3 of Hopkins et al.\ 2005b and Figure 2 of Hopkins et
al.\ 2005d), representing both the heavily obscured growth phase and
the ``blowout'' phase we have identified here as the ``broad line''
phase.  Over most of a simulation, we find the general trend shown in
Figure~\ref{fig:NH.systematics} and discussed above, namely that
typical column densities increase with intrinsic (unobscured)
luminosity.  This is because the total time at moderate to large
luminosities is dominated by black holes growing in the
obscured/starburst stages; here, the same gas inflows fueling black
hole growth also give rise to large column densities and starbursts
which obscure the black hole activity.  However, when the quasar nears
its final, peak luminosity, there is a rapid ``blowout'' phase as
feedback from the growing accretion heats the surrounding gas, driving
a strong wind and eventually terminating rapid accretion, leaving a
remnant with a black hole satisfying the $M_{\rm BH}-\sigma$
relation. This can be identified with the traditional bright optical
quasar phase, as the final stage of black hole growth with a rapidly
declining density (allowing the quasar to be observed in optical
samples), giving typical luminosities, column densities, and lifetimes
of optical quasars.  In these stages, larger luminosities imply more
violent ``blowout'' events, i.e.\ a brighter peak luminosity quasar
more effectively expels the nearby gas and dust, rendering a dramatic
decrease in column density at these bright stages (see Hopkins et al.
2005f).

We are essentially modeling this bimodality in more detail by directly
determining the ``broad-line'' phase from our simulations.  However,
the broad line fraction-luminosity relation we predict is also a
consequence of the more complicated relationship between column
density, peak luminosity, and bolometric and observed luminosity, as
opposed to the predictions from a model with correlation between $\nh$
and only observed luminosity.  The key point is that we find, near the
{\em peak} luminosity of the quasar, as feedback drives away gas and
slows down accretion, the typical column densities fall rapidly with
luminosity in a manner similar to that observed. In our model for the
luminosity function, quasars below the observed ``break'' are either
accreting efficiently in early stages of growth or are in
sub-Eddington phases coming into or out of their peak quasar
activity. Around and above the break, the luminosity function becomes
dominated by sources at high Eddington ratio at or near their peak
luminosities. Based on the above calculation, we then {\em expect}
what is observed, that in this range of luminosities, the fraction of
objects observed with large column densities will rapidly decrease
with luminosity as the observed sample is increasingly dominated by
sources at their peak luminosities in this blowout phase. This also
further emphasizes that the evolution of quasars dominates over static
geometrical effects in determining the observed column density
distribution at any given luminosity.

Finally, if host galaxy contamination were not a factor, we would
expect from our column density model that, at low luminosities
($L\lesssim\Lcut{10}$, well below the range of most observations
shown), the broad-line fraction would again increase (i.e.\ the
obscured fraction would decrease), as the lack of gas to power
significant accretion would also imply a lack of gas to produce
obscuring columns. However, at these luminosities, typical of faint
Seyfert galaxies or LINERs, our modeling becomes uncertain; it is
quite possible, as discussed previously, that cold gas remaining in
relaxed systems could collapse to form a traditional dense molecular
torus on scales $\sim\,$pc, well below our resolution
limits. Furthermore, host galaxy light is likely to overwhelm any AGN
broad-line contribution, and selection effects will also become
significant at these luminosities.

\subsection{The Distribution of Active Broad-Line Quasar Masses}
\label{sec:BLmasses}

Our determination of the ``broad-line'' or optical phase in quasar
evolution allows us to make a further prediction, namely the mass
distribution of currently active broad-line quasars.  At some
redshift, the total number density of observed, currently active
broad-line quasars with a given $\Lp$ will be (in the absence of
selection effects)
\begin{equation}
n_{\rm BL}(\Lp)\approx\nLP\, t_{\rm BL}(\Lp), 
\end{equation}
where $t_{\rm BL}(\Lp)$ is the total integrated time that a quasar
with peak luminosity $\Lp$ spends as a ``broad-line'' object (using
our criterion for the ratio of the observed quasar B-band luminosity to
that of the host galaxy), given by integrating our formulae in
\S~\ref{sec:whenBL} or directly calculated from the simulations.
Since we have determined roughly that a quasar should be observable as
a ``broad-line'' object at times with $L\gtrsim 0.2\,\Lp$ primarily
just after it reaches its peak luminosity, in the ``blowout'' phase of
its evolution, we expect the instantaneous black hole mass at the time
of observation as a broad-line quasar to be, on average, $M^{\rm
BL}_{\rm BH}\approx\mbhf(\Lp)$, where $\mbhf\sim M_{\rm Edd}(\Lp)$
modulo the order unity corrections described in
\S~\ref{sec:detailsCompare}. Using our fitted $\nLP$ distribution from
the luminosity function, extrapolated to low redshift ($z\sim0$), and
combining it with the integrated ``broad-line'' lifetimes from our
simulations as above, we obtain the differential number density of
sources in a logarithmic interval in $\Lp$.  Finally, we use our
Equation~\ref{eq:Mbhf} for $\mbhf(\Lp)$ determined from our fitted
quasar lifetimes (demanding that $E_{\rm rad}=\epsilon_{r}M_{\rm
BH}^{f}c^{2}$) to convert this to a distribution in black hole mass.

%\clearpage
\begin{figure*}
    \centering
    \plotone{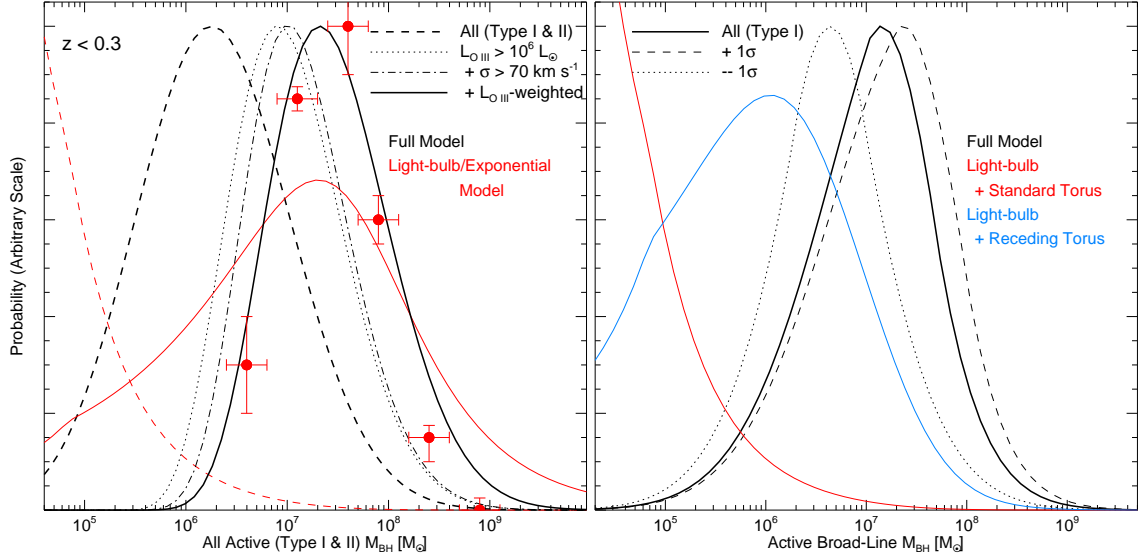}
    \caption{Predicted distribution of currently active black hole masses, 
    both considering all types (Type I \&\ II; left) and only those visible as broad-line quasars (Type I; right), 
    at low $z\lesssim0.3$ redshift, from our $\nLP$ distribution and the estimation of the ``broad-line'' 
    phase directly from the simulations. 
    In the left panel (all quasar types), we consider the result with 
    arbitrarily faint luminosity limits (dashed line), 
    and with the luminosity completeness limit (dotted) and both luminosity limit and velocity dispersion limit (dash-dot) 
    of the SDSS sample of \citet{Heckman04}. We then consider the mass distribution with these limits, 
    weighted by OIII luminosity, for direct comparison to the mass function of \citet{Heckman04}, shown as
    red circles (vertical errors represent the range in different parameterizations of the luminosity-weighted mass function 
    from \citet{Heckman04}, their Fig.~1, horizontal errors a $\sim0.2$\,dex uncertainty in the black hole mass). 
    Black lines show this for our full model, red lines show the full distribution (dashed) and 
    distribution with the same weighting and selection effects as \citet{Heckman04} (solid) for a 
    light-bulb or exponential light curve model of quasar evolution.
    At right, the distribution of active ``broad-line'' quasar masses (solid, where 
    an object is a ``broad-line'' quasar if the observed quasar B-band luminosity is above 
    a factor $f_{\rm BL}=1$ of that of the host galaxy -- dotted and dashed lines 
    show the result if $f_{\rm BL}=0.3$ or 3, respectively). Black lines show the prediction of the full 
    model, red and blue lines the predictions of a light-bulb/exponential light curve model with 
    a standard torus model (red) and receding torus model (blue) used to determine the broad-line fraction. 
    %agrees well with the active broad-line 
    %mass function from the SDSS, with some expected incompleteness in the SDSS sample at low $M_{\rm BH}$
    %(Greene et al., in preparation).
    \label{fig:BL.massfunct}}
\end{figure*}
%\clearpage

Our predicted $n(M_{\rm BH})$, i.e.\ the number of observed {\em
active} quasars at low redshift in a logarithmic interval of black
hole mass, is shown in Figure~\ref{fig:BL.massfunct}.  We consider the
complete distribution of active quasar masses, for both broad-line and
non broad-line objects, in the left panel of the figure, and the
distribution of broad-line objects only, $n(M^{\rm BL}_{\rm BH})$, in
the right panel. On the left, we show the complete distribution which
would be observed without any observational limits (dashed line). We
calculate this from the distributions of Eddington ratios in our
simulations, as a function of current and peak luminosity, and our fit
to $\nLp$ (as, e.g.\ for our Monte Carlo realizations).  We also
consider the observed distribution if we apply the luminosity limit
for completeness from the SDSS sample of \citet{Heckman04} (dotted),
$L_{\rm [O\,III]}>\Lcut{6}$, which using their bolometric corrections
yields $L>3.5\times\Lcut{9}$, and then additionally applying their
minimum velocity dispersion $\sigma>70\,{\rm km\,s^{-1}}$
(dot-dashed). Finally, we can weight this distribution by luminosity
(solid line) to compare directly to that determined in their
Fig.~1. The red points are taken from the luminosity-weighted black
hole mass function of \citet{Heckman04}, which serves as a rough
estimate of the {\em active} black hole mass distribution given their
selection effects.  Vertical error bars represent the range in
parameterizations of the mass function from \citet{Heckman04},
including whether or not star formation is corrected for and limiting
the sample to luminosities $L\gtrsim\Lcut{10}$ or Eddington ratios
$>0.01$. Horizontal errors represent an uncertainty of $0.2\,$dex in the
black hole mass estimation (representative of uncertainties in the
$M_{\rm BH}-\sigma$ relation used). The agreement is good, especially
given the significant effects of the selection criteria and
luminosity-weighting.

We also consider the predictions of a ``light-bulb'' or ``exponential
/ fixed Eddington ratio'' model of the quasar lifetime for the active
black hole mass distribution (red lines). For purposes of the active
black hole mass function, the two predictions are identical and
independent of the assumed quasar lifetime (modulo the arbitrary
normalization), as both assume that all observed quasars are accreting
at a fixed Eddington ratio, giving the distribution of active black hole
masses. The dashed line shows the prediction for the complete active
black hole mass function, which rises sharply to lower luminosities,
as it must given a luminosity function which increases monotonically
to lower luminosities.  The solid line shows the prediction of such a
model with the complete set of selection effects from
\citet{Heckman04} described above applied, as with the solid black
line showing the prediction of our modeling. Here, we chose the
characteristic Eddington ratio $\approx1.0$ by fitting the predicted
curve to the \citet{Heckman04} observations. Note that both the
characteristic Eddington ratio and lifetime (normalization) of the
curve are fitted, so the relative normalization of this curve and our
full model prediction are not the same; for example, the predicted
total absolute number of active $M_{\rm BH}>10^{9}$ quasars is higher
in the full model than in the light-bulb or exponential models. Still,
it is clear that these models produce too broad a distribution of
active black hole masses, in disagreement with the observations.  We
could, of course, obtain an arbitrarily close agreement with the
observations if we fit to the {\em distribution} of accretion rates,
but such a model would recover a quasar lifetime and accretion rate
distribution quite similar to ours, as is evident from the agreement
between the predictions of our simulations and the observations.  A
purely empirical model of this type is considered by e.g.\
\citet{Merloni04}, who finds that similar qualitative evolution in the
quasar lifetime and anti-hierarchical black hole assembly to that
predicted by our modeling is implied by the combination of quasar
luminosity functions and the black hole mass function.

On the right of the figure, we show our predicted mass distribution
for low-redshift, active ``broad-line'' quasars (solid black lines),
where we estimate that an object is a ``broad-line'' quasar if the
observed quasar B-band luminosity is above a factor $f_{\rm BL}=1$ of
that of the host galaxy -- dotted and dashed lines show the result if
$f_{\rm BL}=0.3$ or 3, respectively, parameterizing the range of
different observed samples. As discussed above, the range of $f_{\rm
BL}$ shown can be, alternatively, thought of as a parameterization of
uncertainty in the host galaxy gas fraction, if (in an observed
sample), the sensitivity to seeing quasar broad lines against host
galaxy contamination is known. Therefore, the location of the peak in
the active broad-line black hole mass function can be used, just as
the mean broad line fraction vs.\ luminosity, as a test of the typical
gas fractions of bright quasar host galaxies, and can constrain
potential evolution in these gas fractions with redshift.  

The prediction shown is testable, but appears to be in good
agreement with preliminary results for the distribution of active
broad-line black hole masses from the SDSS \citep[e.g.,][]{MD04}.  The
observations may show fewer low-mass black holes than we predict, but
this is expected, as observed samples are likely incomplete at the low
luminosities of these objects (even at the Eddington limit, a
$10^{5}\,M_{\sun}$ black hole has magnitude $M_{g}\sim-16$). If, in
our model, we were to consider instead a standard torus scenario for
the definition of the broad-line phase, we would predict the same
curve as that shown in the left half of the figure (black dashed; our
prediction for the cumulative active black hole mass function). This
is because the standard torus model predicts that a constant fraction of
objects are broad-line quasars, regardless of mass or luminosity, thus
giving identical distributions of Type I and Type II quasar masses. If
we consider a luminosity-dependent or receding torus model, the
prediction is nearly identical to the black line shown. This is
because, as shown in Figure~\ref{fig:BL.fraction}, our prediction for
the broad line fraction as a function of luminosity is similar to that
of the receding torus model. The differences in the model predictions
for the broad-line fraction as a function of luminosity do manifest in
the prediction for the active broad-line black hole mass function, but
the difference in these models is smaller than the $\sim1\sigma$ range
from different values of $f_{\rm BL}$ shown.  However, if we consider
different models for the quasar light curve or lifetime, the predicted
active broad-line mass function is quite different (as is the
cumulative active black hole mass function).

We show the predictions of a light-bulb or exponential light curve
model for quasar evolution in the figure, adopting either a standard
torus model (red) or receding torus model (blue) to determine the
broad-line fraction as a function of luminosity. For the standard
torus model, this predicts that the broad line mass function should
trace the observed luminosity function, rising monotonically to lower
black hole masses in power-law fashion (just as seen in the red dashed
line in the left half of the figure for the cumulative black hole mass
function). For the receding torus model, the active black hole mass
function shows a peak (because, at lower luminosities, there are more
observed quasars, but a larger fraction of them are
obscured). However, the location of this peak is at roughly an order
of magnitude smaller black hole mass than for our prediction. This
assumes a typical Eddington ratio $\sim1$, which we have fitted to the
cumulative black hole mass function -- the peak in the broad-line
active black hole mass function in these models could be shifted to
larger black hole masses by assuming a smaller typical Eddington
ratio, but this would only worsen the agreement with the cumulative
black hole mass function of \citet{Heckman04}. Furthermore, a robust
difference between the models is that the light bulb or
Eddington-limited/exponential models predict, for the standard torus
case, no turnover in the active broad-line black hole mass function,
and for the receding torus case, a broader distribution in active
broad-line quasar black hole masses than is predicted in our
modeling. Roughly, the lognormal width of this distribution in our
model is $\sim0.6$\,dex, whereas the light-bulb or exponential light
curve models have a distribution with width $\sim1.0$\,dex. As noted
above, we obtain a similar prediction if we adopt our full obscuration
model instead of the receding torus model here.  A determination of
the range of active, broad-line quasar masses can, therefore,
constrain quasar lifetimes and light curves.

Our model makes an accurate prediction for the distribution of {\em
active} black hole masses, even at $z\sim0$ where our extrapolation of
the luminosity function is uncertain.  It is important to distinguish
this from the predicted relic black hole mass distribution, derived in
\S~\ref{sec:smbh}, which must account for all quasars, i.e.\ $\nLP$
integrated over redshift.  We additionally find for broad-line
quasars, as we expect from our prediction of the broad-line phase,
that these objects are primarily radiating at large Eddington ratios,
$l\sim0.2-1$, but we address this in more detail in
\S~\ref{sec:eddington}. The success of this prediction serves not only
to support our model, but also implies that we can extrapolate to
fairly low luminosities, even bright Seyfert systems at $z\sim0$.
This suggests that many of these systems, at least at the bright end,
may be related to our assumed quasar evolution model,
fueled by similar mechanisms and either exhibiting weak interactions
among galaxies or relaxing from an earlier, brighter stage in their
evolution.  As we speculate in \S~\ref{sec:discussion}, our
description of self-regulated black hole growth may also be
relevant to fainter Seyferts, even those that reside in
apparently undisturbed galaxies.

\section{The Distribution of Eddington Ratios}
\label{sec:eddington}

In traditional models of quasar lifetimes and light curves, the
Eddington ratio, $l\equiv L/L_{\rm Edd}$ is generally assumed to be
constant.  Even complex models of the quasar population which allow
for a wide range of Eddington ratios according to some probability
distribution $P(l)$ implicitly associate a fixed Eddington ratio with
each individual quasar, and do not allow for $P(l)$ to depend on
instantaneous luminosity or host system properties. However, this is a
misleading assumption in the context of our model, as the Eddington
ratio varies in a complicated manner over most of the quasar light
curve. Furthermore, the integrated time at a given Eddington ratio is
different in different systems, with more massive, higher peak
luminosity systems spending more time at large ($l \sim1$) Eddington
ratios.

The probability of being at a given Eddington ratio should properly be
thought of as a conditional joint distribution $P(l\,|\,L,\,\Lp)$ in
both instantaneous and peak luminosity, just as the quasar
``lifetime'' is more properly a conditional distribution
$t_{Q}(L\,|\,\Lp)$. Rather than adopting a uniform Eddington ratio or
Eddington ratio distribution, empirical estimates must consider more
detailed formulations such as the framework presented in \citet{SW03},
which allows for a conditional bivariate Eddington ratio distribution
and can therefore incorporate these physically motivated dependencies
and complications in de-convolving observations of the quasar
luminosity function to determine e.g.\ Eddington ratio distributions,
active black hole mass functions, and other physical quantities.

%\clearpage
\begin{figure}
    \centering
    \plotone{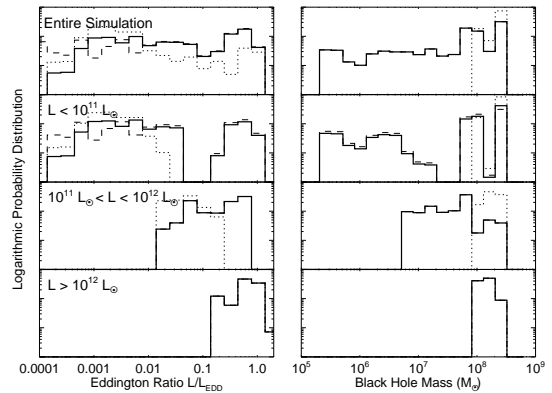}
    \caption{Distribution of Eddington ratios (left panels) and
    instantaneous black hole mass (right panels) as a function of quasar
    bolometric luminosity for our fiducial Milky Way-like A3 simulation, 
    with $\vvir=160\,{\rm km\,s^{-1}}$ and
    $\Lp\sim5\times\Lcut{13}$. The trend of an increasingly narrow
    Eddington ratio and mass distribution (concentrated at higher values)
    with increasing luminosity is clear. The result of applying an
    ADAF-type radiative efficiency correction at low accretion rates is
    shown (dashed) as well as the result of considering only times after
    the final merger, with $M_{\rm BH}\sim\mbhf$ (dotted).
    \label{fig:Pofl.A3}}
\end{figure}
%\clearpage

Figure~\ref{fig:Pofl.A3} shows the distribution of Eddington ratios as
a function of luminosity for the fiducial, Milky Way-like A3
simulation ($\vvir=160\,{\rm km\,s^{-1}}$). Over the course of the
simulation, the system spends a roughly comparable amount of time at a
wide range of Eddington ratios from $l\sim0.001-1$. At high
luminosities, $L>\Lcut{12}$ for a system with $\Lp\approx\Lcut{13}$,
the range of Eddington ratios, is concentrated at high values
$l\sim0.5-1$ with some time spent at ratios as low as
$l\sim0.1$. Note, however, that the y-axis of the plot is scaled
logarithmically, so the time spent at $l\sim0.1$ in this luminosity
interval is a factor $\sim5$ smaller than the time spent at
$l\gtrsim0.5$. Considering lower luminosities $\Lcut{11}<L<\Lcut{12}$,
the distribution of Eddington ratios broadens down to
$l\sim0.01$. Going to lower luminosities still, $L<\Lcut{11}$, the
distribution broadens further, with comparable time spent at ratios as
low as $l\sim0.001$, and becomes somewhat bimodal.  At large
luminosities near $\Lp$, the system is primarily in Eddington-limited
or near-Eddington growth. However, as we consider lower luminosities,
we include both early times when the black hole is growing efficiently
(high $l$) and late or intermediate times when the black hole is more
massive but the accretion rate falls (low $l$). As we go to lower
luminosities, the {\em total} time spent in sub-Eddington states
increasingly dominates the time spent at $l\sim1$, although the time
spent at any given value of $l$ is fairly flat with $\log(l)$.

Roughly, at some luminosity $L$, there is a constant probability of
being in some logarithmic interval in $l$,
\begin{equation}
P(l | L,\Lp) \sim \left[\log\Bigl( \frac{\Lp}{L}\Bigr)\right]^{-1},\
\frac{L}{\Lp} < l < 1,
\end{equation}
and $P(l | L,\Lp)=0$ otherwise.  This is especially clear if we
compare the distribution of Eddington ratios in each luminosity range
obtained if we consider only times after the final merger of the
black holes (dotted histograms). At the highest
luminosities, the distribution is identical to that obtained
previously, since all the time at these luminosities is during the
final merger. However, as we move to lower luminosities, the
characteristic $l$ move systematically lower, as we are seeing only
the relaxation after the final ``blowout'' near $\Lp$, with
characteristic Eddington ratio $l=L/\Lp$ at any given luminosity
$L$. These trends are also clear if we consider the distribution of
{\em instantaneous} black hole masses in each luminosity interval
shown in the figure, which is trivially related to the Eddington ratio
distribution at a given luminosity $L$ as
\begin{equation}
M_{\rm BH} = M_{0}\,\frac{L}{l\,L_{\rm Edd}(M_{0})} = \frac{L\,t_{S}}{l\,\epsilon_{r}c^{2}}. 
\end{equation}
Of course, it is clear here that $M_{\rm BH}\approx
\mbhf=3\times10^{8}\,M_{\sun}$ if we consider only times after the
final merger.

It has also been argued from observations of stellar black hole
binaries that a transition between accretion states occurs at a
critical Eddington ratio $\dot{m}\equiv \dot{M}/\dot{M_{\rm Edd}}$,
from radiatively inefficient accretion flows at low accretion rates
\citep[e.g.,][]{EMN97} to radiatively efficient accretion through a
standard \citet{SS73} disk. Although the critical Eddington ratio for
supermassive black holes is uncertain, observations of black hole
binaries \citep{Maccarone03} as well as theoretical extensions of
accretion models \citep[e.g.,][]{MLMH00} suggest $\dot{m}_{\rm
crit}\sim0.01$. We can examine whether this has a large impact on our
predictions for the luminosity function and $\nLp$ distribution, by
determining whether the distribution of Eddington ratios is
significantly changed by such a correction.  Because we assume a
constant radiative efficiency $L=\epsilon_{r}\,\dot{M}\,c^{2}$ with
$\epsilon_{r}=0.1$, we account for this effect by multiplying the
simulation luminosity at all times by an additional ``efficiency
factor'' $f_{\rm eff}$ which depends on the Eddington ratio
$l=L/L_{\rm Edd}$,
\begin{equation}
  f_{\rm eff} = \left\{ \begin{array}{ll}
      1  & \mathrm{ if\ } \eEdd > 0.01 \\
      100\,\eEdd & \mathrm{ if\ } \eEdd \leq 0.01.
\end{array}
    \right.
\end{equation}
This choice for the efficiency factor follows from ADAF models
\citep{NY95} and ensures that the radiative efficiency is continuous
at the critical Eddington ratio $\eEdd_{\rm crit}=0.01$. Applying this
correction and then examining the distribution of Eddington ratios as
a function of luminosity (dashed histograms in
Figure~\ref{fig:Pofl.A3}), we see that the distribution of Eddington
ratios is essentially identical, with only a slightly higher
probability of observing extremely low Eddington ratios
$l\lesssim0.001$.  Of course, our modeling of accretion processes does
not allow us to accurately describe ADAF-like accretion at these low
Eddington ratios, but such low values are not relevant for the
observed luminosity functions and quantities with which we make our
comparisons. This implies that such a transition in the radiative
efficiency with accretion rate should not alter our conclusions
regarding the luminosity function and the $\nLp$ distribution
(essentially, the corrections are important only at luminosities well
below those relevant in constructing the observed luminosity
functions; see also Hopkins et al.\ 2005c for a calculation
of the effects of such a correction on the fitted quasar lifetime and
$\nLp$ distributions, which leads to the same conclusion).

Despite the broad range of Eddington ratios in the simulations, this
entire distribution is unlikely to be observable in many samples. The
effect of this can be predicted based on the behavior seen in
Figure~\ref{fig:Pofl.A3}.  For example, we consider the distribution
of Eddington ratios that would be observed if the B-band luminosity
$\LBo\geq\Lcut{11}$, comparable to the selection limits at high
redshift of many optical quasar samples.  As expected from the change
in $l$ with luminosity, this restricts the observed range of Eddington
ratios to large values $l\sim0.1-1$, in good agreement with the range
of Eddington ratios actually observed in such samples. Essentially, it
has reduced the observed range to a bolometric luminosity
$L\gtrsim\Lcut{12}$ in the case shown, giving a similar distribution
to that seen in the lower panel of the figure.

%\clearpage
\begin{figure}
    \centering
    \plotone{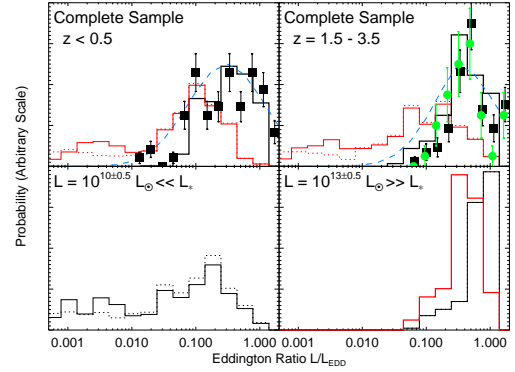}
    \caption{Predicted distribution of Eddington ratios based on
    the luminosity function and the quasar evolution in our
    simulations, in two redshift intervals $z<0.5$ (upper left) and
    $1.5<z<3.5$ (upper right). The observed distributions for radio
    loud (black squares) and radio quiet (green circles) quasars are
    shown from \citet{Vestergaard04} with Poisson errors. Thick black
    lines show the predicted distribution given the same minimum
    observed luminosity as the observed sample. Thin red lines show
    the predicted distributions for a sample extending to arbitrarily
    faint luminosities, dotted lines show the same, with the ADAF
    correction of \S~\ref{sec:eddington} applied at low accretion rates.  
    Blue dashed lines show the prediction for a fixed (luminosity-independent) Eddington 
    ratio distribution in a light-bulb or exponential light curve model, fitted to the 
    $z<0.5$ data and used to predict the $1.5<z<3.5$ Eddington ratio distribution 
    given the observational luminosity limit. Lower panels
    show the predicted distributions for $z\lessim1$ in two luminosity
    intervals, above and below the ``break'' luminosity in the
    observed luminosity function (red lines here correspond to an
    observed (attenuated) B-band luminosity $\LBo>\Lcut{11}$).
    \label{fig:Pofl.all}}
\end{figure}
%\clearpage

We compare our predicted distribution of Eddington ratios to
observations in Figure~\ref{fig:Pofl.all}. Using the distribution of
peak luminosities $\nLp$ determined from the luminosity function, we
can integrate over all luminosities to infer the observed Eddington
ratio distribution,
%\begin{equation}
%      \begin{split}
%       P(l)\propto & \int{{\rm d}\log L}\int{{\rm d}\log\Lp} \\
%       & \times P(l | L, \Lp)\,\frac{{\rm d}t(L,\Lp)}{{\rm d}\log L} \nLP .
%      \end{split} 
%\end{equation}
      \begin{eqnarray}
       P(l)&\propto& \int{{\rm d}\log L}\int{{\rm d}\log\Lp} \nonumber\\
       && \times P(l | L, \Lp)\,\frac{{\rm d}t(L,\Lp)}{{\rm d}\log L} \nLP .
      \end{eqnarray} 
As our estimate of $P(l | L, \Lp)$ above is rough, we do this by
binning in $\Lp$ and averaging the binned $P(l | L, \Lp)\,\dtdL$ for
each simulation in the range of $\Lp$, then weighting by $\nLp$ and
integrating. We consider both the entire distribution that would be
observed in the absence of selection effects (red histograms), and
the distribution observed demanding a B-band luminosity above some
reference value, $\LBo>L_{\rm min}$ (black histograms).  The results are shown
for redshifts $z<0.5$ and $z=1.5 - 3.5$, along with the observed
distribution from \citet{Vestergaard04}, with assumed Poisson
errors. The observations should be compared to the black histograms,
which have luminosity thresholds $L_{\rm}=\Lcut{10}\ {\rm and}\
\Lcut{11}$ for $z<0.5$ and $z=1.5 - 3.5$, respectively, corresponding
approximately to the minimum observable luminosities in the observed
samples in each redshift interval.

The agreement is good, given the observational uncertainties, and it
suggests that the observed Eddington ratio distribution can be related
to the non-trivial nature of quasar lifetimes and light curves we
model, rather than some arbitrary distribution of fixed $l$ across
sources. However, the selection effects in the observed samples are
quite significant -- the complete distribution of Eddington ratios is
similar in both samples, implying that the difference in the observed
Eddington ratio distribution is primarily a consequence of the higher
luminosity limit in the observed samples -- and a more detailed test
of this prediction requires fainter samples.

Still, there is a systematic offset in the observed samples at $z<0.5$
and $z=1.5-3.5$ which does not owe to selection effects.  At
progressively lower redshifts, more quasars with luminosities further
below the ``break'' in the luminosity function are observed, and
therefore the observed Eddington ratio is broadened to lower Eddington
ratios $l\sim0.1$, whereas at high redshift the distribution is more
peaked at slightly higher Eddington ratios.  This difference, although
not dramatic, is a prediction of our model not captured in ``light
bulb'' or ``fixed Eddington ratio'' models, even when allowing for a
distribution of Eddington ratios, if such a distribution is static. We
demonstrate this by fitting the low-redshift Eddington ratio
distribution to a Gaussian (blue dashed lines in upper left), and then
assuming that this distribution of accretion rates is unchanged with
redshift, giving (after applying the same selection effects which
yield the black histograms plotted) the blue dashed line in the upper
right panel.  Although the agreement may appear reasonable, the
difference is significant -- such a fit overpredicts the fraction of
high redshift objects at Eddington ratios $\lessim0.1$ and
underpredicts the fraction at $\sim0.3$, giving a somewhat poor fit
overall ($\reducechi=2.7$, but with typical $\gtrsim3\sigma$
overpredictions for Eddington ratios $\lesssim0.1$).

Furthermore, without being modified to allow for a distribution of
Eddington ratios, such models are clearly inconsistent with the
observations, as they would predict a single, constant Eddington
ratio. However, models which fit the observed evolution in the quasar
luminosity function with a non-static distribution of accretion rates
do recover the broadening of the Eddington ratio distribution at low
redshift, so long as strong evolution in the distribution of accretion
rates for systems of a given black hole mass is not allowed
\citep{SW03}, giving a qualitatively similar picture of the evolution
we model.  Regardless of the evolution in accretion rates, an
advantage of our modeling is that it provides a physically motivated
predicted distribution of accretion rates, as opposed to being forced
to adopt the distribution of accretion rates from observational input
(which can be, as demonstrated in the figure, significantly biased by
observational selection effects).  The dotted histograms show the
distribution if we apply our ADAF correction to the intrinsic
distribution, and demonstrate that this does not
significantly change the result.  We note that our model for black
hole accretion employs the Eddington limit as a maximum accretion
rate; if we remove this restriction, we find that the simulations
spend some small but non-negligible time with $l\sim1-2$, which is
also consistent with the observations.

Furthermore, we can make a prediction of this model which can be
falsified, namely that the Eddington ratio distribution at
luminosities well below the break in the luminosity function should be
broader and extend to lower values than the distribution at
luminosities above the break luminosity. We quantify this in the lower
panels of Figure~\ref{fig:Pofl.all}, for the distribution at low
redshifts $z\lesssim1$. Here we consider two bins in luminosity,
$L=10^{9.5}-10^{10.5}\,L_{\sun}$ and
$L=10^{12.5}-10^{13.5}\,L_{\sun}$, for redshifts where the break in
the luminosity function is at approximately
$L\sim10^{11}-10^{12}\,L_{\sun}$. Clearly, the distribution is
broader and extends to lower Eddington ratios in the former
luminosity interval, whereas in the latter it is strongly peaked about
$l\sim0.2-1$, for both the complete distribution (black) and that with
$\LBo\geq\Lcut{11}$ (red). The distribution obtained applying the ADAF
correction described above is shown as dotted histograms.  Despite the
fact that the Eddington ratio distribution at low luminosities will be
strongly biased by selection effects, a reasonably complete sample
should be able to test this prediction, at least qualitatively.

%\clearpage
\begin{figure}
    \centering
    \plotone{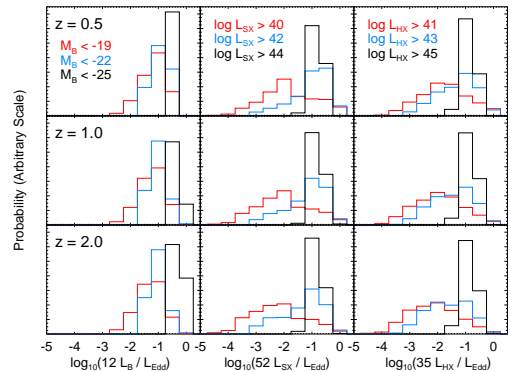}
    \caption{Predicted distribution of Eddington ratios based on
    the luminosity function and the quasar evolution in our
    simulations, at three redshifts $z=0.5$ (top panels), $z=1.0$ (middle), and $z=2.0$ (bottom). 
    The inferred distribution of Eddington ratios, adopting a constant bolometric 
    correction from the observed (attenuated) luminosity in each of three bands is shown, 
    i.e.\ assuming $L=12\,L_{B}^{\rm obs}$ ($4400\,$\AA; left), 
    $L=52\,L_{SX}^{\rm obs}$ (0.5-2 keV; middle), and
    $L=35\,L_{HX}^{\rm obs}$ (2-10 keV; right). For each waveband, results are shown 
    for three reference luminosities. In B-band, $M_{B}<-19$ (red), $M_{B}<-22$ (blue), 
    and $M_{B}<-25$ (black). In soft X-rays, $\log(L_{SX} [{\rm erg\,s^{-1}}])>40\ (\rm red),\ 
    42\ (\rm blue),\ 44\ (\rm black)$. In hard X-rays, $\log(L_{HX} [{\rm erg\,s^{-1}}])>41\ (\rm red),\ 
    43\ (\rm blue),\ 45\ (\rm black)$. 
    \label{fig:Pofl.grid}}
\end{figure}
%\clearpage

We illustrate the effects of changing observed waveband, redshift, and
luminosity thresholds on the observed Eddington ratio distribution in
Figure~\ref{fig:Pofl.grid}.  Here, we plot the predicted distribution
of Eddington ratios determined as in Figure~\ref{fig:Pofl.all}, from
our fitted $\nLp$ distribution at each redshift and the distribution
of Eddington ratios as a function of instantaneous and peak luminosity
in each of our simulations (specifically, these are drawn from the
Monte Carlo realizations of the quasar population described in
\S~\ref{sec:discussion}).  We show the predictions at three redshifts
$z=0.5$ (top panels), $z=1.0$ (middle), and $z=2.0$ (bottom).  For
each redshift, results are shown in three wavebands, and with three
reference luminosities.  In B-band, we require $M_{B}<-19$ (red),
$M_{B}<-22$ (blue), and $M_{B}<-25$ (black). In soft X-rays,
$\log(L_{SX} [{\rm erg\,s^{-1}}])>40\ (\rm red),\ 42\ (\rm blue),\ 44\
(\rm black)$. In hard X-rays, $\log(L_{HX} [{\rm erg\,s^{-1}}])>41\
(\rm red),\ 43\ (\rm blue),\ 45\ (\rm black)$.  The observationally
inferred distribution of Eddington ratios at each redshift is loosely
estimated by adopting a constant bolometric correction from the
observed (attenuated) luminosity in each of three bands shown, i.e.\
assuming $L=12\,L_{B}^{\rm obs}$ ($4400\,$\AA; left),
$L=52\,L_{SX}^{\rm obs}$ (0.5-2 keV; middle), and $L=35\,L_{HX}^{\rm
obs}$ (2-10 keV; right). This follows common practice in many
observational estimates of the Eddington ratio distribution and allows
for the effects of attenuation, but we caution that it
can be misleading.

If we instead use the luminosity-dependent bolometric corrections of
\citet{Marconi04} which we adopt throughout, even given that we are
calculating from the observed (attenuated) luminosities, we do not see
the large population of highly sub-Eddington (Eddington ratios
$\lesssim10^{-3}$) quasars in soft and hard X-ray samples with low
luminosity thresholds. This is because these are actually reasonably
high-Eddington ratio quasars, %, albeit dim ($L\lessim10^{11}\,L_{\sun}$), quasars, 
but our bolometric corrections imply that a larger fraction
of the bolometric luminosity is radiated in the X-ray at low
bolometric luminosity, meaning that assuming a constant bolometric
correction will underestimate the Eddington ratios of high-bolometric 
luminosity sources. Regardless, the figure illustrates both the importance
of different wavelengths (i.e.\ the ability to observe more
low-Eddington ratio sources in X-ray as compared to optical samples)
and luminosity/magnitude limits on the inferred distribution of
Eddington ratios.  For example, even for relatively deep B-band quasar
samples complete to $M_{B}<-23$ (i.e.\ complete to essentially all
objects traditionally classified as having ``quasar-like''
luminosities), the expected observed Eddington ratio distribution at
$z\sim0.5-2$ is quite sharply peaked about $\sim0.1-0.3$, in good
agreement with recent observational results \citep{Kollmeier05}.

We do not compare to the $z=0$ distribution of black hole accretion
rates, as this is dominated by objects at extremely low Eddington
ratios $l\sim10^{-5}-10^{-4}$
\citep[e.g.,][]{Ho02,Marchesini04,Jester05}, which are well below the
range we model, and are not likely to be driven by merger activity
(many of these objects are quiescent, low-luminosity Seyferts in
normal spiral galaxy hosts); furthermore, many of these objects are
not accreting at the Bondi rate
\citep{FC88,BB99,DiMatteo00,NIA00,QG00,DCF01,Loewenstein01,Bower03},
clearly showing that our simulations must incorporate more
sophisticated models for accretion in quiescent, low-luminosity states
(when gravitational torques cannot provide a mechanism to drive large
amounts of gas to the central regions of the galaxy) in order to
describe such phases.

However, it has been suggested that the rapid ``blowout'' phase and
subsequent decay in accretion rates seen in our simulations, coupled
with spectral modeling of radiatively inefficient accretion modes, can
explain the apparently bimodal distribution of low-redshift accretion
rates \citep{CX05}.  Moreover, present-day, relaxed ellipticals are
observed to have mass accretion rates $\sim10^{-4}$ implying a long
relaxation time at moderate and low accretion rates, qualitatively
similar to that seen after the ``blowout'' in our modeling (Hopkins et
al. 2005f).  A pure exponential decay in accretion rate after the peak
quasar phase would give $\dot{m}=\dot{M}/\dot{M}_{\rm
Edd}\sim\exp{(-t_{H}/t_{Q})}$ at present, where $t_{H}$ is the Hubble
time and $t_{Q}$ is the quasar lifetime of order e.g.\ the Salpeter
time $t_{S}=4\times10^{7}$\,yr, yielding an unreasonably low expected
accretion rate $\dot{m}\sim10^{-145}$. Even assuming an order of
magnitude larger quasar lifetime, this gives $\dot{m}\sim10^{-15}$,
far below observed values, implying that regardless of the fueling
mechanisms at low luminosities, the basic key point of our modeling
must be true to some extent, namely that quasars spend long times
relaxing at moderate to low Eddington ratios.

\section{The Mass Function of Relic Supermassive Black Holes from Quasars}
\label{sec:smbh}

From the $M_{\rm BH}$-$\sigma$ relation and other host galaxy-black 
hole scalings,
estimates of bulge and spheroid velocity dispersions have been used to
determine the total mass density ($\rhobh$) and mass distribution of
local, primarily inactive supermassive black holes
\citep[e.g.,][]{Salucci99,MS02,YT02,Ferrarese02,AR02,Marconi04,Shankar04}.
These estimates, along with others based on X-ray background synthesis
\citep[e.g.,][]{FI99,ERZ02}, have compared these quantities to those
expected based on the mass distribution of `relic' black holes grown
in quasars.  It appears that most, and perhaps nearly all of the
present-day black hole mass density was accumulated in bright quasar
phases, and the $M_{\rm BH}-\sigma$ and $M_{\rm BH}-L_{\rm bulge}$
correlations yield estimates of the local mass function in good
agreement with those from hard X-ray AGN luminosity functions
\citep{Marconi04}.

However, this modeling is dependent on several assumptions.  Namely,
the average radiative efficiency $\epsilon_{r}$, Eddington ratio $l$,
and average quasar lifetime $t_{Q}$ are generally taken to be
constants and either input into the model or constrained by demanding
agreement with the local mass function.  In our simulations, we find
the quasar lifetime and Eddington ratio to be complex functions of
both luminosity and host system properties (as opposed to being
constants). We also find that quasars spend a large fraction of their
lives in obscured growth phases, suggesting some mass gain outside of
the bright quasar phase. It is thus of interest to determine the relic
black hole mass function expected from our model for quasar evolution.

Using our estimate for the birthrate of quasars with a given peak
luminosity at a particular redshift, $\nLp$, obtained from the
luminosity function in \S~\ref{sec:fullLF}, we can estimate the total
number density of relic quasars accumulated by a particular redshift
that were born with a given $\Lp$ (per logarithmic interval in $\Lp$)
from
\begin{equation}
n(\Lp)=\int{\,\nLp\,{\rm d}t}=\int{\frac{\dot{n}(\Lp,z)\,{\rm d}z}{(1+z)H(z)}}.
\end{equation}
By redshift $z=0$, most of these quasars will be ``dead,'' with only
a small residual fraction having been activated in the recent past.

Using our log-normal form for $\nLp$, with normalization $\nstar$ and
dispersion $\sstar$ held constant and only the median
$\lstar=\lstar^{0}\,\exp{(k_{L}\tau)}$ evolving with redshift, this
integral can be evaluated numerically to give the space density of
relic quasars $n(\Lp)$.  Finally, we use $\mbhf(\Lp)$, roughly the
Eddington mass of the given peak luminosity (but determined more
precisely in \S~\ref{sec:detailsCompare}) to convert from ${\rm
d}n(\Lp)/{\rm d}\log\Lp$ to ${\rm d}n(M_{\rm BH})/{\rm d}\log M_{\rm
BH}$.  This formulation implicitly assumes that black holes do not
undergo subsequent mergers after the initial quasar-producing event.
However, this effect should be small (a factor $\lesssim2$) as
subsequent mergers would be dry (gas poor).
We explicitly calculate the effects of dry mergers on the
spheroid mass function (essentially a rescaling of the black hole mass
function calculated here) in \citet{H05e}, and show that this is a
small effect (significantly less than the uncertainties owing to our
fit to the quasar luminosity function) even assuming the maximum dry
merger rates of e.g.\ \citet{vanDokkum05}.

This mass function can then be integrated over $d M_{\rm BH}$ to give
the total present-day black hole mass density, $\rhobh$. Neglecting
temporarily the small corrections to $\mbhf(\Lp)$ from
\S~\ref{sec:detailsCompare}, we expect
\begin{equation}
\mbhf\approx M_{\rm Edd}(\Lp)=\frac{\Lp\,t_{S}}{\epsilon_{r}c^{2}}
%2.95\times 10^{-5} M_{\sun} \, \Bigl( \frac{\Lp}{L_{\sun}}\Bigr), 
\end{equation}
where $t_{S}/\epsilon_{r}c^{2}\approx2.95\times10^{-5}\,M_{\sun}/L_{\sun}$,
so therefore, 
\begin{equation}
\rhobh=\frac{t_{S}}{\epsilon_{r}c^{2}}\int{\Lp\,n(\Lp)\,{\rm d}\log \Lp}.
\end{equation}
This can be combined with the integral over redshift for $n(\Lp)$,
giving, at each $z$, a pure Gaussian integral over $\log{(\Lp)}$, 
in the form
%\begin{equation}
%\begin{split}
%\rhobh &=\frac{\lstar^{0}\,t_{S}}{\epsilon_{r}c^{2}}\frac{\nstar}{H_{0}}\,
%e^{\frac{1}{2}(\sstar \ln{10})^{2}}\,\int{\frac{e^{k_{L}\tau}\,dz}{(1+z)\,\hat{H}(z)}}\\
%&= \frac{\lstar^{0}\,t_{S}}{k_{L}\epsilon_{r}c^{2}}\frac{\nstar}{H_{0}}\,
%e^{\frac{1}{2}(\sstar \ln{10})^{2}}\,\Bigl(e^{k_{L}\tau_{f}}-e^{k_{L}\tau}\Bigr), 
%\end{split}
%\end{equation}
\begin{eqnarray}
\rhobh &=&\frac{\lstar^{0}\,t_{S}}{\epsilon_{r}c^{2}}\frac{\nstar}{H_{0}}\,
e^{\frac{1}{2}(\sstar \ln{10})^{2}}\,\int{\frac{e^{k_{L}\tau}\,dz}{(1+z)\,\hat{H}(z)}}\nonumber\\
&=& \frac{\lstar^{0}\,t_{S}}{k_{L}\epsilon_{r}c^{2}}\frac{\nstar}{H_{0}}\,
e^{\frac{1}{2}(\sstar \ln{10})^{2}}\,\Bigl(e^{k_{L}\tau_{f}}-e^{k_{L}\tau}\Bigr), 
\end{eqnarray}
where $\hat{H}(z)\equiv H(z)/H_{0}$ and $\tau_{f}$ is the fractional
lookback time at some upper limit.  We must modify this integral above
$z\sim2$ to account for the decreasing space density of bright
quasars, applying either our density or peak-luminosity evolution
turnover from \S~\ref{sec:fullLF}, but quasars at these high redshifts
contribute only a small fraction to the present-day density.  Thus, in
this formulation, the evolution of the total supermassive black hole
mass density, i.e.\ $\rhobh(z)/\rhobh(z=0)$, is given approximately by
the dimensionless integral above, and depends only on how
$\lstar$ evolves, essentially the rate at which the break
in the quasar luminosity function shifts.  Although this is not strictly
true if we include corrections to $\mbhf(\Lp)$ based on $\Lp$, the 
difference is small and this behavior is essentially preserved.  Note 
that the total supermassive black hole mass density is independent of
corrections from subsequent dry mergers, which (being gas poor) conserve
total black hole mass.

%\clearpage
\begin{figure*}
    \centering
    \plotone{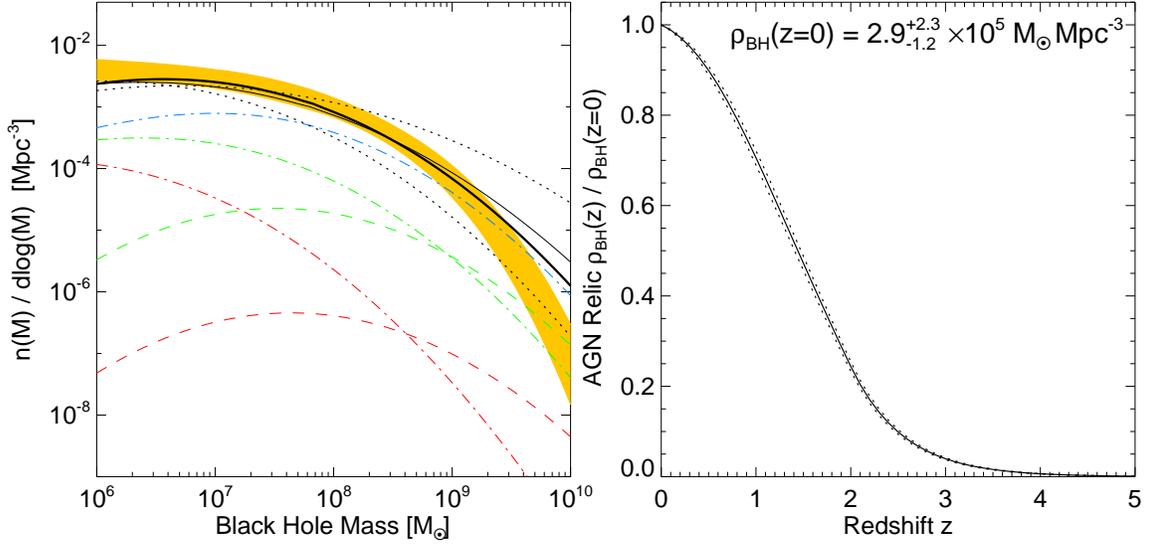}
    \caption{Right: Total predicted quasar relic black hole mass
    density and evolution of the fractional black hole mass
    density with redshift. Dotted lines show the difference resulting
    from $1\sigma$ deviation in fitted $\nLp$ from the luminosity
    function.  Left: Predicted present $z=0$ relic mass function
    (thick black line), for comparison with the $1\sigma$ range
    (yellow) of the inferred supermassive black hole mass function
    from \citet{Marconi04}. Also shown are the results given $1\sigma$
    errors in the fitted $\nLp$ distribution (dotted lines), or ignoring
    the small corrections to $\mbhf(\Lp)$ from \S~\ref{sec:detailsCompare}
    (thin black line). Dot-dashed lines show the predicted mass
    function at $z=1.5,\,3.0,\,5.0$ (blue, green, and red,
    respectively). The extensions to $z>2$ includes the turnover (pure
    peak luminosity evolution form) in the quasar space density above
    $z=2$ from high-redshift luminosity functions described in 
    \S~\ref{sec:fullLF}, except for the dashed green and red lines 
    which use the pure density evolution form. 
    \label{fig:SMBH.distrib}}
\end{figure*}
%\clearpage

Figure~\ref{fig:SMBH.distrib} shows our prediction for the mass
distribution of supermassive black holes, as well as the total density
$\rhobh$ and its evolution with redshift. We find a total relic black
hole mass density of
$\rhobh=2.9^{+2.3}_{-1.2}\times10^{5}\,M_{\sun}\,{\rm Mpc^{-3}}$, in
agreement with the observational estimate of
$\rhobh=2.9\pm0.5\,h^{2}_{0.7}\times10^{5}\,M_{\sun}\,{\rm Mpc^{-3}}$,
by \citet{YT02} ($h_{0.7}\equiv H_{0}/70\,{\rm km\,s^{-1}\,Mpc^{-1}}$;
their result is converted from $h=0.65$), and within $1\sigma$ of the
value $\rhobh=4.6^{+1.9}_{-1.4}h^{2}_{0.7}\times10^{5}\,M_{\sun}\,{\rm
Mpc^{-3}}$, of \citet{Marconi04}, based on the observations of
\citet{Marzke94}, \citet{Kochanek01}, \citet{Nakamura03},
\citet{Bernardi03}, and \citet{Sheth03}.  The fractional evolution of
$\rhobh$ with redshift is quite well constrained, and we find, as with
previous estimates, that most of the present-day black hole mass
density accumulates at moderate to low redshifts 
$z\approx 0.5 - 2.5$. The $1\sigma$
errors are shown as dotted lines in the figure, and are close to our
best-fit estimate, as we have demonstrated that this quantity depends
only on $k_{L}$, the rate of evolution of the break in
the luminosity function
with redshift, which is fairly well-constrained by
observations (from our fitting to the luminosity functions,
$k_{L}=5.61\pm0.28$). The difference in $\rhobh$ if we include or
neglect the small corrections to $\mbhf$ is negligible compared to our
errors ($\sim5\%$).

Our estimate for the relic black hole mass distribution (thick black
line) also agrees well with observational estimates, with all
observations within the range allowed by the $1\sigma$ errors of our
fitting to the luminosity function (dotted lines). The observations
shown are again from \citet{Marconi04}, based on the combination of
observations by \citet{Marzke94}, \citet{Kochanek01},
\citet{Nakamura03}, \citet{Bernardi03}, and \citet{Sheth03}.  The high
mass end of the black hole mass function $M_{\rm BH}>10^{9}\,M_{\sun}$
is relatively sensitive to whether or not we apply the $\mbhf(\Lp)$
corrections of \S~\ref{sec:detailsCompare}, instead of taking
$\mbhf=M_{\rm Edd}(\Lp)$ (thin line), as well as to our fitting
procedure.  However, the agreement is still good, and this is also
where the observational estimates of the mass distribution are most
uncertain, as they are generally extrapolated to these masses, and are
sensitive to the assumed intrinsic dispersions in the $M_{\rm
BH}-\sigma$ and $M_{\rm BH}-L_{\rm bulge}$ relations \citep{YT02}.

If, instead, we adopt a light-bulb, constant Eddington ratio, or
exponential light curve model for quasar evolution, we would have
$M_{\rm BH}^{f}\propto \Lp$, and thus the prediction would be similar
to the thin black line shown, a somewhat worse fit at high black hole
masses. However, in these models this can be remedied by adjusting the
typical Eddington ratios, quasar lifetimes, or radiative
efficiencies. We do not show the range of predictions of these models
for the relic supermassive black hole mass function, as they have been
examined in detail previously
\citep[e.g.,][]{Salucci99,MS02,YT02,Ferrarese02,AR02,Marconi04,Shankar04}.
These works demonstrate that the observed quasar luminosity functions
are consistent with the relic supermassive black hole mass function,
given typical radiative efficiencies $\epsilon_{r}\sim0.1$ and
Eddington ratios $\sim0.5-1.0$, and that most of the mass of black
holes is accumulated in bright, observed phases, or else the required
radiative efficiency would violate theoretical limits.

That our model of quasar lifetimes and obscuration reproduces the
observed $z=0$ supermassive black hole mass function explicitly
demonstrates that we are consistent with these constraints.  By
choice, the radiative efficiency in our simulations is
$\epsilon_{r}=0.1$, and accretion rates are not allowed to exceed
Eddington. As noted in \S~\ref{sec:BLqso}, most of the black hole mass
is accumulated and radiant energy released in the final, ``blowout''
phase of quasar evolution, and here our black hole mass function and
cumulative black hole mass density demonstrate that our modeling is
consistent with integrated energy and mass arguments such as that of
\citet{Soltan82}, despite the fact that quasars spend more {\em time}
in obscured phases than they do in bright optical quasar phases.  In
fact, comparison of our predicted total black hole mass density with
estimates from the $z=0$ black hole mass distribution allows some
latitude for significant mass gain in radiatively inefficient growth
or black holes in small, disky spheroids, although we emphasize that
this is mainly because the uncertainty in our prediction is large, it
is not inherent or necessary in our modeling.

%\clearpage
\begin{figure*}
    \centering
    %\epsscale{1.1}
    \plotone{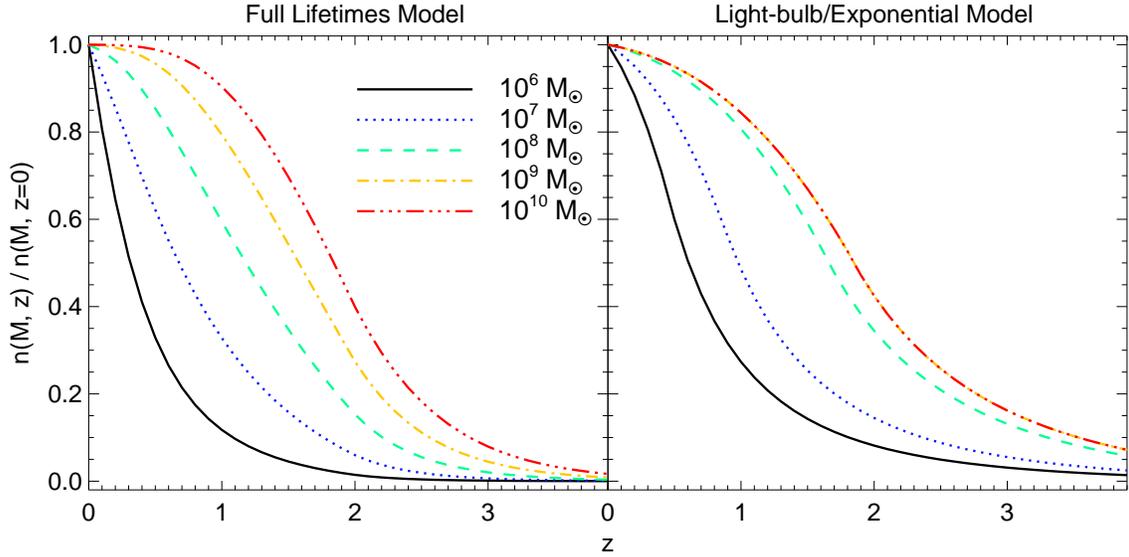}
    \caption{Fractional number
    density $n(M,\,z)/n(M,\,z=0)$ of black holes of a given mass as a function of
    redshift, for several different black hole masses as shown. For
    $z>2$ this includes the turnover (pure density evolution
    form) in the quasar space density above $z=2$ from high-redshift
    luminosity functions described in \S~\ref{sec:fullLF}.
    Left panel shows the results using our full model of quasar lifetimes, 
    right panel assuming a ``light-bulb'' or exponential (constant 
    Eddington ratio) light curve model. 
    The yellow dot-dash ($10^{9}\,M_{\sun}$) and red triple-dot-dash
    ($10^{10}\,M_{\sun}$) curves are nearly identical in the right panel. 
    \label{fig:SMBH.evol}}
\end{figure*}
%\clearpage

The anti-hierarchical nature of black hole formation, where less
massive black holes are formed at lower redshift, is reflected in our
modeling by the shift of the break in the quasar luminosity function
to lower values with decreasing redshift.  This can be seen in
Figure~\ref{fig:SMBH.distrib}, where the black hole mass distributions
are shown at redshifts $z=1.5,\,3.0\ {\rm and}\ 5.0$, assuming either
pure peak luminosity evolution or pure density evolution for $z>2$
(dot-dashed and dashed, respectively).  While the choice for the
turnover in the $z>2$ quasar density matters little for the $z<2$
black hole mass functions, the low-$M_{\rm BH}$ distribution at high
redshift (where observations do not constrain $\nLp$ well) is quite
different between the two models.  Figure~\ref{fig:SMBH.evol} plots
the fractional number density of black holes of a given mass as a
function of redshift, i.e.\ $n(M,\,z)/n(M,\,z=0)$, where $n(M)={\rm
d}n/{\rm d}\log{(M)}$ is just the number density at mass $M$.  This
figure demonstrates that higher-mass black holes originated over a
larger range of redshifts, and that they mostly formed
at higher redshift, compared to lower-mass black holes.

The right panel of Figure~\ref{fig:SMBH.evol} compares our prediction
to that of a light-bulb or exponential light curve model for quasar
lifetimes.  In these models, the anti-hierarchical nature of black
hole assembly is dramatically suppressed.  At the high-mass end, there
is no measurable difference in the distribution of formation redshifts
(i.e.\ the $M_{\rm BH}=10^{9}\,M_{\sun}$ and $M_{\rm
BH}=10^{10}\,M_{\sun}$ curves are indistinguishable), and there is
little change in the formation times at $M_{\rm BH}=10^{8}\,M_{\sun}$.
The shift in formation redshift at lower masses, although significant,
is smaller than that predicted in our model.  If spheroids and black
holes are produced together, as in our picture, these models of the
quasar lifetime would imply that spheroids of masses $M_{\rm
vir}\sim10^{11}-10^{13}\,M_{\sun}$ all formed over nearly identical
ranges of redshifts, which is inconsistent with many observations
indicating anti-hierarchical growth of the red, elliptical galaxy
population \citep[e.g.,][]{Treu01,vanDokkum01,Treu02,
vDS03,Gebhardt03,Rusin03,vandeVen03,Wuyts04,Treu05,Holden05,
vdW05,SA05,Nelan05}.  Implications of our model for the red galaxy
sequence are considered in \citet{H05e}, where we show that this
weaker anti-hierarchical black hole (and correspondingly, spheroid)
evolution is inconsistent with observed luminosity functions,
color-magnitude relations, and mass-to-light ratios of elliptical
galaxies.

Our modeling reproduces the observed total density and mass
distribution of supermassive black holes at $z=0$ with black holes
accreting at the canonical efficiency $\epsilon_{r}=0.1$ expected for
efficient accretion through a \citet{SS73} disk. Presumably, a large
change in $\epsilon_{r}$ would give a significantly different relation
between peak luminosity and black hole mass (for the same $\Lp$,
$M_{\rm BH}^{f}\propto 1/\epsilon_{r}$), and thus if the quasar
lifetime remained similar as a function of peak luminosity, this would
translate to a shift in the black hole mass function.  The long
obscured stage in black hole evolution does not generate problems in
reproducing the black hole mass density, and the final phases of
growth are still in bright optical quasar stages.  However, a large
Compton-thick population of black holes at all luminosities (or even
at some range of luminosities at or above the break in the luminosity
function) \citep[e.g.,][]{Gilli01,Ueda03}, or a large population
accreting in a radiatively inefficient ADAF-type solution, 
as invoked to explain discrepancies in the X-ray
background produced by synthesis models \citep{DEFN99}, would result in a significant
over-prediction of the present-day supermassive black hole density.
As we demonstrate in \S~\ref{sec:xrb.tot}, invoking such populations
is unnecessary, as our picture for quasar lifetimes and evolutionary
obscuration self-consistently reproduces the observed X-ray
background.

Finally, we note that we reproduce the $z=0$ distribution of black
hole masses {\em inferred} from the distribution of spheroid velocity
dispersions \citep{Sheth03} and luminosity functions
\citep{Marzke94,Kochanek01,Nakamura03}, based on the observed $M_{\rm
BH}-\sigma$ relation and fundamental plane for galaxy properties
\citep[e.g.,][]{Bernardi03,Gebhardt03}. Therefore, since our modeling
also reproduces the observed $M_{\rm BH}-\sigma$
\citep{DSH05,Robertson05b} and fundamental plane (Robertson et al., in
preparation) relations, we implicitly reproduce the $z=0$ distribution
of spheroid velocity dispersions and spheroid luminosity functions,
given our basic assumption that the mergers that produce these
spheroids also give rise to luminous quasar activity.

\section{The Cosmic X-Ray Background}
\label{sec:xrb}
\subsection{The Integrated Spectra of Individual Quasars}
\label{sec:xrb.indiv}

Unresolved extragalactic sources, specifically obscured AGN, have been
invoked to explain the cosmic X-ray background \citep[e.g,][]{SW89}.
This picture has been confirmed as deep surveys with {\it Chandra} and
{\it XMM-Newton} have resolved most or all of the X-ray background
into discrete sources, primarily obscured and unobscured AGN
\citep{Brandt01,Hasinger01,Rosati02,Giacconi02,Baldi02}.  The X-ray
background, however, has a harder X-ray spectrum than typical quasars,
with a photon index $\Gamma\sim1.4$ in the $1-10$\,keV range
\citep{Marshall80}.  Therefore, obscured AGN are important in
producing this shape, as absorption in the ultraviolent and soft
X-rays hardens the observed spectrum.  Indeed, population synthesis
models based on observed quasar luminosity functions and involving
large numbers of obscured AGN have been successful at matching both
the X-ray background intensity and spectral shape
\citep{Madau94,Comastri95,Gilli99,Gilli01}. However, these models make
arbitrary assumptions about the ratio of obscured to unobscured
sources and its evolution with redshift, choosing these quantities to
reproduce the X-ray background.  Furthermore, as X-ray surveys have
been extended to higher redshifts, it has become clear that both the
observed redshift distribution of X-ray sources and the ratio of
obscured to unobscured sources is inconsistent with that required by
these models \citep{Hasinger02,Barger03}.  Even synthesis models based
on higher-redshift X-ray surveys and using observationally derived
ratios of obscured to unobscured sources \citep[e.g.,][]{Ueda03} have
invoked ad hoc assumptions about additional populations of obscured
sources to reproduce the X-ray background shape and intensity.

%\clearpage
\begin{figure*}
    \centering
    \plotone{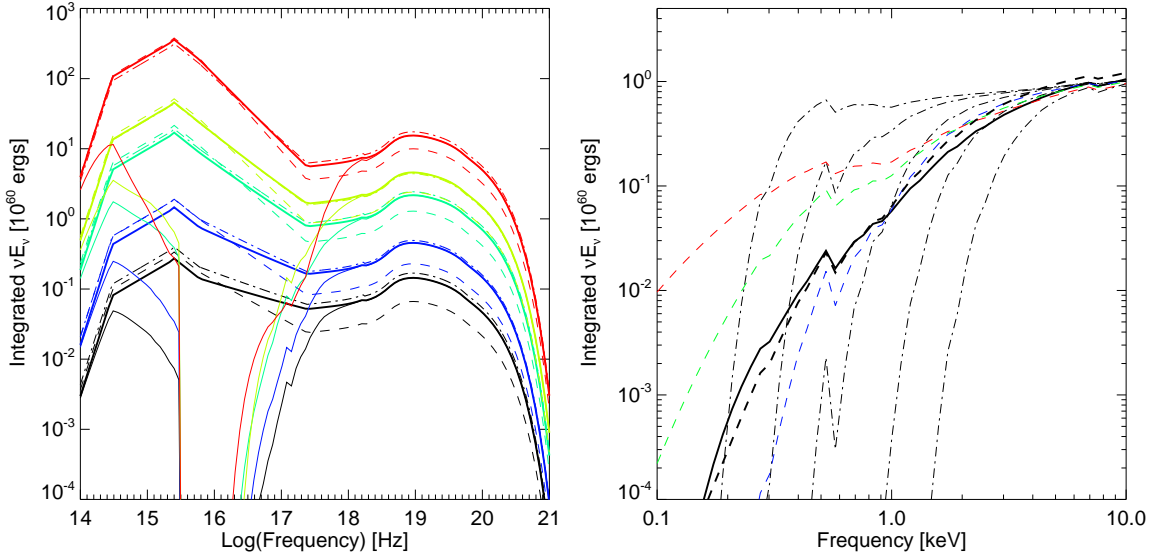}
    \caption{Left: Integrated intrinsic spectra (thick solid lines)
    from simulations A1, A2, A3, A4, and A5 (black, blue, green,
    yellow, red, respectively), with virial velocities $\vvir=80,\, 113,\,
    160,\, 226,\, {\rm and}\ 320\,{\rm km\,s^{-1}}$. The predicted
    integrated spectra from our model for quasar lifetimes are shown
    as dot-dashed lines, and the prediction of a ``light bulb''
    model, where the same total energy is radiated at $L=\Lp$, as
    dashed lines. Integrated observed spectra are shown as thin solid
    lines.  Right: Integrated observed X-ray spectrum from the A3
    simulation (thick black line), compared with the integrated
    intrinsic spectrum, reddened by various column density
    distributions: our fitted \NH\ distributions from \S~\ref{sec:NHfunction}
    (thick black dashed line), constant (luminosity-independent) lognormal
    \NH\ distribution with $\meanNH=10^{22}\ {\rm cm^{-2}}$ and
    $\sigNH=0.4,\,0.7,\,1.0$ (blue, green, and red dashed lines,
    respectively), and constant
    $\nh=10^{21},\,10^{21.5},\,10^{22},\,10^{22.5},\ {\rm and}\ 10^{23}\
    {\rm cm^{-2}}$ (thin dot-dashed lines).
    \label{fig:integ.spectra}}
\end{figure*}
%\clearpage

We can test our model by examining whether the quasar luminosity
function, relic AGN mass distribution, and X-ray background can be
simultaneously reproduced in a self-consistent manner.  Because our
formulation describes the birthrate of quasars with a peak luminosity
$\Lp$, it is most useful to consider the integrated energy spectrum of
such a quasar over its lifetime,
\begin{equation}
\nu E_{\nu} = \int{{\rm d}t\,\nu L_{\nu}(t)} = \int{\nu f_{\nu}(L)
  L\,\frac{{\rm d}t(L,\Lp)}{{\rm d}\log L}\,{\rm d}\log L},
\end{equation}
where $f_{\nu}(L)$ is the bolometric correction ($L_{\nu}\equiv
f_{\nu}\,L$).  As an example,
Figure~\ref{fig:integ.spectra} shows the integrated
intrinsic spectra (thick solid lines) from the simulations
A1, A2, A3, A4, and A5,
described in \S~\ref{sec:sims}.  The final black hole masses for these
simulations are
$\mbhf=7\times10^{6},\,3\times10^{7},\,3\times10^{8},\,7\times10^{8},\,{\rm
and}\ 2\times10^{9}\, M_{\sun}$,
respectively.  The integrated spectral shape in the
X-ray, in particular, is ultimately determined by the observationally
motivated bolometric corrections of \citet{Marconi04}, with a
reflection component in the X-ray determined following \citet{MZ95},
and, in the case of the observed spectrum, the distribution of column
densities calculated from the simulations.  Using our fits
to the lifetime $\dtdL$ as a function of instantaneous and peak
luminosities, we can calculate the expected $\nu E_{\nu}$ from the
integral above. These integrated spectra are shown as the dot-dashed
lines in the figure, and agree well with the actual integrated spectra
of the simulations, demonstrating the self-consistency of our model
and applicability of our fitted lifetimes.  

This can be compared to idealized models for the quasar lifetime,
where we allow the quasar to radiate just at its peak luminosity
$\Lp\approx L_{\rm Edd}(\mbhf)$ for some fixed lifetime
$t_{Q}^{0}$. We determine $t_{Q}^{0}$ by demanding that the total
energetics be correct, $\Lp t_{Q}^{0} = \epsilon_{r}\mbhf c^{2}$.  The
predicted integrated energy spectra are shown as the dashed lines, and
under-predict the soft and hard X-ray energy output by a factor
$\sim1.5-2$. This is because higher-luminosity quasars tend to have a
larger fraction of their energy radiated in the UV-optical rather than
the X-ray \citep[e.g.,][]{Wilkes94,Green95,VBS03,Strateva05},
reflected in our bolometric corrections. Thus, assuming that the
quasar spends all its time at $\Lp$ does not account for extended
times at lower luminosity, where the ratio of X-ray to total
luminosity is higher, which would generate an integrated spectrum with
a larger fraction of its energy in the X-ray. Assuming that the quasar
undergoes pure Eddington-limited growth to its peak luminosity
produces an almost identical integrated spectrum to this light-bulb
model, as it is similarly dominated by $L\sim\Lp$.

Of course, the intrinsic integrated energy spectrum of the simulations
is not what determines the X-ray background, but rather the integrated
{\em observed} spectrum is the critical quantity. This is shown as the
thin lines in the left panel of Figure~\ref{fig:integ.spectra}, and in
detail for our fiducial A3 simulation in the right panel of the figure
(thick solid line).  Along a given sightline, the observed integrated
spectrum will be
\begin{equation}
\nu \frac{{\rm d} E_{\nu}}{{\rm d}\Omega}=\int{{\rm d}t\,\nu \frac{L_{\nu}(t)}{4\pi}\,e^{-\tau_{\nu}(\Omega,\,t)}}, 
\end{equation}
where $\tau_{\nu}$ is the optical depth at a given frequency.
We can integrate over solid angle and obtain
\begin{equation}
\nu E_{\nu,\,{\rm obs}} = \int{\nu f_{\nu}\EV{e^{-\tau_{\nu}}}\,
  L\,\frac{{\rm d}t(L,\Lp)}{{\rm d}\log L}\,{\rm d}\log L},
\end{equation}
where $\EV{e^{-\tau_{\nu}}}$ is the averaged $e^{-\tau_{\nu}}$ over
the column density distribution $P(\nh | L,\Lp)$.  Using our fits to
the column density distribution and quasar lifetimes and calculating
$\nu E_{\nu,\,{\rm obs}}$ as above, we reproduce the integrated
observed spectrum quite well (black dashed line). For
comparison, we show that it is not a good approximation to redden the
spectrum with a constant \NH, giving the results for
$\nh=10^{21},10^{21.5},10^{22},10^{22.5},\ {\rm and}\ 10^{23}\ {\rm
cm^{-2}}$ (thin dot-dashed lines).  Even allowing for a distribution
of \NH\ values, the resulting spectrum is a poor match to the observed
one if that distribution is taken to be static (i.e.\
luminosity-independent, as in traditional torus models, for
example).  We show the results of reddening the intrinsic spectrum by
such a (Gaussian) distribution, varying the dispersion $\sigNH=0.4,\,0.7,\,1.0$
(blue, green, and red dashed lines, respectively), for a median column
density $\meanNH=10^{22}\ {\rm cm^{-2}}$, the median column density
expected around $\Lp$ in this simulation. Therefore, the luminosity and host system
property dependence of both quasar lifetimes and the column density
distribution must be accounted for in attempting to properly predict
the X-ray background spectrum from observations of the quasar
luminosity function.  Finally, note that the hard cutoff in the
observed UV spectra at 912\AA\ owes to our calculated cross-sections
being incomplete in the extreme UV.  Properly modeling the escape
fraction and observed emission at these frequencies, while not
important for the X-ray background, is critical to calculating the
contribution of quasars to reionization, and requires a more
detailed modeling of scattering and absorption, especially in the
bright optical quasar phase.

\subsection{The Integrated X-Ray Background}
\label{sec:xrb.tot}

Given the volume emissivity $j_{\nu}(z)$ (per unit {\em
comoving} volume) of some isotropic process at a given frequency at
redshift $z$, the resulting background specific intensity at frequency
$\nu_{0}$ at $z=0$ is \citep{Peacock99}
\begin{equation}
I_{\nu_{0}} = \frac{c}{4\pi}
\int{\frac{j_{\nu}[(1+z)\nu_{0},z]}{(1+z)H(z)}} {\rm d}z.
\end{equation}
If we were to consider the emissivity $j_{\nu}$ per unit
physical volume, there would be an extra factor of $(1+z)^{-3}$ in the
integral above.  In \S~\ref{sec:xrb.indiv}, we determined the
integrated observed energy $E_{\nu,\, \rm obs}(\Lp)$ produced by a
quasar with peak luminosity $\Lp$. We have also inferred $\nLp(z)$
in \S~\ref{sec:fullLF}, the rate at which quasars of peak luminosity
$\Lp$ are created per unit comoving volume per unit cosmological
time. Therefore, the comoving volume emissivity is just
\begin{equation}
j_{\nu}(z) = \int{E_{\nu,\, \rm obs}(\Lp)\,\nLp\,d\log \Lp}, 
\end{equation}
or, expanding $E_{\nu,\, \rm obs}$, 
%\begin{equation}
%\begin{split}
%j_{\nu}(z) = & \int{d\log \Lp\int{d\log L}} \\
% & \times f_{\nu}\EV{e^{-\tau_{\nu}}}\, L\,\frac{dt(L,\Lp)}{{\rm d}\log L}\,\nLp.
%\end{split}
%\end{equation}
\begin{eqnarray}
j_{\nu}(z)& = & \int{d\log \Lp\int{d\log L}} \nonumber\\
 && \times f_{\nu}\EV{e^{-\tau_{\nu}}}\, L\,\frac{dt(L,\Lp)}{{\rm d}\log L}\,\nLp.
\end{eqnarray}
If the column density distribution were independent of
$\Lp$, as is assumed in even luminosity-dependent torus models or
observationally determined \NH\ functions used for X-ray background
synthesis \citep[e.g.,][]{Ueda03}, then we could combine terms in
$\Lp$ and integrate over them. This simplification, along with the
definition of the luminosity function in terms of $\Lp$, gives the
more traditional formula for the X-ray background in terms of only the
observed column density distribution and luminosity function,
\begin{equation}
j_{\nu}(z) = \int{d\log L\,\frac{d\Phi}{d\log L} L_{\nu}\EV{e^{-\tau_{\nu}}}}. 
\label{eq:simple.XRB}
\end{equation}
However, as we showed in \S~\ref{sec:NHdistrib} and
\S~\ref{sec:xrb.indiv}, neglecting the dependence on $\Lp$ is not a
good approximation at all luminosities and gives an inaccurate
estimate of the integrated quasar spectrum; therefore, ``purely
observation-based'' synthesis models of the X-ray background will be
inaccurate in a similar manner to synthesis models with an
inappropriate model for the quasar lifetime. Essentially, this
``averages out'' the varying distribution of column densities with
$\Lp$, which changes the shape of the spectrum in a non-linear manner,
especially when integrated over varying bolometric corrections as
shown above.

%\clearpage
\begin{figure}
    \centering
    %\epsscale{.7}
    \plotone{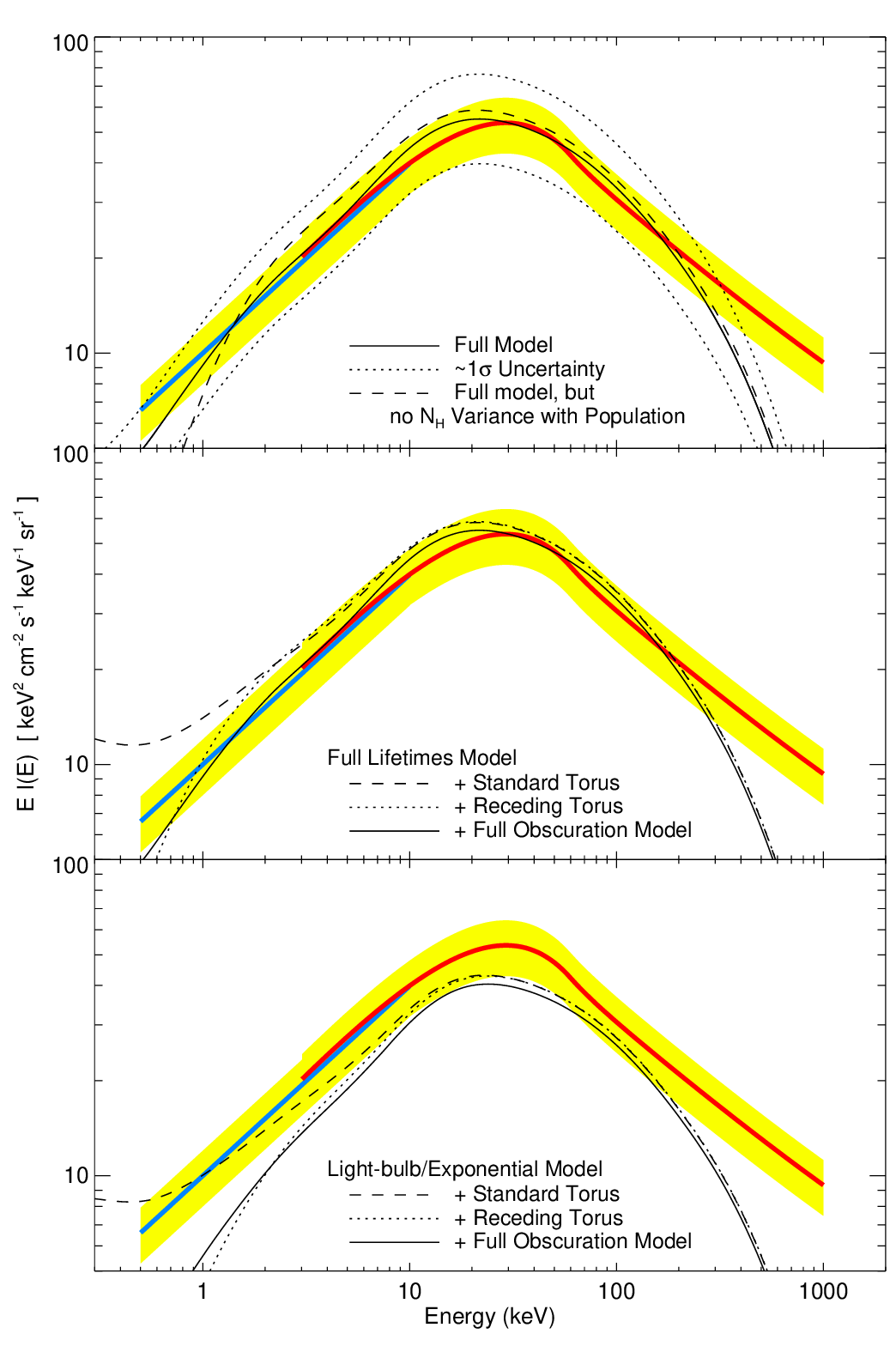}
    \caption{Predicted integrated X-ray background spectrum (solid
    black line) from our model of quasar lifetimes and attenuation,
    with the peak luminosity distribution $\nLp$ determined from the
    luminosity function. Blue and red thick lines show the observed spectrum
    from \citet{Barcons00} and \citet{Gruber99}, respectively. The
    shaded yellow area illustrates the uncertainty in normalization
    between both samples (alternatively, $2\sigma$ errors in the
    Barcons et al.\ 2000 normalization). The predictions given
    $1\sigma$ deviations in the fitted $\nLp$ distribution (dotted
    lines) and given the $\nLp$ distribution determined from hard
    X-ray data only (dashed line) are shown in the upper panel. Middle 
    panel shows the prediction using our modeling of quasar lifetimes 
    but different models of obscuration, lower panel the prediction with a 
    ``light-bulb'' or exponential (constant Eddington ratio) model
    and different obscuration models.
    \label{fig:XRB.spectrum}}
\end{figure}
%\clearpage

Figure~\ref{fig:XRB.spectrum} (upper panel) shows the predicted X-ray
background spectrum from our full modeling of quasar lifetimes and
obscuration (solid lines). We use our analytical fits to the quasar
lifetime and column density distributions as in \S~\ref{sec:xrb.indiv}
above, as Figure~\ref{fig:integ.spectra} demonstrates that they
accurately reproduce the actual integrated quasar X-ray spectra of the
simulations, and the analytical forms are integrated over all
luminosities and redshifts.  The dotted lines show the deviation
resulting from shifting the parameters describing our fitted $\nLp$
distribution by $1\sigma$ in either direction, although degeneracies
in the parameters suggest that the actual uncertainty in the
background prediction is smaller.  The dashed line shows the predicted
X-ray background if we ignore the broadening of the \NH\ distribution
across simulations ($\sigNH=1.2$) and instead consider only the
dispersion of an individual simulation at a given luminosity
($\sigNH=0.4$).

These can be compared to the observations of \citet{Gruber99} (red
curve, for $E\geq3\,{\rm keV}$) and \citet{Barcons00} (cyan curve, for
$E\leq10\,{\rm keV}$). We increase the normalization of the
\citet{Gruber99} spectrum to match that of the best estimate from
\citet{Barcons00} over the range of overlap, determined from combined
{\it ASCA, BeppoSAX,} and {\it ROSAT} data to be
$10.0^{+0.6}_{-0.9}\,{\rm keV\,cm^{-2}\,s^{-1}\,sr^{-1}\,keV^{-1}}$ at
1\,keV. The uncertainty in the normalization between the two samples,
$\sim20\%$, is shown as the shaded yellow range (alternatively, this
represents the $\sim2\sigma$ errors in the {\it ROSAT} normalization).

In the middle panel of the figure, we calculate the predicted X-ray
background using our full model of the quasar lifetime, but with
different models for quasar obscuration. The solid black line shows
the prediction using our full model of quasar obscuration, and is
identical to the solid black line in the upper panel. The observations
are likewise shown in an identical manner to the upper panel. The
dashed black line is the prediction adopting the standard torus model
for quasar obscuration, and the dotted line adopts the receding
(luminosity-dependent) torus model.  These models produce the same
overall $\sim30\,$keV normalization, as this is relatively unaffected
by obscuration, but they predict a slightly ($\sim20\%$) higher
background at low energies, giving a slightly softer spectrum. This
may appear counterintuitive, given that in Figure~\ref{fig:NH.distrib}
these models tend to overpredict the number of high-column density
sources, but this is because these models predict a strongly {\em
bimodal} column density distribution, with unobscured sightlines
encountering negligible column densities. These unobscured sightlines
dominate the soft X-ray integrated spectrum, where the large column
densities through the torus attenuate the quasar spectrum heavily.
However, this net offset in the predicted background spectrum is
generally within the range of the systematic theoretical and
observational uncertainties, and can further be alleviated by tuning
the parameters of the torus model to fit the X-ray background spectrum
(e.g.\ Treister \& Urry 2005, although their fits require a larger
fraction of Compton-thick $\nh\sim10^{25}\,{\rm cm^{-2}}$ sources than
shown for even the receding torus model in
Figure~\ref{fig:NH.distrib}).  The feature at
$\lesssim5\,$keV in the standard torus model prediction is a
consequence of assuming that ``unobscured'' lines of sight encounter
negligible column density, and does not appear if such sightlines
encounter moderate ($\sim10^{21}\,{\rm cm^{-2}}$) columns.

The lower panel of the figure shows the predicted X-ray background
spectrum if we instead consider a light-bulb or exponential light
curve (fixed Eddington ratio) model for the quasar lifetime, again
with various descriptions of quasar obscuration.  In such models, the
predicted X-ray background spectrum is independent of the quasar
lifetime or characteristic Eddington ratio assumed (see
Equation~\ref{eq:simple.XRB}). However, as shown in
Figure~\ref{fig:integ.spectra}, these models do imply a different
integrated spectrum for quasars; i.e.\ different effective bolometric
corrections for predicting the X-ray background. In particular, in
this model, the observed quasar spectrum at a given luminosity
(averaged over the quasar population at that luminosity) is the same
as the ``effective'' quasar spectrum one would use to calculate the
{\em total} contribution to the X-ray background from quasars of the
corresponding observed or peak luminosity, whereas this is not true in
our model of quasar lifetimes.  The observations are shown in the same
manner as the preceding panels.  The black solid line shows the
prediction with this simplified model for the quasar lifetime, but
still adopting our full model for obscuration as a function of
instantaneous and peak luminosity, the dashed line assumes instead a
standard torus model for obscuration, and the dotted line assumes a
receding torus for the obscuration. The variations among different
obscuration models are relatively small at most energies, and similar
to those discussed above adopting our full model of quasar
lifetimes. 

In all three cases, however, this model for the quasar lifetime
significantly under-predicts the X-ray background, particularly at the
$\sim30\,$keV peak.  This shortfall is well-known, and earlier attempts
\citep[e.g.,][]{Madau94,Comastri95,Gilli99,Gilli01,Pompilio00,Ueda03}
have generally had to invoke additional assumptions about large
obscured populations or a strong increase in the obscured fraction
with redshift, neither of which is consistent with observations
\citep[e.g.,][]{Hasinger02,Barger03,Ueda03,Szokoly04,Barger05}.  The
difference between the predictions of various quasar lifetime models
is, as explained above, attributable to the difference between the
integrated quasar spectrum produced in our full model of the quasar
lifetime (in which quasars spend long periods of time at low
luminosities, with harder X-ray spectra), and the integrated spectrum
in these simplified quasar lifetime models, which is proportional to
the {\em instantaneous} quasar spectrum, and therefore underpredicts
the hard X-ray portion of the spectrum by as much as $\sim50\%$.

Our prediction of the X-ray background agrees well with the observed
spectrum over the range $\sim1-100\,{\rm keV}$.  (At energies above
$100\,{\rm keV}$ it is likely that processes we have not included,
such as those involving magnetic fields, contribute significantly to
the background.) Unlike previous synthesis models for the X-ray
background, we are able to do so without invoking assumptions about
large Compton thick populations or larger obscured populations at
different redshifts.  In part, this is because our modeling allows us
to predict, based on $\nLp$ and our column density formulation, the
population of Compton thick sources (see
Figure~\ref{fig:NH.distrib}). However, as we have demonstrated, it is
primarily because the deficit in previous synthesis models can be
attributed to their inability to properly account for the dependence
of quasar lifetimes and attenuation on both the instantaneous quasar
luminosity and the host system properties (peak luminosity).  Our
picture, on the other hand, yields an estimate for the X-ray
background spectrum that is simultaneously consistent with the
observed supermassive black hole mass distribution and total density,
as well as the ``luminosity-dependent density evolution'' observed in
X-ray samples \citep{H05f}. The background is primarily built up from
$z\sim2.5$ to $z\sim0.5$, as is evident from the evolution of the black
hole mass density in Figure~\ref{fig:SMBH.distrib}, although a harder
spectrum at low luminosities will weight this slightly towards lower
redshifts (where more low-luminosity quasars are forming).  Compton
thick and relaxing, low-luminosity sources are accounted for, not as
large, independent populations, but as evolutionary phenomena
continuously connected to the ``normal'' quasar population.

\section{Discussion}
\label{sec:discussion}

\subsection{General Implications of our Model}
\label{sec:genimpl}

Our modeling suggests two important paradigm shifts in interpreting
quasar populations and evolution:

{\bf (1)} First, as proposed in \citet{H05c}, a proper accounting of
the luminosity dependence of quasar lifetimes (as opposed to models in
which quasars grow in a pure exponential fashion or turn on and off as
``light bulbs'') implies a novel interpretation of the luminosity
function.  The steep bright end (luminosities above the ``break'' in
the luminosity function) consists of quasars radiating near their
Eddington limits and is directly related to the distribution of
intrinsic peak luminosities (or final black hole masses) as has been
assumed previously.  However, the shallow, faint end of the luminosity
function describes black holes either growing in early stages of
activity or in extended, quiescent states going into or coming out of
a peak bright quasar phase, with Eddington ratios generally between $l
\sim0.01$ and 1. The ``break'' luminosity in the luminosity function
corresponds directly to the {\em peak} in the birthrate of quasars
as a function of peak luminosity $\nLP$.

This interpretation resolves inconsistencies in a number of previous
theoretical studies. For example, semi-analytical models of the quasar
luminosity functions \citep[e.g.,][]{KH00,HM00,WL03} assume, based on
simplified models for the quasar lifetime, that quasars at the faint
end of the luminosity function correspond to low final-mass black
holes (low $\Lp\sim L$), presumably in small halos.  Consequently,
these models overpredict the number of active low-mass black holes (as
estimated from radio source counts), especially at high redshift, by
orders of magnitude \citep{HQB04}, and overpredict the number of
low-mass spheroids and red galaxies observed \citep{H05e}.

Moreover, both observations \citep{MD04} and comparison of the
present-day black hole mass function with radio and X-ray luminosity
functions \citep[e.g.][]{Merloni04} suggest anti-hierarchical
evolution for the growth of supermassive black holes, where the most
massive black holes were produced mainly at high ($z\gtrsim2$)
redshift, and low-mass black holes mostly formed later, which does
not follow from idealized descriptions of quasar lifetimes and the
luminosity function (for a review, see e.g.\ Combes 2005).

A one-to-one correspondence between observed luminosity and black hole
mass does produce anti-hierarchical behavior in some sense at the
high-mass end, because the most massive black holes are formed at
$z\sim2-3$ during the peak of bright quasar activity and the quasar
luminosity function evolves to lower luminosities at lower redshifts
(as is also the case for our model because the bright end of the
luminosity function is dominated by sources near their peak
luminosities). However, at black hole masses equal to or below
$\sim10^{8}\,M_{\sun}$ (i.e.\ galaxies of stellar mass
$\lesssim10^{11}\,M_{\sun}$), the evolution in the quasar luminosity
function implies a roughly constant production of black holes with
these masses at all redshifts, which is inconsistent with observations
of galaxy spheroids indicating that typical ages increase with mass,
ruling out a large population of low-mass spheroids with ages equal to
or older than those of high-mass spheroids
\citep[e.g.,][]{Treu01,vanDokkum01,Treu02,
vDS03,Gebhardt03,Rusin03,vandeVen03,Wuyts04,Treu05,Holden05,vdW05,SA05,
Nelan05}.  As demonstrated in Figure~\ref{fig:SMBH.evol}, such a model
does not produce anti-hierarchical growth or any age gradients within
the high-mass spheroid population, also inconsistent with
observations.  Even given observed ``luminosity-dependent density
evolution''
\citep[e.g.][]{Page97,Miyaji00,Miyaji01,LaFranca02,Cowie03,
Ueda03,Fiore03,Hunt04,Cirasuolo05,HMS05}, implying that the densities
of lower redshift quasars peak at lower redshift, the inferred
anti-hierarchical evolution if observed luminosity directly
corresponds to black hole mass (i.e.\ as in ``light-bulb'' or ``fixed
Eddington ratio'' models) is not strong enough to account for observed
anti-hierarchical growth of the corresponding galaxy spheroids
\citep{H05e}.

Furthermore, in these earlier models, a ``break'' in the luminosity
function is not necessarily reproduced \citep{WL03}, and the faint-end
slope has no direct physical motivation.  The break may be caused by
feedback mechanisms which set a characteristic turnover in both the
galaxy mass function and quasar luminosity function (e.g., Scannapieco
\& Oh 2004; Dekel \& Birnboim 2004), as in our modeling.  The $\nLp$
distributions in our model and ``light bulb'' or ``fixed Eddington
ratio'' models are comparable at and above the break in the quasar
luminosity function, and therefore make similar predictions for some
observations at these luminosities.  However, the {\em faint-end}
slope has a different physical motivation in our model.  Unlike the
bright-end slope, which is determined directly by the active final
black hole mass function or peak luminosity distribution (in
essentially all models of the quasar lifetime), the faint-end slope in
our modeling is a consequence of the quasar lifetime as a function of
luminosity, and is a prediction of our simulations and modeling almost
independent of the underlying faint-end slope of the active black hole
mass function or peak luminosity distribution. In \citet{H05f} we
examine this in more detail, and demonstrate that it predicts well the
evolution in the faint-end quasar luminosity function slope with
redshift and the observed ``luminosity-dependent density evolution''
in many samples \citep{Page97,Miyaji00,Miyaji01,LaFranca02,Cowie03,
Ueda03,Fiore03,Hunt04,Cirasuolo05,HMS05}.

Other observational evidence for our picture exists; for example in
the observed distribution of Eddington ratios (see
\S~\ref{sec:eddington}), the distribution of low-redshift, active
black hole masses (see \S~\ref{sec:BLmasses}), and the turnover in the
expected distribution of black hole masses in early-type galaxies at
$\sim10^{8}\,M_{\sun}$ \citep[e.g.,][]{Sheth03}.  Total (integrated)
quasar lifetimes estimated from observations are inferred to increase
with increasing black hole mass as we predict \citep{YT02}, and
furthermore, the Eddington ratios of observed quasar samples are seen
to increase systematically with redshift, as the sample becomes
increasingly dominated by luminosities above the break in the
luminosity function \citep{MD04}.

Moreover, observations show that the evolution of the luminosity
function with decreasing redshift is driven by a decrease in the
characteristic mass scale of actively accreting black holes
\citep[e.g.,][]{Heckman04}, which can be explained in our model by the
relation of the observed luminosity function to the {\em peak} in the
distribution of active black hole masses $\nLp$. This observation,
however, has caused considerable confusion, as observations of both
radio-quiet \citep{WooUrry02} and radio-loud \citep{ODowd02} local
(low redshift) AGN indicate that nuclear and host luminosities are
uncorrelated, implying that nuclear luminosity does not depend on
black hole mass \citep{Heckman04}, and therefore that the primary
variable determining the nuclear luminosity is the Eddington ratio,
with the luminosity function spanning a broad range in Eddington
ratios \citep{Hao05}.  Furthermore, observations show that this is
{\em not} true of high redshift quasars, as both direct estimates of
accretion rates \citep[e.g.,][]{Vestergaard04,MD04} and the fact that
their high luminosities would yield unreasonably large black hole
masses rule out substantially sub-Eddington accretion rates for most
objects.  Many previous empirical and semi-analytical models could not
simultaneously account for these observations.  To explain just the
low-redshift observations, such models adopt tunable distributions of
Eddington ratios fitted to the data.  However, both these observations
are consequences of our interpretation of the luminosity function, as
observations of local AGN and the low-redshift luminosity function are
dominated by quasars below the break in the luminosity function, which
are undergoing sub-Eddington growth and span a wide range of Eddington
ratios, while observations at high redshift are dominated by bright
objects at or above the break in the luminosity function, which are
undergoing Eddington-limited (or near Eddington-limited) growth near
their peak luminosity (see \S~\ref{sec:eddington}).

{\bf (2)} The second paradigm shift indicated by our modeling is that
quasar obscuration is not a static or quasi-static geometric effect,
but is primarily an {\em evolutionary} effect. The physical reasoning
for this is simple: the massive gas inflows required to fuel quasar
activity produce large obscuring columns, and so column densities are
correlated with quasar luminosity. The basic picture of buried quasar
activity associated with the early growth of supermassive black holes
and starburst activity has been proposed previously and studied for
some time \citep[e.g.,][]{SM96,Fabian99}, but our modeling allows us
to describe the evolution of obscuration in a self-consistent manner,
defining obscured and unobscured phases appropriately and identifying
dynamical correlations between the column density distribution and
instantaneous and peak luminosities.

There is substantial observational support for this picture.  Point-like X-ray
sources have been observed in many bright sub-millimeter or infrared
and starburst sources, with essentially all very luminous infrared
galaxies showing evidence of buried quasar activity
\citep[e.g.,][]{SM96,Komossa03,Ptak03}, indicating simultaneous buried
black hole growth and star formation at redshifts corresponding to
peak quasar activity ($z\gtrsim1$) \citep{Alexander05a,Alexander05b}.
The buried black holes in high-z starbursting galaxies appear to be
active but undermassive compared to the quiescent galaxy black
hole-stellar mass relation \citep{Borys05}, implying that they are
rapidly growing in the starburst but have not yet reached their final
masses, presumably set in the subsequent ``blowout'' phase.
Similarly, observations suggest that obscured AGN are significantly
more likely to exhibit strong sub-millimeter emission characteristic
of star formation, implying both that obscured black hole growth and
star formation are correlated and that obscuration mechanisms
(responsible for re-radiation in the submm and IR) may be primarily
isotropic in at least some cases \citep[e.g.,][]{Page04,Stevens05}.
Evidence from quasar emission line structure
\citep[e.g.,][]{Kuraszkiewicz00,Tran03}, directly related to the inner
broad-line region, suggests that isotropic obscuration of quasars can
be important, in contradiction to angle-dependent models.  Finally,
many observations
\citep[e.g.,][]{Steffen03,Ueda03,Hasinger04,GRW04,sazrev04,Barger05,Simpson05}
indicate that the fraction of broad-line or obscured quasars is a
function of luminosity, which cannot be accounted for in traditional
static ``torus'' models \citep[e.g.,][]{Antonucci93} or reproduced
even by modified luminosity-dependent torus models \citep{Lawrence91},
an observation that is explained by our model (see
\S~\ref{sec:BLqso} for a detailed discussion).

Much of the obscuration in our modeling comes from large scales,
arising from the inner regions of the host galaxy on scales
$\sim50\,$pc or larger.  While our resolution limits prevent our
ruling out the possibility of gas collapse to a dense, $\sim\,$pc
scale torus surrounding the black hole, during the peak obscured
phases of the final merger, our simulations indicate that these large
scales dominate the contribution to the column density, with quite
large columns, which should be observationally testable.  Indeed, this
is suggested by the typical scales of obscuration in starbursting
systems (e.g.\ Soifer et al.\ 1984a,b; Sanders et al.\ 1986, 1988a,b;
for a review, see e.g.\ Soifer et al.\ 1987), given that, as discussed
above, the dominant obscured phase of growth is closely associated
with a starburst as implied observationally
\citep{Alexander05a,Alexander05b,Borys05}.  

Observations of polarized light in intrinsically bright Type II AGN
with unobscured luminosities typical of quasars (as opposed to local,
dim Seyfert II objects in relaxed hosts) show scattering on large
scales $\sim$kpc, and in some cases obscuration clearly generated over
scales extending beyond the host galaxy in the form of distortions,
tidal tails, and streams from interactions and major mergers
\citep{Zakamska04,Zakamska05}.  The angular structure seen in these
observations is consistent with our modeling.  Moreover, in optically
faint X-ray quasars \citep[e.g.][]{Donley05} it appears that
obscuration is generated by the host galaxies, and is directly related
to host galaxy morphologies and line-of-sight distance through the
host.  The critical point is that, regardless of the angular structure
of obscuration, typical column densities are strongly evolving
functions of time, luminosity, and host system properties, and the
observed distribution of column densities is dominated by these
effects, not by differences in viewing angle across a uniform
population.  This is the case in our modeling as the lognormal
dispersion (across different lines of sight) in column densities is
$\sigNH\sim0.4$ for a given simulation at some instant, whereas
typical column densities across simulations, as a function of
instantaneous and peak luminosities, span several orders of magnitude
from $\nh\sim10^{18}-10^{26}\ {\rm cm^{-2}}$.

\subsection{Specific Predictions of our Model}
\label{sec:specpred}

Our predictions include:

\begin{itemize}

\item Quasar Lifetimes: We find that for a particular source, the
quasar lifetime depends sensitively on luminosity, with the observed
lifetime in addition depending on the observed waveband. Intrinsic
quasar lifetimes vary from $t_{Q}\sim10^{6}-10^{8}$\,yrs, with
observable lifetimes $\sim10^{7}$\,yrs in optical B-band
\citep{H05a,H05b}, in good agreement with observational estimates
\citep[for a review, see][]{Martini04}.

\item Luminosity Functions: Using a parameterization of the intrinsic
distribution of peak luminosities (final quasar black hole masses) at
a given redshift, our model of quasar lifetimes allows us to reproduce
the observed luminosity function at all luminosities and redshifts
$z=0-6$.  Although this is an empirical determination of the peak
luminosity distribution, it implies a new interpretation of the
luminosity function \citep{H05c}, which provides a physical basis for
the observed ``break'' corresponding to the peak in the peak
luminosity distribution.  Moreover, the faint end slope is not
determined by our empirical fitting procedure, but instead by the
dependence of the quasar lifetime on luminosity, with its value and
redshift evolution predicted by our modeling \citep{H05f}.  The
evolution of typical column densities in different stages of merger
activity produces a significant population of obscured quasars,
accounting for the difference between hard X-ray \citep[e.g.,][]{Ueda03},
soft X-ray \citep[e.g.,][]{Miyaji01}, and optical B-band
\citep[e.g.,][]{Croom04} luminosity functions (\S~\ref{sec:fullLF}).

\item Column Density Distributions: The evolution of the column
densities in our simulations reproduces the observed distribution of
columns in optically-selected quasar samples, when the appropriate
selection criteria are applied \citep{H05b}, as well as complete
column distributions in hard X-ray selected samples
(\S~\ref{sec:NHdistrib}). Column density evolution over the course of
a merger yields a wider observed distribution of columns than that
produced across different viewing angles at a given point in a
merger.

\item Broad Line Luminosity Function and Fraction: Using our
simulations to estimate when quasars will be observable as broad-line
objects (either based on the ratio of quasar to host galaxy optical
B-band luminosity or the obscuring column density), we reproduce the
luminosity function of broad-line quasars in hard X-ray selected
samples as well as optical broad-line quasar surveys, and the fraction
of broad-line quasars in a given sample as a function of luminosity,
to better precision than traditional or luminosity-dependent (but
non-dynamical) torus models which are fitted to the data
(\S~\ref{sec:BLfraction}).  By providing an a priori prediction of the
broad-line fraction as a function of luminosity and redshift which
depends systematically on the typical quasar host galaxy gas fraction,
we propose that observations of the broad line fraction at different
redshifts can be used to constrain the gas fraction of quasar hosts
and its evolution with redshift.

\item Active Black Hole Mass Functions: Using our prescription for
deciding when objects will be visible as ``broad-line'' quasars, we
predict the distribution of low-redshift, broad-line and non-broad
line active quasar masses, in good agreement with observations from
the SDSS, with expected incompleteness in the observed sample at low
$M_{\rm BH}\lesssim10^{6}\,M_{\sun}$ black hole masses
(\S~\ref{sec:BLmasses}).  This is a new prediction which can be tested
in greater detail by future observations, and our calculations allow
us to model the differences in active black hole mass functions of the
Type I and Type II populations.  The width of the expected broad-line
black hole mass function depends significantly on the model of quasar
lifetimes, enabling such measurements to probe the statistics of
quasar evolution.

\item Eddington Ratios: We determine Eddington ratio distributions
from our simulations, given the peak luminosity distribution implied
by the observed quasar luminosity function. The predicted
distribution, once the appropriate observed magnitude limit is
imposed, agrees well with observations at both low ($z<0.5$) and high
($1.5<z<3.5$) redshifts (\S~\ref{sec:eddington}).  As noted above, our
interpretation of the luminosity function explains seemingly
contradictory observations of Eddington ratios at different redshifts.
There is even a suggestion \citep{CX05} that the evolution of quasars
seen in our simulations (with bright phases in mergers and extended
relaxation after) can account for observations of bimodal Eddington
ratio distributions at $z\sim0$ \citep{Marchesini04}, when coupled
with an appropriate description of radiatively inefficient accretion
phases, although it is possible that many of these low-redshift black
holes are not fueled by mergers, especially in e.g.\ low-luminosity
Seyferts.

\item Relic Black Hole Mass Function: With our model for quasar
lifetimes, the luminosity function at a given redshift implies a
birthrate of sources with given peak luminosities, $\nLp$, which
translates to a distribution in final black hole masses.  Integrating
this over redshift, we predict the present-day mass distribution and
total mass density of supermassive black holes. They agree well with
observational estimates inferred from local populations of galaxy
spheroids.  In our picture, these spheroids are produced
simultaneously with the supermassive black holes they harbor
(\S~\ref{sec:smbh}).  We demonstrate that the integrated supermassive
black hole density, quasar flux density, and number counts in
different wavebands can be reconciled with a radiative efficiency
$\epsilon_{r}=0.1$, satisfying the constraints of counting arguments
such as that of \citet{Soltan82}. Further, we show in
\S~\ref{sec:detailsCompare} and \S~\ref{sec:whenBL} that the
corrections to such observational arguments based on optical quasar
samples are small (order unity) when we account for the luminosity
dependence of quasar lifetimes, despite an extended obscured phase of
quasar growth.  In other words, although a quasar spends more time
obscured than it does as a bright optical source, the total mass
growth and radiated energy are dominated by the final ``blowout''
stage visible as a bright optical quasar.

\item X-ray Background: The integrated quasar spectrum from our models
of quasar lifetimes and column densities as a function of
instantaneous and peak luminosities can be combined with the birthrate
of quasars with a given peak luminosity to give the integrated cosmic
background in any frequency range. We predict both the normalization
and shape of the X-ray background from $\sim1-100$\,keV, with our
modeling accounting for quasar obscuration as an evolutionary process
(with a corresponding population of Compton-thick objects), avoiding
any need for arbitrary assumptions about additional obscured
populations (\S~\ref{sec:xrb.tot}). For any model in which the quasar
spectrum depends on luminosity or accretion rate, we demonstrate that
a proper modeling of the quasar lifetime is critical to reproducing
observed backgrounds.

\item Correlation Functions: In \citet{Lidz05}, we predict the quasar
correlation function and bias as a function of redshift and luminosity
using our model, and compare it to that expected using ``light bulb''
or exponential light curves.  As most quasars in our modeling have a
characteristic peak luminosity or final black hole mass corresponding
to the peak of the $\nLp$ distribution, they reside in hosts of
similar mass, and there is little change in bias with luminosity at a
given redshift, in contrast to idealized models for the quasar
lifetime and luminosity function.  Our predicted bias agrees well with
the observations of \citet{Croom05}, who also find no evidence for a
dependence of the correlation on quasar luminosity at a given
redshift, as we expect. In fact, \citet{PMN04} and \citet{Croom05}
find that their observations can be explained if quasars lie in hosts
with a constant characteristic mass $\sim 2\times10^{12}\,M_{\sun}$
($h=0.7$).  If we consider their redshift range $z\sim1-2$, we predict
the quasar population will be dominated by sources with
$\Lp=\lstar(z)$, which given $\mbhf(\Lp)$ and using the $M_{\rm
BH}-M_{\rm halo}$ relation of \citet{WL03} yields a nearly constant
characteristic host halo mass $\sim1-2\times10^{12}\,M_{\sun}$, in
good agreement.  Similarly, \citet{Adelberger05} find that the
quasar-galaxy cross-correlation function does not vary with
luminosity, implying with $\sim90\%$ confidence that faint and bright
quasars reside in halos with similar masses and that fainter AGN are
longer lived, strongly disfavoring traditional ``light bulb'' and
exponential light curve models.  Furthermore, \citet{Hennawi05} find
an order of magnitude excess in quasar clustering at small scales
$\lesssim 40\,h^{-1}\,{\rm kpc}$, with the correlation function
becoming progressively steeper at sub-Mpc scales, suggesting that
quasar activity is triggered by interactions and mergers.

\item Host Galaxy Properties: Because black
hole growth and spheroid formation occur together in our
picture, our modeling allows us to describe
relationships between black hole and galaxy properties.  For example, we
reproduce both the observed $M_{\rm BH}-\sigma$ relation
\citep{DSH05,Robertson05b} and the fundamental plane of elliptical
galaxies (Robertson et al., in preparation).  Since we also reproduce
the distribution of relic black holes inferred from the $z=0$
distribution of spheroid velocity dispersions or luminosity functions
using the observed versions of these relations, our match to these
relations indicates that we also reproduce these distributions of host
spheroid properties.  We consider this in detail in \citet{H05e}, and
find that we are able to account for a wide range of host galaxy
properties, including luminosity and mass functions, color-magnitude
relations, mass-to-light ratios, and ages as a function of size, mass,
and redshift.  With our modeling of the quasar lifetime as motivated
by our simulations, the evolution and distribution of properties of
red-sequence galaxies and the quasar population are shown to be
self-consistent, which is not the case for idealized models of quasar
evolution.

\end{itemize}

Aside from an empirical estimate of the distribution of peak quasar
luminosities $\nLp$, we determine all of the quantities summarized
above self-consistently from the input physics of our simulations,
including a physically motivated dynamic accretion and feedback model
in which black holes accrete at the Bondi rate determined from the
surrounding gas, and $\sim5\%$ of the radiant energy couples thermally
to that gas. Beyond this, our simulations enable us to calculate the
various predictions above {\em a priori}, without the need for
additional assumptions or tunable parameters.

We compare each of these predictions to those obtained using idealized
descriptions of the quasar lifetime, i.e.\ ``light-bulb'' and
exponential light curve (constant Eddington ratio) models, and the
column density distribution, i.e.\ standard and ``receding''
(luminosity-dependent) torus models. We fit all these (along with our
full model) to the observed luminosity function in the same manner
(allowing the same degree of freedom to ensure that they all yield the
same observed luminosity function), and we fit the free parameters of
these tunable models (e.g.\ typical Eddington ratios and quasar
lifetimes for the ``light-bulb'' or exponential models, typical column
densities and torus scalings for the torus models) {\em independently}
to each observation to maximize their ability to reproduce
observations. However, we still find better agreement between our
model (with no parameters tuned to match observations) and the
observations in nearly every case where the tunable phenomenological
model is not guaranteed to reproduce the observation by construction.
The one exception is the relic supermassive black hole mass
function, for which the predictions of our modeling and idealized
lifetime models are essentially identical, reflecting the fact that in
both cases black hole growth is dominated by bright, optically
observable, high Eddington ratio phases.

Moreover, the best-fit parameters for the idealized models, when
fitted independently to each observation, are not self-consistent.
For example, calculations of the black hole mass function imply high
Eddington ratios $l\sim0.5-1$ \citep[e.g.,][]{YT02}, and our fit to
the active black hole mass function \citep{Heckman04} suggests
$l\sim1$, but the observed distribution of accretion shows a typical
$l\sim0.3$ \citep{Vestergaard04}, and fitting to the broad-line
fraction as a function of luminosity with our full obscuration model
but these lifetime models implies a lower $l\sim0.05$.  Likewise,
fitting torus models to the X-ray background suggests typical column
densities through the torus of $\nh\sim10^{25}\,{\rm cm^{-2}}$
\citep[e.g.,][]{TU05}, while fitting to the observed column density
distributions \citep{Treister04,Mainieri05} suggests equatorial
columns $\nh\lesssim10^{24}\,{\rm cm^{-2}}$.  Clearly then,
reproducing the observations listed above, and in particular doing so
self-consistently, is not implicit in any model which successfully
reproduces the quasar luminosity function, even at multiple
frequencies.

\subsection{Further Testable Predictions of our Model}
\label{sec:testpred}

Our model for quasar evolution makes a number of observationally
testable predictions:

\begin{itemize}

\item Quasar lifetimes are only weakly constrained by observations
\citep[e.g.][]{Martini04}, but future studies may be able to measure
both the lifetime of individual quasars and the statistical lifetimes
of quasar populations as a function of luminosity.  We describe in
detail our predictions for the evolution of individual quasars and
quantify their lifetimes in \S~\ref{sec:methods}, and further predict
the distribution of both integrated and differential lifetimes in an
observed sample as a function of luminosity. This should provide a
basis for comparison with a wide range of observations, with the most
important prediction being that the quasar lifetime should increase
with decreasing luminosity.

\item For a reasonably complete, optically selected sample above some
luminosity, the distribution of observed column densities should
broaden to both larger and smaller \NH\ values as the minimum observed
luminosity is decreased, as both intrinsically faint periods with low
column density and intrinsically bright periods with high column
density become observable.

\item Similarly, the Eddington ratio distribution should be a function
of observed luminosity, with a broad distribution of Eddington ratios
down to $l\sim0.01-0.1$ at luminosities well below the break in the
observed luminosity function, and a more strongly peaked distribution
about $l\sim0.2-1$ for luminosities above the break
(Figure~\ref{fig:Pofl.all}).

\item In our interpretation, the bright and faint ends of the
luminosity function correspond statistically to similar mixes of
galaxies, but in various stages of evolution; whereas in all other
competing scenarios, the quasar luminosity is directly related to the
mass of the host galaxy.  Therefore, any observational probe that
differentiates quasars based on their host galaxy properties such as,
for example, the dependence of the clustering of quasars on
luminosity, or the host stellar mass and size as a function of
luminosity (although we caution that this is somewhat dependent of the
modeling of star formation in mergers), can be used to discriminate
our picture from older models. We present a detailed prediction of the
quasar correlation function based on our modeling for comparison with
observations in \citet{Lidz05}.

\item Our distribution $\nLp$ directly translates to a black hole
merger rate, as a function of mass, in our modeling, allowing a
detailed prediction of the gravitational wave signal from black hole
mergers as a function of redshift.

\item The broad line fraction as a function of luminosity, defined by
requiring that ``broad-line'' objects have an observed B-band
luminosity above a fraction $f_{\rm BL}$ of that of their host galaxy,
is a prediction of our model quasar and galaxy light curves. However,
the uncertainties are large, primarily because different observational
samples have varying sensitivity to quasar vs.\ host galaxy optical
light.  Furthermore, the host galaxy gas fraction and $f_{\rm BL}$ are
degenerate in these predictions -- a well-defined observational sample
complete to some $f_{\rm BL}$ can constrain our modeling of quasar
fueling and the relation between quasar and host galaxy light
curves. In particular, such observations, either by measuring the
faint-end shape of the ``broad-line'' quasar luminosity function or
the mean ``broad-line'' fraction at a given luminosity as a function
of redshift, can constrain the gas fractions of quasar host galaxies
and their evolution, essentially a free parameter in our empirical
modeling.

\item We also predict the distribution of active, low-redshift black
hole masses in \S~\ref{sec:BLqso}. These predictions can be compared
to mass functions for active black holes from numerous quasar surveys,
which should include improved mass functions of the entire quasar
population complete to lower luminosities as well as future mass
functions for the population of active broad-line AGN. We provide
predictions for the black hole mass function of all active quasars,
and for just the ``broad-line'' population (as a function of the
survey selection).

\item Because the evolution of spheroids and supermassive black holes
is linked in our modeling, with each affecting the evolution of the
other, we can also use the distribution of observed quasar properties
to predict galaxy properties such as number counts, spheroid masses
and luminosities, and colors as a function of redshift.  For the
calculation and discussion of these predictions, see \citet{H05e}.

\item In our model, the growth of supermassive black holes is
dominated by galaxy mergers.  Therefore, at any given redshift, the
mass (and as a consequence, luminosity) function of galaxy mergers
should have a similar shape to our distribution of quasar birthrates,
$\nLp$, distinct from the shapes of either the quasar or total galaxy
luminosity functions.  Indeed, preliminary observational estimates of
both the merger luminosity function
\citep[e.g.,][]{Xu04,Conselice03,Wolf05} and quasar host galaxy
luminosity function \citep{bkss97,Hamilton02}, primarily at low
redshifts, appear be consistent with this expectation.  Theoretically,
it may be possible to predict the merger luminosity function using
either cosmological simulations or semi-analytical models; we discuss
this further in \S~\ref{sec:finis}.

\end{itemize}

\subsection{Mock Quasar Catalogs}
\label{sec:mockcat}

In principle, our modeling can be used to predict the distributions of
quasar luminosities in various wavebands, column densities, active
black hole masses, and peak luminosities for a wide range of
observational samples, but it is impractical for us to plot
predictions of these quantities for all possible sample selection
criteria.  To enable comparison with a wider range of observations, we
have used our modeling and the conditional probability distributions
for these quantities from our simulations to generate Monte Carlo
realizations of quasar populations, which we provide publicly via ftp\footnote{
\url{ftp://cfa-ftp.harvard.edu/pub/phopkins/qso\_catalogs/}}.  

At a particular redshift, we use our fitted $\nLp$ distribution and
our suite of simulations to generate a random population of mock
``quasars.''  We first generate the peak luminosities of each
``quasar'' according to the fitted $\nLp$ at that redshift.  For each
object, we then use the probability of being at a given instantaneous
luminosity in simulations with a similar peak luminosity to generate a
current bolometric luminosity.  In practice, we calculate the
$P(L\,|\Lp)$ distribution by summing $w(\Lp ,\ L_{\rm peak,\ i})\times
P(L\,| L_{\rm peak,\ i})$, where $\Lp$ is the mock quasar peak
luminosity, $L_{\rm peak,\ i}$ is the peak luminosity of each
simulation and $w(\Lp ,\ L_{\rm peak,\ i})$ is a Gaussian weighting
factor ($\propto\exp(-\log^{2}(\Lp/L_{\rm peak,\ i})/2 (0.05)^{2})$).
Knowing the instantaneous bolometric luminosity $L$ and peak
luminosity $\Lp$, we then follow an identical procedure to determine
the joint distribution $P(X\,|\,L,\,\Lp)$ of each subsequent quantity
$X$, from simulations with similar $L$ and $\Lp$.  We have compared
this with Monte Carlo realizations based on our fitted probability
distributions in this paper, and find that essentially identical
results are achieved for e.g.\ the distribution of $L$ and $\Lp$, and
column densities in phases of growth not near peak luminosity.
However, this modeling is not identical for e.g.\ the distribution of
Eddington ratios and column densities around $L\sim\Lp$, which
reflects the fact that our fits to the Eddington ratio distribution
(\S~\ref{sec:eddington}) are rough and that our fits to the column
density distribution do not apply to the final ``blowout'' phase of
quasar evolution (as discussed in detail in \S~\ref{sec:BLqso}).

For each mock quasar, we generate a peak luminosity, final
(post-merger) black hole mass, instantaneous bolometric luminosity,
intrinsic (un-attenuated) B-band ($\nu L_{\nu}$ at $\nu=4400$\AA),
soft X-ray (0.5-2 keV), and hard X-ray (2-10 keV) luminosity, observed
(attenuated using the generated column density and the reddening/dust
extinction modeling described in \S~\ref{sec:NH}, with SMC-like
reddening curves and extinction following e.g.\ Pei 1992, Morrison \&
McCammon 1983) B-band, soft X-ray, and hard X-ray luminosities, column
density of neutral hydrogen, column density of neutral+ionized
hydrogen, and instantaneous black hole mass. The intrinsic
luminosities in each band are calculated using the bolometric
corrections described in \citet{Marconi04}, which account for the
luminosity dependence of the optical-to-X-ray luminosity ratio
$\alpha_{\rm OX}$ (as discussed in \S~\ref{sec:fullLF}), and then
attenuated to give the observed luminosities. We also provide
intrinsic and attenuated luminosities in each waveband using the
constant bolometric corrections of \citet{Elvis94}, but we caution
that these are not calculated in a completely self-consistent manner,
as our assumed bolometric luminosity function to which we fit the
$\nLp$ distribution is based on using the luminosity-dependent
bolometric corrections. We do not directly calculate Eddington ratios,
as these are defined differently in many observed samples, but they
should be calculable with the given luminosities and black hole
masses.

We calculate these quantities for a mock sample of $\sim10^{9}$
quasars at each redshift $z=0.2,\ 0.5,\ 1,\ 2,\ {\rm and}\ 3$. Most of
these quasars are at luminosities orders of magnitude below those
observed, therefore for space considerations and because our
predictions become uncertain at low luminosities, we retain only the
$10^{6}$ quasars with brightest bolometric luminosities at each
redshift.  This introduces some uncertainty in our statistics at the
lowest luminosities in any given band, but these luminosities are
generally still well below those observed in most samples.  At any
luminosity, but especially at the brightest luminosities, there is
also a significant amount of effective ``noise'' owing to our
incomplete sampling of the enormous parameter space of possible
mergers, and decreasing total time across simulations spent at large
luminosities, which can be estimated from e.g.\
Figures~\ref{fig:LF.hists} and \ref{fig:Ueda.BL}.  Finally, at each
redshift, we generate two distributions, reflecting the $\sim1\sigma$
range in $\nLp$, and roughly parameterizing the degeneracies in our
fit to the observed luminosity functions and uncertainty in the
faint-end of $\nLp$ -- ``Fit 1'' has a lower $\lstar$ (lower peak in
$\nLp$), with a larger $\sstar$ (broader $\nLp$ distribution), and
``Fit 2'' has a higher $\lstar$ and smaller $\sstar$ (more narrowly
peaked $\nLp$ distribution).  We show a few example ``quasars'' from
our $z=0.2$ mock distribution in Table~\ref{tbl:montecarlo}, to
demonstrate the format and units used.

\subsection{Starburst Galaxies}
\label{sec:relburst}

Although we do not yet model the re-radiation of absorbed light by
dust or the contribution of stellar light to quasar host IR
luminosities, including these in our picture for quasar evolution will
enable us to predict luminosity functions in the IR and sub-mm and
their evolution with redshift.  We can at this point, however,
estimate if our model for quasar lifetimes and merger-driven evolution
with $\nLp$ is consistent with the observed distribution of
ultraluminous infrared galaxies.  Naively, we might expect that since
the obscured quasar phase has a duration up to $\sim10$ times that of
the optically observable quasar phase, there should be $\sim10$ times
as many ULIRGs as bright optical QSOs.  But, this neglects the
complicated, luminosity dependent nature of quasar lifetimes.

Given that the bright quasars we simulate attain, during their peak
growth phase, an intrinsic luminosity comparable to that of the host
starburst, and that this period of peak growth has a similar duration
to the starburst phase \citep[see Figure~\ref{fig:BL.in.sims}
and][]{DSH05,SDH05b}, we can estimate (roughly) the ULIRG bolometric
luminosity function from our bolometric quasar luminosity function.
Thus, the more accurate comparison to the ULIRG luminosity function is
with the hard X-ray quasar luminosity function, as this recovers (and
at some luminosities can be dominated by) ``buried'' quasars in
starburst phases.  This is only applicable {\em above} the break in
the luminosity function, where quasars are undergoing peak quasar
growth.  Below the break, quasars are, on average, sub-Eddington and
can have luminosities well below that of their star-forming hosts (see
Figure~\ref{fig:BL.in.sims}), so we expect our quasar luminosity
function to be significantly shallower than the ULIRG luminosity
function at these luminosities.  Note also that this does not imply
that ULIRGs are all AGN-dominated, as the starburst and peak AGN
activity can be (and generally are) somewhat offset, but only says
that the lifetime curves at the bright end should be similar.

Considering the luminosity function at $z=0.15$, then, we expect ULIRG
densities $d\Phi/{\rm d} M_{\rm bol}\sim 3\times10^{-7}\ {\rm and}\
9\times10^{-8}\ {\rm Mpc^{-3}\,mag^{-1}}$ at $L\sim1.6\times\Lcut{12}$
and $2.5\times\Lcut{12}$, respectively.  These estimates are
consistent with the observed density in the {\it IRAS} 1 Jy Survey
\citep{Kim98} at a mean redshift $z=0.15$, with ${\rm d}\Phi/{\rm d}
M_{\rm bol}\sim 5\times10^{-7},\ 7\times10^{-8}\ {\rm
Mpc^{-3}\,mag^{-1}}$ (rescaled to our cosmology), and as expected, our
quasar luminosity function slope becomes significantly shallower than
the observed 1 Jy survey luminosity function slope below $L\sim
10^{11}-10^{12}\,L_{\sun}$, roughly the break luminosity of the
luminosity function.  We predict these densities to change with
redshift according to the evolution of $\nLp$, decreasing by a factor
$\sim1.5$ at $z=0.04$, in good agreement with the evolution of IRAS
ULIRG luminosity functions \citep{Kim98}. Likewise, at $z\sim1-3$, we
predict a mean space density $\Phi(L>\Lcut{11})\sim1-3\times10^{5}\
{\rm Mpc^{-3}}$, in agreement with the $\sim5\times10^{5}\ {\rm
Mpc^{-3}}$ density of such sources expected to reproduce the observed
cumulative source density $4\times10^{4}\,{\rm deg^{-2}}$ of 1\,mJy
$850\,\mu{\rm m}$ SCUBA sources \citep{Barger99}. Furthermore, our
prediction of the fraction of buried AGN and its evolution with
redshift agrees well with determinations from X-ray samples
\citep{Barger05} and recent Spitzer results in the mid and
near-infrared at $z\sim2$ \citep{Martinez05}.

\subsection{AGN not Triggered by Mergers}
\label{sec:nonmerger}

Some low redshift quasars (e.g.\ Bahcall et al.\ 1996) and many
nearby, low-luminosity Seyferts appear to reside in ordinary,
relatively undisturbed galaxies.  Our picture for quasar evolution
does not immediately account for these objects because we suppose that
nuclear activity is mainly triggered by tidal torques during a merger.

This work is primarily concerned with the origin of the
majority of the mass in spheroids and supermassive black holes, and as
a consequence, the relation of this to the abundance and
evolution of quasars and the cosmic X-ray background.  Based on our
present analysis, we believe that a merger-driven picture can account
for the main part of each of these, and, as described earlier, that the
most relevant phase in the history of the Universe to these phenomena
appears to have been at moderate redshifts, $z\sim2.5$ to $z\sim0.5$.

Our model does not exclude other mechanisms for triggering AGN and it
is likely that a variety of stochastic or continuous processes are
relevant to nuclear activity in undisturbed disks and residual
low-level accretion in relaxed systems.  This is not contrary to our
picture because most of the total black hole mass density in the
Universe is in spheroid-dominated systems. The principal requirement
of our model is that AGN activity in undisturbed galaxies should not
contribute a large fraction of the black hole mass density in the
Universe, to avoid spoiling tight correlations between the black hole
and host galaxy properties and producing too large a present-day black
hole mass density in violation of the Soltan (1982) constraint.

For example, if a molecular cloud passed through the center of our Galaxy near Sgr
A$^{\ast}$, it is possible that the Milky Way would resemble a Seyfert
for some period of time.  Alternatively, it has long been recognized
that mass loss from normal stellar evolution of bulge stars or stellar
clusters near the centers of galaxies can provide a continuous supply
of fuel for low-level accretion \citep[e.g.,][]{MLC81,MB81,Shull83}.
Typical galactic stellar mass loss rates
($\dot{M}\sim1\,M_{\sun}\,{\rm yr^{-1}}\,(10^{11}\,M_{\sun})^{-1}$)
yield Bondi-Hoyle accretion rates $\sim10^{-5}-10^{-4}$ of Eddington
in relaxed, dynamically hot systems; and mass loss rates from O and
W-R stars ($\dot{M}\sim10^{-6}\,M_{\sun}\,{\rm
yr^{-1}}\,(10\,M_{\sun})^{-1}$) in young, dense star clusters near the
centers of galaxies with sufficient cold gas for continued star
formation can yield rates as high as $\sim10^{-2}$ of Eddington.

Even though these fueling mechanisms do not involve mergers, the scenario
we have discussed might still be relevant to the origin of these black holes. Of
course, the black holes and spheroids in disk-dominated systems may
have produced in a manner that did not involve mergers.
Alternatively, most of the black hole mass in these objects (which is
small compared to that in spheroid-dominated galaxies) could have been
assembled long ago in mergers with bright quasar phases and then these
``dead'' quasars are resurrected sporadically by other fueling
mechanisms.

Independent of how these black holes were formed, elements of our
modeling may still account for certain observed properties of Seyferts.  The
observed Seyfert luminosity function appears to join smoothly onto the
quasar luminosity function \citep{Hao05}. It is not obvious that this
would be the case if the two types of objects are produced by entirely
distinct mechanisms. In addressing this, it is useful to
separate the process by which gas is delivered to the black hole from
the subsequent evolution that determines the observed activity. In
our picture, gas is delivered to the black hole by gravitational
torques during a merger, but other mechanisms, like bar-induced
fueling may be important for objects such as Seyferts. Regardless,
the induced activity may be generic, if black hole growth is
self-regulated in the way we describe it in our simulations.

In Hopkins et al. (2005f) we show that the faint end slope of the
quasar luminosity function in our model is partly determined by the
time dependence of the ``blowout'' phase of black hole growth.  We
derive an analytical model for this using a Sedov-Taylor type analysis
and show that the impact of this feedback depends on the mass of the
host. This analysis does not depend on the fueling mechanism, only on
the subsequent evolution. If this self-regulated growth applies to
Seyferts as well (for example if Seyfert growth is regulated by a
balance between accretion feedback and the spheroid potential, as
expected if these objects obey a similar $M_{\rm BH}-\sigma$
relation), we would expect the Seyfert luminosity function to smoothly
join onto the quasar one, even if the fuel is delivered in a different
manner.

\section{Conclusions}
\label{sec:finis}

We have studied the evolution of quasars in simulations of galaxy
mergers spanning a wide region of parameter space.  In agreement with
earlier work \citep{H05a}, we find that the lifetime of a particular
source depends on luminosity and increases at lower luminosities, and
that quasar obscuration is time-dependent.  Our new, large set of
simulations shows that the lifetime and obscuration can be expressed
in terms of the instantaneous and peak luminosities of a quasar and
that these descriptions are robust, with no systematic dependence on
simulation parameters.  We have combined these results with a
semi-empirical method to describe the cosmological distribution of
quasar properties, allowing us to predict a large number of
observables as a function of e.g.\ luminosity and redshift. This
approach also makes it possible to compare our picture to simpler
models for quasar lifetimes and obscuration.

In the model we examine, quasars are triggered by mergers of gas-rich
galaxies, which produce inflows of gas through gravitational torquing,
fueling starbursts and rapid black hole growth.  The large gas
densities obscure the central black hole at optical wavelengths until
feedback energy from the growth of the black hole ejects gas and
rapidly slows further accretion (``blowout'').  Quasar lifetimes and
light curves are non-trivial, with strong accretion activity during
first passage of the merging galaxies and extended quiescent
(sub-Eddington) phases leading into and out of the phase of peak
quasar activity associated with the final merger.  The ``blowout''
phase in which the quasar is visible as a bright, near-Eddington
optical source has a lifetime related to the dynamical time in the
inner regions of the merging galaxies, which characterizes the
timescale over which obscuring gas and dust are expelled, but the
quasar spends a longer time at lower luminosities before and after
this stage.  These evolutionary processes have important consequences
which cannot be captured in models of pure exponential or ``on/off''
quasar growth.

Our work emphasizes several goals for quasar and galaxy observations
and theory. Observationally, it is important to constrain the faint
end of the {\em peak} luminosity distribution; i.e.\ the low-mass
active black hole distribution.  Unfortunately, our modeling of quasar
lifetimes implies that the faint-end quasar luminosity function is
dominated by quasars with peak luminosities around the break in the
luminosity function, and can provide only weak constraints on the
faint-end $\Lp$ distribution.  However, there is still hope, as for
example broad-line quasar activity is more closely associated with
near-peak luminosities, and thus probing the faint-end of broad-line
luminosity functions may in particular improve the estimates.
Moreover, studies of the black hole mass distribution (or the
distribution of galaxy spheroids) as a function of redshift, extending
to small spheroid masses/velocity dispersions probes the faint end of
$\nLp$.  Other techniques, such as studies of faint radio sources at
high redshift \citep{HQB04} can similarly constrain these populations.
Furthermore, the calculations in this paper can be combined to better
determine $\nLp$, as, given a model for the quasar lifetime and
obscuration, they all derive from this fundamental quantity.
Additional observational tests of the modeling we have presented will
provide an important means of constraining models for AGN accretion
and feedback; for example, the faint-end slope of the quasar lifetime
depends on how the ``blowout'' phase occurs and could provide a
sensitive probe of feedback models, enabling the adoption of more
realistic and sophisticated feedback prescriptions than we have thus
far employed.  Of course, improved constraints on the luminosity
function at all luminosities at high redshift remains a valuable means
of testing theories of quasar evolution.

Our simulations are based on isolated galaxy mergers, and thus do not
provide a cosmological prediction for the distribution of peak
luminosities $\nLp$, merger rates, or mass functions - we instead have
adopted a semi-empirical model, in which we use our modeling of quasar
evolution to determine these distributions from the observed
luminosity function.  While this allows us to predict a large number
of observables and to demonstrate that a wide range of quasar and
galaxy properties are self-consistent in a model of merger-driven
quasar activity with realistic quasar lifetimes, future theoretical
work in these areas should predict the distribution of peak
luminosities $\nLp$ and its evolution with redshift. These quantities
are to be distinguished from the distribution of observed
luminosities, as the two are not trivially related in our model or any
other with a non-trivial quasar lifetime.

Although the quasar birthrate as a function of peak luminosity will
be, in general, a complicated function of galaxy merger rates, gas
fractions, morphologies, and other factors, we have parameterized it
for comparison with the results of future cosmological simulations and
semi-analytical models. This distribution is particularly valuable as
a theoretical quantity because it is more directly related to physical
galaxy properties than even the complete (intrinsic) luminosity
function, and additionally because theoretical modeling which
successfully reproduces this $\nLp$ distribution is guaranteed to
reproduce the large number of observable quantities we have discussed
in detail in this work.  We cannot determine the cosmological context
in our detailed simulations of the relatively small-scale physics of
galaxy mergers, and conversely, cosmological simulations and
semi-analytical models cannot resolve the detailed physics driving
quasar activity in mergers. However, our determination of quasar
evolution as a function of peak luminosity or final black hole mass
can be grafted onto these cosmological models to greatly increase the
effective dynamic range of such calculations.  Combined with our
modeling, this would remove the one significant empirical element we
have adopted, and allow for a complete prediction of the above
quantities from a single theoretical framework.

In these efforts, we emphasize that the mergers relevant to our
picture are of a specific type. First, the merging galaxies must
contain a supply of cold gas in a rotationally supported disk.  Hot,
diffuse gas, as in the halos of elliptical galaxies, will not be
subject to the gravitational torques which drive gas into galaxy
centers and fuel black hole growth. Clearly, gas-poor mergers are also
not important for this process. Second, the mergers will likely
involve galaxies of comparable, although not necessarily equal, mass,
so that the gravitational torques excited are strong enough and
penetrate deep enough into galaxy centers to drive substantial inflows
of gas. The precise requirement for the mass ratio is somewhat
ill-defined because it also depends on the orbit geometry, but mergers
with a mass ratio larger than $10 : 1$ are probably not generally
important to our model. Simulations of minor mergers involving
galaxies with mass ratios $\lesssim 10:1$ (e.g.\ Hernquist 1989; Hernquist
\& Mihos 1995) have shown that for particular orbital geometries,
these events can produce starbursts similar to those in major mergers,
leaving behind disturbed remnants with dynamically heated disks (e.g.\
Quinn et al.\ 1993; Mihos et al.\ 1995; Walker et al.\ 1996).  It is
of interest to establish whether black hole growth can also be
triggered in minor mergers, as these events may be relevant to weak
AGN activity like that in some Seyfert galaxies or LINERs.

In summary, the work presented here supports the conjecture that many
aspects of galaxy formation and evolution can be understood in terms
of the ``cosmic cycle'' in Figure~\ref{fig:cosmiccycle}.  To be sure,
much of what is summarized in Figure~\ref{fig:cosmiccycle} has been
proposed elsewhere, either in the context of observations or
theoretical models.  Our modeling of galaxy formation and evolution
emphasizes the possibility that supermassive black holes could be {\it
responsible} for much of what goes on in shaping galaxies, rather than
being bystanders, closing the loop in Figure~\ref{fig:cosmiccycle}.
In this sense, black holes may be the ``prime movers'' driving galaxy
evolution, as has been proposed earlier for extragalactic radio
sources (e.g.\ Begelman, Blandford \& Rees 1984; Rees 1984).  It may
seem counterintuitive that compact objects with masses much smaller
than those of galaxies could have such an impact, but it is precisely
the concentrated nature of matter in black holes that makes this idea
plausible.

Consider a black hole of mass $M_{\rm BH}$ at the center of a
spherical galaxy of mass $M_{\rm sph}$ with a characteristic velocity
dispersion $\sigma$.  The energy available to affect the galaxy
through the growth of the black hole will be some small fraction of
its rest-mass, $E_{\rm feed} \sim \epsilon_f M_{\rm BH} c^2$.  This
can be compared with the binding energy of the galaxy, $E_{\rm bind}
\sim M_{\rm sph} \sigma^2$.  Observations indicate that $M_{\rm BH}$
and $M_{\rm sph}$ are correlated and that, roughly $M_{\rm BH} \sim
(0.002 - 0.005) M_{\rm sph}$ (Magorrian et al.\ 1998; Marconi \& Hunt
2003).  Therefore, the ratio of the feedback energy to the binding
energy of the galaxy is $E_{\rm feed} / E_{\rm sph} > 10
\epsilon_{f,-2} \, \sigma_{300}^{-2}$, for an assumed efficiency of
1\%, $\epsilon_{f,-2} \equiv \epsilon /0.01$ and scaling the velocity
dispersion to $\sigma_{300} \equiv \sigma / 300$ km/sec, as for
relatively massive galaxies.  This result demonstrates that the
supermassive black holes in the centers of spheroidal galaxies are by
far the largest supply of potential energy in these objects, exceeding
even the galaxy binding energy.  When viewed in this way, if even a
small fraction of the black hole radiant energy can couple to the
surrounding ISM, then black hole growth is not an implausible
mechanism for regulating galaxy formation and evolution; in fact, it
appears almost inevitable that it should play this role.

\acknowledgments We thank our referee, David Weinberg, for many
comments and suggestions that greatly improved this paper.  We
thank Paul Martini, for 
helpful discussion, and Gordon Richards and Alessandro Marconi, for
generously providing data for observational comparisons.  This work
was supported in part by NSF grants ACI 96-19019, AST 00-71019, AST
02-06299, and AST 03-07690, and NASA ATP grants NAG5-12140,
NAG5-13292, and NAG5-13381.  The simulations were performed at the
Center for Parallel Astrophysical Computing at the Harvard-Smithsonian
Center for Astrophysics.

\clearpage
%\begin{deluxetable}{llllllllllll}
\begin{deluxetable}{cccccccccccc}
\tabletypesize{\scriptsize}
\tablecaption{Mock Quasar Distribution Examples \label{tbl:montecarlo}}
\tablewidth{0pt}
\tablehead{
\colhead{\tablenotemark{1}$\Lp$} & 
\colhead{\tablenotemark{2}$M_{\rm BH}^{f}$} & 
\colhead{\tablenotemark{3}$L$} & 
\colhead{\tablenotemark{4}$M_{\rm BH}$} & 
\colhead{\tablenotemark{5}$\nh$} &
\colhead{\tablenotemark{6}$\nhi$} &
\colhead{\tablenotemark{7}$L_{B}^{\rm i}$} & 
\colhead{\tablenotemark{8}$L_{SX}^{\rm i}$} & 
\colhead{\tablenotemark{9}$L_{HX}^{\rm i}$} & 
\colhead{\tablenotemark{10}$L_{B}^{\rm obs}$} & 
\colhead{\tablenotemark{11}$L_{SX}^{\rm obs}$} & 
\colhead{\tablenotemark{12}$L_{HX}^{\rm obs}$} 
}
\startdata
10.6 & 6.1 & 8.5 & 6.1 & 20.5 & 20.1 & 7.2\ \ 7.5 & 7.4\ \ 6.8 & 7.6\ \ 7.0 & 7.2\ \ 7.4 & 7.4\ \ 6.8 & 7.6\ \ 7.0  \\
10.4 & 6.4 & 8.7 & 6.0 & 22.2 & 22.0 & 7.4\ \ 7.6 & 7.6\ \ 7.0 & 7.8\ \ 7.2 & 5.8\ \ 6.0 & 7.5\ \ 6.9 & 7.8\ \ 7.2  \\
... & & & & & & & & & & & 
\enddata
\tablenotetext{1}{Peak quasar bolometric luminosity, $\log_{10}(\Lp/L_{\sun})$}
\tablenotetext{2}{Final (post-merger) black hole mass, $\log_{10}(M_{\rm BH}^{f}/M_{\sun})$}
\tablenotetext{3}{Current (at time of ``observation'') intrinsic (no attenuation) 
bolometric luminosity, $\log_{10}(L/L_{\sun})$}
\tablenotetext{4}{Current black hole mass, $\log_{10}(M_{\rm BH}/M_{\sun})$}
\tablenotetext{5}{Total (neutral+ionized) hydrogen 
column density along the ``observed'' sightline, $\log_{10}(N_{\rm H}/{\rm cm}^{-2})$}
\tablenotetext{6}{Neutral hydrogen 
column density along the ``observed'' sightline, $\log_{10}(N_{\rm H\, I}/{\rm cm}^{-2})$}
\tablenotetext{7}{Intrinsic (no attenuation) B-band luminosity, $\log_{10}(L_{B}^{\rm i}/L_{\sun})$, 
where $L_{B}=\nu_{B}L_{\nu_{B}}$ at $\nu_{B}=4400$\AA.} 
\tablenotetext{  \   }{Calculated with the luminosity-dependent 
bolometric corrections from (Marconi et al.\ 2004; left), }
\tablenotetext{  \   }{and constant (luminosity-independent) $L=11.8\,L_{B}$ (Elvis et al.\ 1994; right).}
\tablenotetext{8}{Intrinsic soft X-ray (0.5-2 keV) luminosity, $\log_{10}(L_{SX}^{\rm i}/L_{\sun})$.}
\tablenotetext{   \  }{Calculated with the luminosity-dependent 
bolometric corrections from (Marconi et al.\ 2004; left), }
\tablenotetext{   \  }{and constant (luminosity-independent) $L=52.5\,L_{SX}$ (Elvis et al.\ 1994; right).}
\tablenotetext{9}{Intrinsic hard X-ray (2-10 keV) luminosity, $\log_{10}(L_{HX}^{\rm i}/L_{\sun})$.}
\tablenotetext{  \   }{Calculated with the luminosity-dependent 
bolometric corrections from (Marconi et al.\ 2004; left), }
\tablenotetext{  \   }{and constant (luminosity-independent) $L=35.0\,L_{HX}$ (Elvis et al.\ 1994; right).}
\tablenotetext{10}{``Observed'' (with attenuation) B-band luminosity, $\log_{10}(L_{B}^{\rm obs}/L_{\sun})$.}
\tablenotetext{  \   }{Left and right use luminosity-dependent and luminosity-independent bolometric corrections, respectively, as $L_{B}^{\rm i}$.}
\tablenotetext{11}{``Observed'' soft X-ray luminosity, $\log_{10}(L_{SX}^{\rm obs}/L_{\sun})$.}
\tablenotetext{  \   }{Left and right use luminosity-dependent and luminosity-independent bolometric corrections, respectively, as $L_{SX}^{\rm i}$.}
\tablenotetext{12}{``Observed'' hard X-ray luminosity, $\log_{10}(L_{HX}^{\rm obs}/L_{\sun})$.}
\tablenotetext{  \   }{Left and right use luminosity-dependent and luminosity-independent bolometric corrections, respectively, as $L_{HX}^{\rm i}$.}
\tablenotetext{  \   }{ \ }
\tablenotetext{  \   }{The complete tables 
can be downloaded at \url{ftp://cfa-ftp.harvard.edu/pub/phopkins/qso\_catalogs/}}
\end{deluxetable}

\end{document}